\documentclass[12pt,a4paper]{report}

\usepackage[T1]{fontenc}
\usepackage{epsfig}
\usepackage{natbib}

\usepackage{setspace} 
\usepackage{sysbio}
\usepackage{graphicx}
\usepackage{url}
\usepackage[none]{hyphenat}

\begin{document}

\begin{center}
\centering
\textsc{Evolutionary Placement of Short Sequence Reads}

\vspace{2cm}

Performance \& Accuracy of Evolutionary Placement Algorithms for Short Sequence Reads under Maximum Likelihood

\vspace{4cm} 

Simon A.~Berger, Alexandros Stamatakis

\textit{The Exelixis Lab, Dept.~of Computer Science Technische Universit\"at M\"unchen Boltzmannstr.~3, 85748 Garching b.~M\"unchen, Germany}


\end{center}

\newpage

\raggedright
\parindent=1cm
\section*{Abstract}

We present an Evolutionary Placement Algorithm (EPA) for the rapid assignment of sequence fragments (short reads) to branches of 
a given phylogenetic tree under the Maximum Likelihood (ML) model. 
The accuracy of the algorithm is evaluated on several real-world data sets and 
compared to placement by pair-wise sequence comparison, using edit distances and BLAST. 

We test two versions of the placement algorithm, one slow and more accurate where branch length optimization is conducted for each short read insertion
and a faster version where the branch lengths are approximated at the insertion position. 
For the slow version, additional heuristic techniques are explored that almost yield the same run time as the fast version, with only a small loss of accuracy. 
When those additional heuristics are employed the run time of the more accurate algorithm is comparable to that of a simple BLAST search for data sets with a 
high number of short query sequences. Moreover, the accuracy of the Evolutionary Placement Algorithm is significantly higher, in particular when the 
taxon sampling of the reference topology is sparse or inadequate. Our algorithm, which has been integrated into RAxML, therefore provides an equally 
fast but more accurate alternative to BLAST for phylogeny-aware analysis of short-read sequence data.

{\textbf Keywords:} Maximum Likelihood, short sequence reads, phylogenetic placement, RAxML, metagenomics

\newpage

\section*{}
Identification of organisms from, e.g., microbial communities, through analysis of their DNA has become an important analysis process in current biology. 
Recently, the advent of new wet-lab techniques such as pyrosequencing (\cite{ronaghi2001psl}) has increased the amount of sequence data available for 
identification and analysis of microbial communities by orders of magnitude. 
This rapid increase in available sequence data poses new challenges for short-read sequence identification tools. 
We can no longer expect that the steady increase in computing power according to Moore's law 
is fast enough to be able to handle this {\em biological data flood} computationally.

A single sequencing run can already generate more than 100,000 short read sequences that comprise 
sequence fragments with a length of approximately 30 to 450 nucleotides (base pairs) depending on the sequencer 
used. Such sequencing runs can be carried out within about an hour. 
Besides rapid full-genome assembly, another important application is the 
sampling of microbial communities, e.g., in permafrost-affected soils (\cite{Ganzert2007}), 
the human and vertebrate gut (\cite{ley05complete,turnbaugh2008cgm,ley2008www}) 
(with important implications on health and nutrition), hypersaline mats (\cite{ley06complete}), or 
on hands (\cite{fierer2008ish}) with respect to hand hygiene and health.

Given the short read sequences, the first step in the analysis of those microbial communities consists in identifying the anonymous reads, i.e.,
a lot of short sequences are available, but we do not know to which known organism they are most closely related to. 
This assignment of the short reads to know organisms, then allows to reason about the composition of the microbial communities,
to determine the microbial diversity, and allows for comparison of the microbial communities between different samples (see~\cite{turnbaugh2008cgm}). 
For instance, in one sample 20\% of the reads might be most closely related to a specific taxonomic group of bacteria, 
while in a different sample (e.g., from a different gut) only 5\% may be associated to this group.

Here we present a novel algorithm, the Evolutionary Placement Algorithm (EPA), for rapid phylogenetic identification of anonymous reads 
(denoted as Query Sequences (QS)), using a set of full length Reference Sequences (RS). 
The most straight-forward approach for identifying the QS is to use tools that are based on sequence similarity such as BLAST. 
However, such a BLAST based approach exhibits an important limitation: 
It can yield misleading assignments of QS to RS, if the RS sample does not contain sequences that are sufficiently closely related to the QS, i.e., 
if the taxon sampling is sparse or inappropriate. 
Any approach based on sequence similarity like BLAST, which is based on pair-wise sequence comparison will not unravel, but silently ignore, 
potential problems in the taxon sampling of the RS. 
For instance, given two RS $a$ and $b$, a QS $q$ may be identified as being most closely related to $a$ by BLAST. 
In reality $q$ might be most closely related to a RS $c$ which is not included in the reference sequence set. 
Since this is a known problem (\cite{koski2001cbh}), many studies of microbial communities employ phylogenetic (evolutionary) methods for QS identification, 
despite the significantly higher computational cost. 
If a QS falls into an inner branch of the phylogenetic reference tree comprising the RS, i.e., it is not located near a leave of the tree,
this indicates that the sampling of the RS is insufficient to identify and capture the diversity of the QS. 
This also provides information about the clades of the reference tree for which the taxon sampling 
is sparse and indicates on which part(s) of the tree sequencing efforts should focus to improve taxon sampling.

To date, phylogeny-based evolutionary identification is conducted as follows: the QS are aligned with respect to a Reference Alignment (RA) for the RS, and then 
inserted into the Reference Tree (RT) either via a complete {\em de novo} tree reconstruction, a constrained tree search, using the RT as a constraint or backbone, 
or some fast and approximate QS addition algorithm as implemented, e.g., in ARB (\cite{arb2004fullref}) using Maximum Parsimony (MP). 
For DNA barcoding, phylogeny-based Bayesian analysis methods have recently been proposed (\cite{munch2008bayesian} and \cite{nielsen2006statistical})
that are however applied to significantly smaller trees with less taxa. 


The current standard approach for analysis of environmental reads yields a fully resolved bifurcating (binary) 
tree that often comprises more than 10,000 sequences (\cite{fierer2008ish,turnbaugh2008cgm}).
The alignments used to reconstruct these trees mostly comprise only a single gene, typically 16S or 18S rRNA. 
The reconstruction of such large trees with thousands of taxa, based on data from a single gene is time-consuming and hard
because of the weak phylogenetic signal, i.e., the reconstruction accuracy decreases for trees with many but relatively 
short sequences (see~\cite{Olaf2000, Moret2002}). Moreover, in metagenomic data sets, a large number of taxa in the alignment (the query sequences),
will only have a length of approximately 200-450 base pairs if a 454 sequencer is used. 
Thus, for identification of short read QS, the lack of phylogenetic signal becomes even more prevalent and critical if a comprehensive tree is reconstructed. 
As an example for the lack of signal and topological stability 
in such hard-to-analyze single gene data sets with many taxa, we may consider the pair-wise topological Robinson Foulds 
distances (\cite{robinson1981cpt}) for a collection of Maximum Likelihood (ML~\cite{felsenstein81}) trees that can not be 
statistically distinguished from each other via the 
standard significance tests (\cite{goldman2000lbt}).  
For a collection of 20 ML trees inferred with RAxML (\cite{AlexandrosStamatakis08232006}) 
on a single-gene rRNA data set with 4,114 taxa the average pair-wise RF distance between the ML trees was approximately 30\%. 
Hence, in order to solve the problems associated to the lack of signal and to significantly accelerate the analysis process, 
we advocate a different approach, that only computes the optimal insertion position for every QS in the RT with respect to its Maximum Likelihood  score.

We introduce a new algorithm for the phylogenetic placement of QS and thoroughly test the placement accuracy on seven previously published DNA and one protein data set. 
We assess the impact of QS length on placement accuracy and also conduct tests on short paired-end reads. 
Because phylogenetic placement is inherently more compute intensive than simple sequence based placement, 
performance optimization is an important factor in the development of such an algorithm if it is to become 
a useful and fast alternative to BLAST. 
Therefore, we have devised several evolutionary placement algorithms and heuristics with varying degrees of computational complexity. 

The algorithm which has been developed and tested in cooperation with microbial biologists is already available in the latest open-source 
code release of RAxML (\cite{AlexandrosStamatakis08232006}) 
(version 7.2.1, released in July 2009, \url{http://wwwkramer.in.tum.de/exelixis/software.html}). 
Our new algorithmic approach represents a useful, scalable, and fast tool for evolutionary (phylogenetic) 
identification of environmental QS. We are not aware of any comparable algorithm that can perform the task described here, in particular
on trees with more than 10,000 taxa.

\section*{Evolutionary Placement Algorithm}
\label{methods_heuristics}

\subsection*{The Maximum Likelihood Model}
\label{ml}

The input of a standard phylogenetic analysis consists of a multiple sequence alignment with $n$ taxa and
$m$ alignment columns (sites). The output is an \emph{unrooted} binary tree.
In order to compute the likelihood on a fixed tree topology one also needs several ML model parameters:
the instantaneous nucleotide substitution matrix $Q$ which contains the transition probabilities for time $dt$ between nucleotide (4x4 matrix), 
the prior probabilities of observing the nucleotides, e.g., $\pi_A, \pi_C, \pi_G, \pi_T$ for DNA data, 
which can be determined empirically from the alignment, and the $\alpha$ shape parameter that forms part of the $\Gamma$ model (\cite{yang94}) of rate heterogeneity.
The $\Gamma$ model accounts for the biological fact that different columns in the alignment evolve at different speeds, and finally the $2n-3$ branch lengths. 
The CAT approximation of rate heterogeneity~(\cite{stamatakis2006cat}) can be used as an efficient and accurate computational work around for $\Gamma$, since
it requires four times less memory and is three to four times faster than phylogenetic inferences under the $\Gamma$ model.
We want to emphasize, that the CAT approximation represents a ``quick and dirty'' work around for $\Gamma$ and should not be confused 
with mixture models (\cite{lartillot2004bayesian}).

Given all these parameters, in order to compute the likelihood of a fixed \emph{unrooted} binary tree topology,  
initially one needs to compute the entries for all internal probability vectors (located at the inner nodes) 
that contain the probabilities $P(A),P(C),P(G),P(T)$ of observing an \verb|A,C,G,| or \verb|T| at each site $c$ of the 
input alignment at the specific inner node, bottom-up from the tips towards a virtual root that can be placed into any branch of the tree. 
This procedure is also know as the Felsenstein pruning algorithm (\cite{felsenstein81}).
Under certain standard model restrictions (time-reversibility of the model) the overall likelihood score will be the 
same regardless of the placement of the virtual root.

Every probability vector entry $\vec{L}(c)$ at a position $c$ ($c=1...m$) $\vec{L}(c)$ at the tips and at the inner nodes
of the tree topology, contains the four probabilities P(A), P(C), P(G), P(T) of observing a nucleotide \verb|A, C, G, T| at  
a specific site $c$ of the input alignment.
The probabilities at the tips (leaves) of the tree for which observed data {\em is} available are set to 1.0 
for the observed nucleotide character at the respective position $c$, 
e.g., for the nucleotide \verb|A|: $\vec{L}(c)=[1.0,0.0,0.0,0.0]$. 
Given a parent node $k$, and two child nodes $i$ and $j$ (with respect to the virtual root), their probability vectors $\vec{L}^{(i)}$ 
and $\vec{L}^{(j)}$, the respective branch lengths leading to the children $b_i$ and $b_j$ and the transition probability matrices 
$P(b_i), P(b_j)$, the likelihood of observing an A at position $c$ of the ancestral (parent) vector $\vec{L}_A^{(k)}(c)$
is computed as follows:

\begin{small}
\begin{equation}\label{formula_lrec}
\vec{L}_A^{(k)}(c) = \big(\sum_{S=A}^T P_{A S}(b_i) \vec{L}_{S}^{(i)}(c)\big)\big(\sum_{S=A}^T P_{A S}(b_j) \vec{L}_{S}^{(j)}(c)\big)
\end{equation}
\end{small}

The transition probability matrix $P(b)$ for a given branch length is obtained from $Q$ by $P(b)=e^{Qb}$.
Once the two probability vectors $\vec{L}^{(i)}$ and $\vec{L}^{(j)}$ to the left and right of the virtual root ($vr$) have been computed, 
the likelihood score $l(c)$ for an alignment column $c, c=1...m$ can be calculated as follows, given the branch length 
$b_{vr}$ between nodes $i$ and $j$:
\begin{small}
\begin{equation}\label{formula_root}
l(c) =  \sum_{R=A}^T \big(\pi_R \vec{L}_{R}^{(i)}(c) \sum_{S=A}^T P_{R S}(b_{vr})\vec{L}_{S}^{(j)}(c)\big)
\end{equation}
\end{small}
The overall score is then computed by summing over the per-column log likelihood scores as indicated in equation~\ref{formula_likelihood}.
\begin{small}
\begin{equation}\label{formula_likelihood}
LnL = \sum_{c=1}^m log(l(c))
\end{equation}
\end{small}

In order to compute the {\em Maximum} Likelihood value for a fixed tree topology
all individual branch lengths, as well as the parameters of the $Q$ matrix and the $\alpha$ shape parameter,  
must also be optimized via an ML estimate. 
For the $Q$ matrix and the $\alpha$ shape parameter the most common approach in state-of-the-art ML implementations consists in using 
Brent's algorithm (\cite{brent1973}). For the optimization of branch lengths the Newton-Raphson method is commonly used. 
In order to optimize the branches of a tree, the branches are repeatedly
visited and optimized one by one until the achieved likelihood improvement (or branch length change) is smaller than some pre-defined $\epsilon$. 
Since the branch length is optimized with respect to the likelihood score, the Newton-Raphson method
only operates on a single pair of likelihood vectors $\vec{L}^{(i)}, \vec{L}^{(j)}$ at a time that define the  branch to be optimized.
Evidently, when a branch of the tree is updated this means that a large number of probability vectors $\vec{L}$ in the 
tree are affected by this change and hence need to be re-computed. 

An important implementation issue is the assignment of memory space for the probability vectors to inner nodes of the tree.
There exist two alternative approaches: a separate vector can be assigned to each of the three outgoing branches of an inner
node, or only one vector can be assigned to each inner node. In the latter case, which is significantly more memory-efficient, 
the probability vectors always maintain a rooted view of the tree, i.e., they are oriented towards the current virtual root of the tree.
In the case that the virtual root is then relocated to a different branch (for instance to optimize the respective branch length), 
a certain number of vectors, for which the orientation to the virtual root has changed need to be re-computed.
If the tree is traversed in an intelligent way for branch length optimization, the number of probability vectors that will 
need to be re-computed can be kept to a minimum. 
RAxML uses this type of rooted probability vector organization.

\subsection*{Evolutionary Placement Algorithm}
\label{algo}

The input for our evolutionary identification algorithm consists of a reference tree comprising the $r$ RS (Reference Sequences), 
and a large comprehensive alignment that contains the $r$ RS {\em and} the $q$ QS (Query 
Sequences).
The task of aligning several QS with respect to a given reference rRNA alignment can for instance be accomplished with
ARB (\cite{arb2004fullref}) or NAST (\cite{desantisjr2006nms}).
One key assumption is, that the Reference Tree (RT) is biologically well-established or that it has been obtained via a preceding
thorough ML analysis.

Initially, the algorithm will read in the RT and reference alignment and mark all sequences of the 
alignment that are {\em not} contained in the reference tree as QS.
Thereafter, the ML model parameters and branch lengths on the reference tree will be optimized using the standard 
procedures implemented in RAxML. 

Once the model parameters and branch lengths have been optimized on the reference tree, the actual identification algorithm is invoked.
It will visit the $2r-3$ branches of the reference tree in depth first-order, starting at an arbitrary branch of the tree 
leading to a tip. At each branch, initially the probability vectors of the reference tree to the left and the right will be re-computed (if they are not already
oriented towards the current branch). Thereafter, the program will successively insert (and remove again) one QS at a time into the current branch and 
compute the likelihood (we henceforth denote this as insertion score) of the respective tree containing $r+1$ taxa. 
The insertion score will then be stored in a $q \times (2r-3)$ table that keeps track of the insertion scores for all $q$ QS 
into all $2r-3$ branches of the reference tree.

\begin{figure}[ht]
  \centering
  \includegraphics[width=0.45\columnwidth]{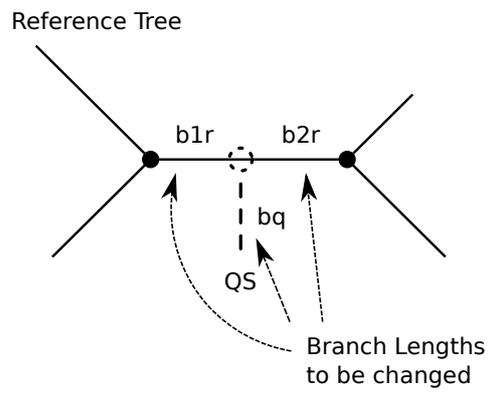}
  \caption{\label{figBranch} Local Optimization of branch lengths for the insertion of a Query Sequences (QS) into the reference tree.}
\end{figure}

In order to more rapidly compute the per-branch insertions of the QS we use an approximation that is comparable
to the Lazy Subtree Rearrangement (LSR) moves that are deployed in the standard RAxML search algorithm (see~\cite{raxml2} for details).
After inserting a QS into a branch of the RT we would normally need to re-optimize all branch lengths of the 
thereby obtained new---extended by one QS---tree topology to obtain the {\em Maximum} Likelihood insertion score. 
Instead, we only optimize the three branches adjacent to the insertion node of the QS (see Figure~\ref{figBranch}) 
before computing the likelihood of the insertion, based
on the same rationale used for the design of the LSR moves. Our experimental results justify this approximation. 
In analogy to the LSR moves we also use two methods to 
re-estimate the three branches adjacent to the insertion branch, a fast method that does not make use of the Newton-Raphson method
and a slow method. The {\em fast} insertion method simply splits the branch of the reference tree $b_r$ into two parts $b_{r1}$ and $b_{r2}$ by setting 
$b_{r1}:=\sqrt{b_r}$, $b_{r2}:=\sqrt{b_r}$, and the branch leading to the QS $b_q:=0.9$, where $0.9$ is the default RAxML value 
to initialize branch lengths.
The {\em slow} method repeatedly applies the Newton-Raphson procedure to all three branches $b_{r1}, b_{r2}, b_q$ until 
no further application of the Newton-Raphson branch length optimization procedure will induce a branch length 
change $>\epsilon$, where $\epsilon := 0.00001$.
Alternatively, our algorithm can also use Maximum Parsimony to pre-score and order promising candidate insertion branches 
in order to further accelerate the placement process. 

\begin{figure}[ht]
  \centering
  \includegraphics[width=0.45\columnwidth]{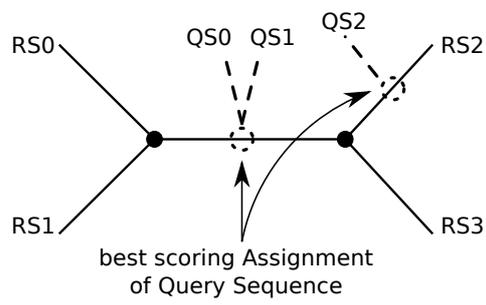}
  \caption{\label{class1} Evolutionary identification of 3 Query Sequences (QS0, QS1, QS2) using a 4-taxon reference tree.}
\end{figure}

The output of this procedure for evolutionary identification of QS consist of
the input RT, enhanced by assignments of the QS to the branches of the RT.
Each QS is attached to the branch that yielded the best insertion score for the specific QS.
Hence, the algorithm will return a potentially multi-furcating tree, if two or or more QS are assigned to the 
same branch. An example output tree for 4 RS and 3 QS is depicted in Figure~\ref{class1}. 

\begin{figure}[ht]
  \centering
  \includegraphics[width=0.45\columnwidth]{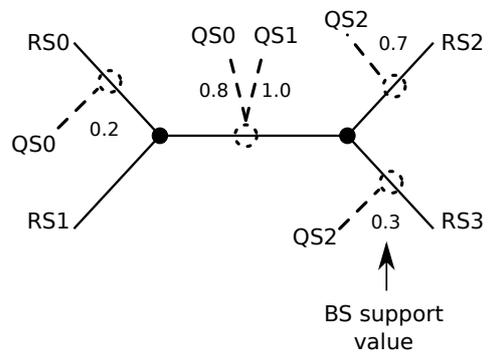}
  \caption{\label{class2} Phylogenetic Placement of 3 query sequences (QS0, QS1, QS2) into a 4-taxon reference tree
with Bootstrap support.}
\end{figure}

Moreover, the EPA algorithm can also conduct a standard phylogenetic bootstrap (\cite{felsenstein1985clp}), i.e., 
repeat the evolutionary identification procedure several times under slight perturbations of the 
input alignment. This allows to account for uncertainty in the placement of the QS as shown in Figure~\ref{class2}.
Thus, a QS might be placed repeatedly into different branches of the reference tree with various levels of 
support. For the Bootstrap replicates we introduce additional heuristics to accelerate the insertion process.
During the insertions on the original input alignment we keep track of the insertion scores for {\em all} QS
into {\em all} branches of the reference tree. For every QS we can then sort the insertion branches by their 
scores and for each Bootstrap replicate only conduct insertions for a specific QS into 10\% of the best-scoring insertion branches on the original alignment. 
This reduces the number of insertion scores to be computed per QS on each bootstrap replicate by 90\% and therefore approximately yields
a ten-fold speedup for the bootstrapping procedure. In a typical application scenario, one may determine the diversity of the environmental sample
for every replicate using for instance UniFrac (\cite{lozupone2005unifrac}), and then compute an average diversity over all replicates.

Alternatively, one could directly use the insertion likelihoods of the QS on the original alignment to compute 
placement uncertainty, i.e., to determine a placement area, rather than a single branch, 
by applying, e.g., the SH-test (\cite{shimodaira1999multiple}).

In analogy to the heuristics used for accelerating the bootstrapping process, we can also 
improve the performance for computing QS insertion scores on the original alignment.
In order to improve the run time of the {\em slow} insertion method we have developed two heuristics, 
that rely on rapid pre-scoring of insertion branches based on {\em fast} likelihood insertion scores or 
Maximum  Parsimony (MP) scores. With those pre-scoring techniques, the number of insertion positions considered for the 
significantly more time consuming {\em slow} insertion process with thorough branch length 
optimization can be reduced to a small fraction of promising candidate branches. 
The proportion of insertion branches suggested by the rapid pre-scoring heuristics for analysis under the slow insertion method 
is determined by a user defined parameter $fh$. As part of our performance evaluation we have tested the fast ML and MP heuristics 
with regard to this parameter setting. Overall, these additional heuristics yield an algorithm that is significantly more accurate, but equally fast 
as BLAST.

\section*{Experimental Setup}
\subsection*{Data Sets}
To test the placement accuracy of the EPA and competing approaches, we used 8 real-world protein (AA) and DNA data alignments containing 140 up to 1,604 sequences. 
The experimental data span a broad range of organisms and include rbcL genes (D500), small subunit rRNA (D150, D218, D714, D855, D1604), 
fungal sequences (D628) as well as protein sequences from {\em Papillomaviridae} (D140).
Table~\ref{table_datasets} provides an overview of the data sets used for evaluating the placement algorithms.
We henceforth denote these data sets as Reference Alignments (RA).
For each data we computed the best-known ML trees, denoted as Reference Trees (RT), including BS support values (\cite{stamatakis2008rba}).
The data sets are available for download at \url{http://wwwkramer.in.tum.de/exelixis/epaData.tar.bz2}.

\subsection*{Generation of QS}
\begin{figure}[ht]\centering
\includegraphics[width=\columnwidth]{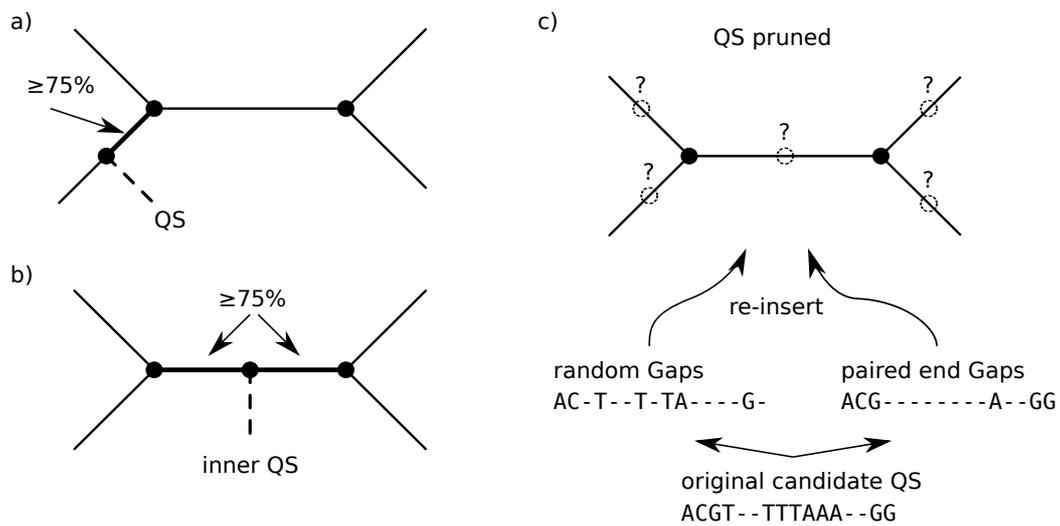}
\caption{Illustration of the criterion for the query sequence (QS) selection and experimental setup.\label{figure_qs} (a) Candidate QS belongs to sub tree of size 2 that is connected to the tree by a well supported branch. It has one other tip as direct neighbor. (b) Candidate QS is connected to the tree by two well supported branches. QS with this property will be referred to as inner QS. (c) Experimental setting: re-insert shortened candidate QS in to pruned reference tree. The QS is shortened by either inserting random gaps or by replacing all but two sub sequences at each end with gaps (paired end reads),}
\end{figure}

For evaluating the accuracy of our algorithm, we pruned one candidate QS at a time from the existing ML trees and then subsequently re-inserted the QS into the tree. 
We only prune and re-insert a subset of candidate QS with good support values from the respective reference trees in order 
to assess placement accuracy for taxa whose position in the original tree is reliable.
A candidate QS is considered to have ``good'' support, when either both (inner) branches to which the taxon is attached have a bootstrap support  
$\ge 75\%$ (Fig.~\ref{figure_qs}b) or if one of the two branches has support $\ge 75\%$ and the other branch leads to a neighboring tip (Fig. \ref{figure_qs}a). 
The threshold setting of $75\%$ reflects the typical empirical cutoff that is widely used to interpret phylogenetic bootstrap results (\cite{hillis1993empirical}).
For each candidate QS, a new, reduced, reference tree is derived by pruning the respective tip from the original tree. 
The QS associated to that taxon is then placed into the reduced tree (Fig. \ref{figure_qs}c) with our EPA algorithm.

In our test data sets, the candidate QS are still full-length sequences.
In a typical application scenario however, the placement algorithm will have to cope with QS that are significantly 
shorter than the sequences of the reference alignment, even for single gene alignments. 
Hence, we carry out a systematic assessment of the placement accuracy depending on query sequence length 
by artificially shortening the candidate QS via insertion of gaps.
We deploy two distinct methods to insert gaps into the candidate QS:

The first method to shorten candidate QS consists of randomly replacing existing characters by gaps. 
In this way we can assess the placement of QS with varying ''virtual read lengths''. 
Multiple placement runs were conducted for query sequences with a relative proportion (with respect to total alignment length) 
of non-gap character of 10\%, 20\%, 30\%,..., up to the full sequence length. 
Because the sequences from which the QS are derived form part of the original multiple alignment, 
the remaining non-gap characters are still in alignment with the RA. 
Because we calculate the amount of gaps relative to the length of the multiple alignment, 
the maximum proportion of available non-gap characters is data set specific and also depends on the individual QS candidate selected. 
Sequences that only contain non-gap characters for every alignment site are an exception for the data sets under study.

In addition to analyzing the accuracy of the EPA with gaps that have been inserted at random, 
we also evaluated accuracy on contiguous subsequences of the candidate QS, which more closely resembles 
the projected application scenario. Typically, a large number of short sequence reads generated by next generation sequencing methods 
will need to be placed into a reference tree. 
We have chosen to shorten the QS, such that they correspond to paired-end reads (see Fig. \ref{figure_qs}c) 
of the gene in the RA (we excluded data set D140 which contains protein sequences of multiple genes in this experiment). 
By using subsequences originating from pre-defined positions in the alignment, we intend to minimize the influence of the contiguous 
subsequence starting position in the alignment on placement accuracy.
Therefore, we do not consider selecting, e.g., contiguous subsequences from the candidate QS with the least amount of gaps 
or randomly selected subsequences that are located at an arbitrary alignment position. If contiguous subsequences 
at arbitrary sites are selected, the placement accuracy assessment may be biased for example 
by positional variability in 16S rRNA~(\cite{chakravorty2007detailed}) such that it will be hard to determine if 
a misplacement occurs because of the algorithm or the data.
While the selection of paired-end subsequences from the beginning and end of the gene may also bias placement accuracy, 
this bias is consistent over all QS. 
Therefore, we have conducted our accuracy assessment on paired-end reads that have been artificially generated from the 
full-length candidate QS by replacing all characters with gaps in the middle of the sequence. 
The artificial paired-end reads are of lengths 2x50 and 2x100 bp. This roughly corresponds 
to the read lengths generated by current high throughput sequencing technologies.

\subsection*{Comparison to Sequence Based Placement}

We conduct our accuracy evaluation by comparison to a typical application scenario, in which appropriate sequence based search 
tools such as BLAST (\cite{altschul1997gba}) are used to assign a QS to the most similar reference sequence. 
In this setting a candidate QS will always be assigned to one of the branches of the phylogenetic tree 
that leads to a reference sequence, i.e., an external branch.

In addition to BLAST, we also use a custom algorithm that is briefly outlined below.
We us a sequence similarity based algorithm as a baseline for comparisons with the EPA. 
Unlike BLAST, our baseline algorithm, also uses the multiple sequence alignment information from the RA to infer placements. 
Extensive experiments have shown, that the best accuracy is obtained, when a simple variation of the edit-distance is used as similarity measure.
For the pair-wise sequence comparisons, only positions are considered, where two non-gap characters are aligned. 
The distance function is the number of misaligned non-gap characters. 
The branch insertion position proposed by this method, will always be a branch that leads to the tip of the reference tree that has the smallest distance to the QS. 
While only a moderate amount of effort was invested to optimize the implementation of this approach, it generated the best results for 
the sequence comparison based methods with respect to placement of QS with random gaps (results not shown). 
However, further tests, also revealed that the distance-function  partially produced very large 
deviations from the correct insertion position for QS with contiguous paired-end reads (results not shown). 
In the latter case, the best results were obtained with a distance function that includes affine gap penalties. 
Character mismatches and gap opening are penalized with a score of 3, while gap extension has a penalty of 1. 
This gap-aware distance function yields less accurate placements than the method without gap penalties on candidate QS with random gaps. 
In the following this approach (using either distance function) will be denoted as SEQuence based Nearest Neighbor (SEQ-NN) placement.

The use of contiguous paired-end sequences also allowed for usage of BLAST sequence queries. 
For the BLAST tests we removed all gaps from the multiple alignment and built a BLAST database for each data set. 
We also removed all gaps from the candidate QS and concatenated two ends of the artificial paired-end reads into one sequence. 
Searches with those sequences were conducted against the corresponding BLAST database. 
The default parameters of the BLAST program from the NCBI C Toolkit were used for character match/mismatch (scores 1 and -3) and gaps (non-affine gap penalty of -1). 
The default values from the NCBI BLAST website with affine gap penalties were also tested, but produced slightly worse placement results and higher run times than the 
default settings.

\subsection*{Accuracy Measures}
\begin{figure}[ht]\centering
\includegraphics[width=0.8\columnwidth]{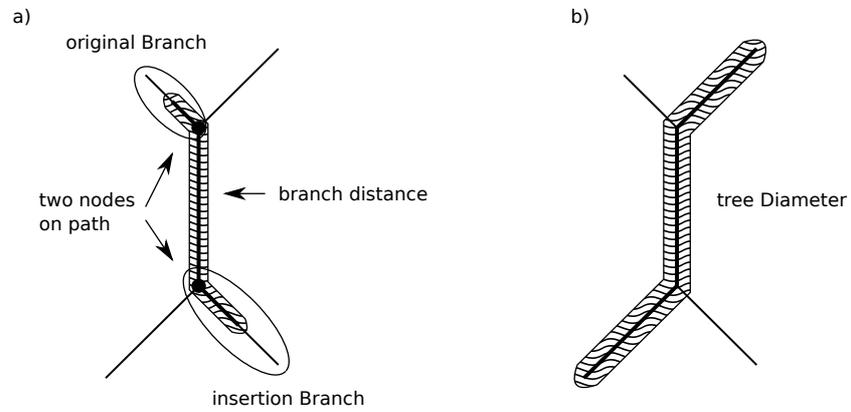}
\caption{Illustration of the tree based distance measures. (a) Example tree with two branches (original and insertion branch) highlighted. There are two nodes on the path, so the node distance is 2. The branch distance corresponds to the length of the connecting path, where of the two end branches only half of the branch length is used. (b) Tree diameter which is used to normalize the branch distance.\label{fig_distances}}
\end{figure}

To quantify placement accuracy, we use two distance measures based on the topology and branch lengths of the original ML tree. 
In all cases we consider an original branch and an insertion branch. 
The original branch is the branch from which the candidate QS was originally pruned in the ML tree, and into which it should ideally be re-inserted. 
The insertion branch is the branch computed by the respective placement algorithm. 
To quantify the distance between the 'correct' original branch and the actual insertion branch we use the following two distance measures: 
The 'Node Distance' (ND), is the unweighted path length in the original tree between the two branches. 
This corresponds to the number of nodes located on the path that connects the two branches (Fig. \ref{fig_distances}a) and represents an absolute distance measure. 
The second measure is the sum of branch lengths on the path connecting the two branches. 
This measure also includes 50\% of the branch length of the insertion-branch and 50\% of the length of the original branch (Fig. \ref{fig_distances}a). 
For comparability between different trees and in order to obtain a relative measure, we normalize the branch path length by  dividing it through 
the maximum tree diameter (Fig. \ref{fig_distances}b). 
The maximum diameter is the branch path of maximum length between two taxa in the RT. 
This distance measure is henceforth denoted as 'Branch Distance Normalized' (BDN\%).

When the EPA is used with bootstrapping, more than one insertion branch can be proposed for each candidate QS. 
For a bootstrap run with $N_{bs}$ replicates, for each QS the output of the EPA contains a set of $i=1...N$, where 
$N \leq N_{bs}$, insertion positions with bootstrap values $S_i$. 
Using this information we derive a set of ND or BDN distances $D_i$ to the original branch for each alternative placement $i$.
We use the $D_i$ to represent the bootstrap placement information as a single quantity for each QS: the Weighted Root Mean Squared Distance (WRMSD), 
$D_{wrms}$:

\begin{small}
\begin{equation}\label{formula_rmsd}
D_{wrms} = \sqrt{\frac{1}{N}\sum_{i=1}^{N} (\frac{S_i}{N_{bs}} D_i)^2}
\end{equation}
\end{small}

\section*{Results and Discussion}
\subsection*{Placement Accuracy for Random Gap QS}
\begin{figure}[ht]\centering

\includegraphics[width=0.8\columnwidth]{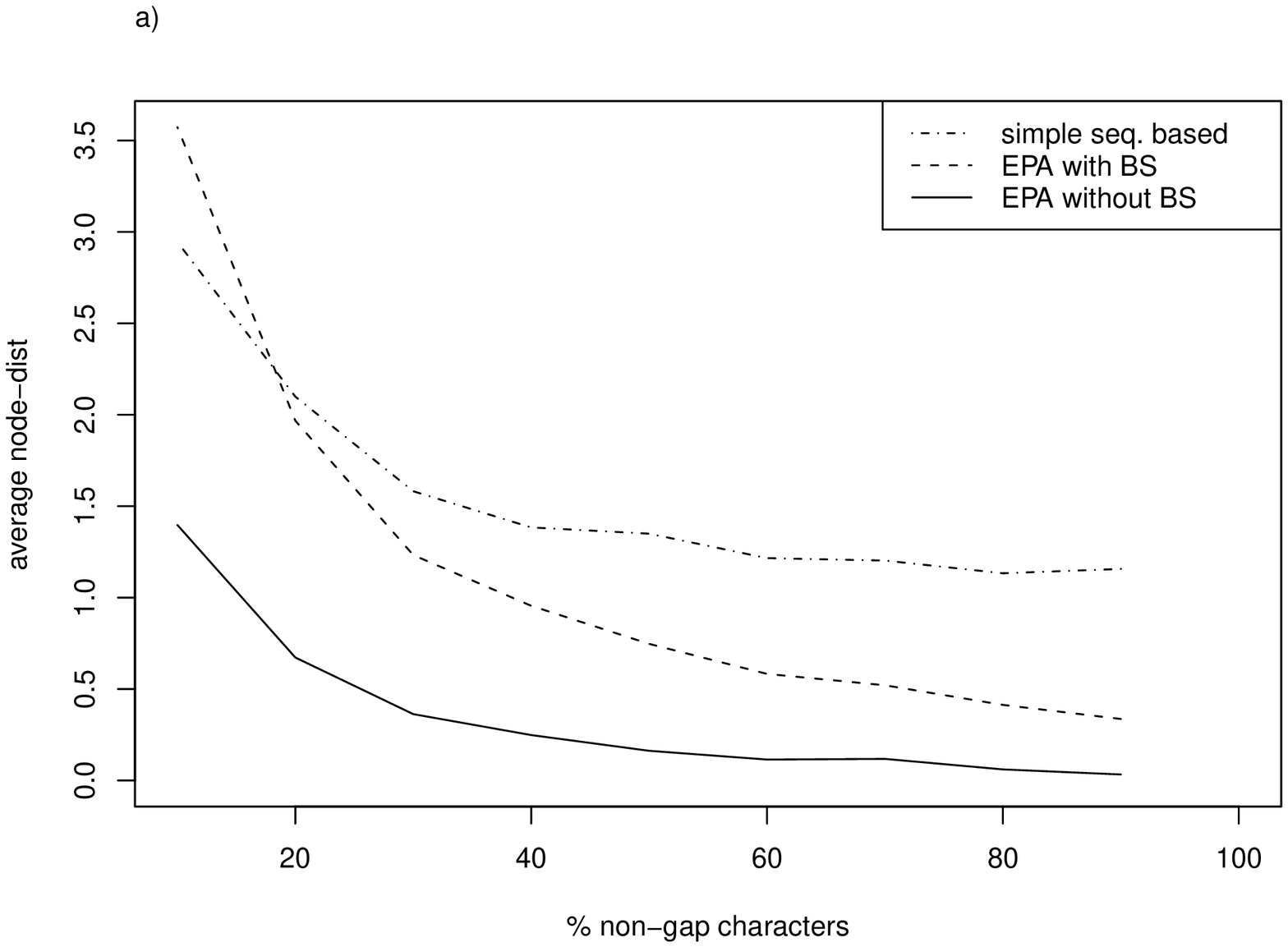}
\includegraphics[width=0.8\columnwidth]{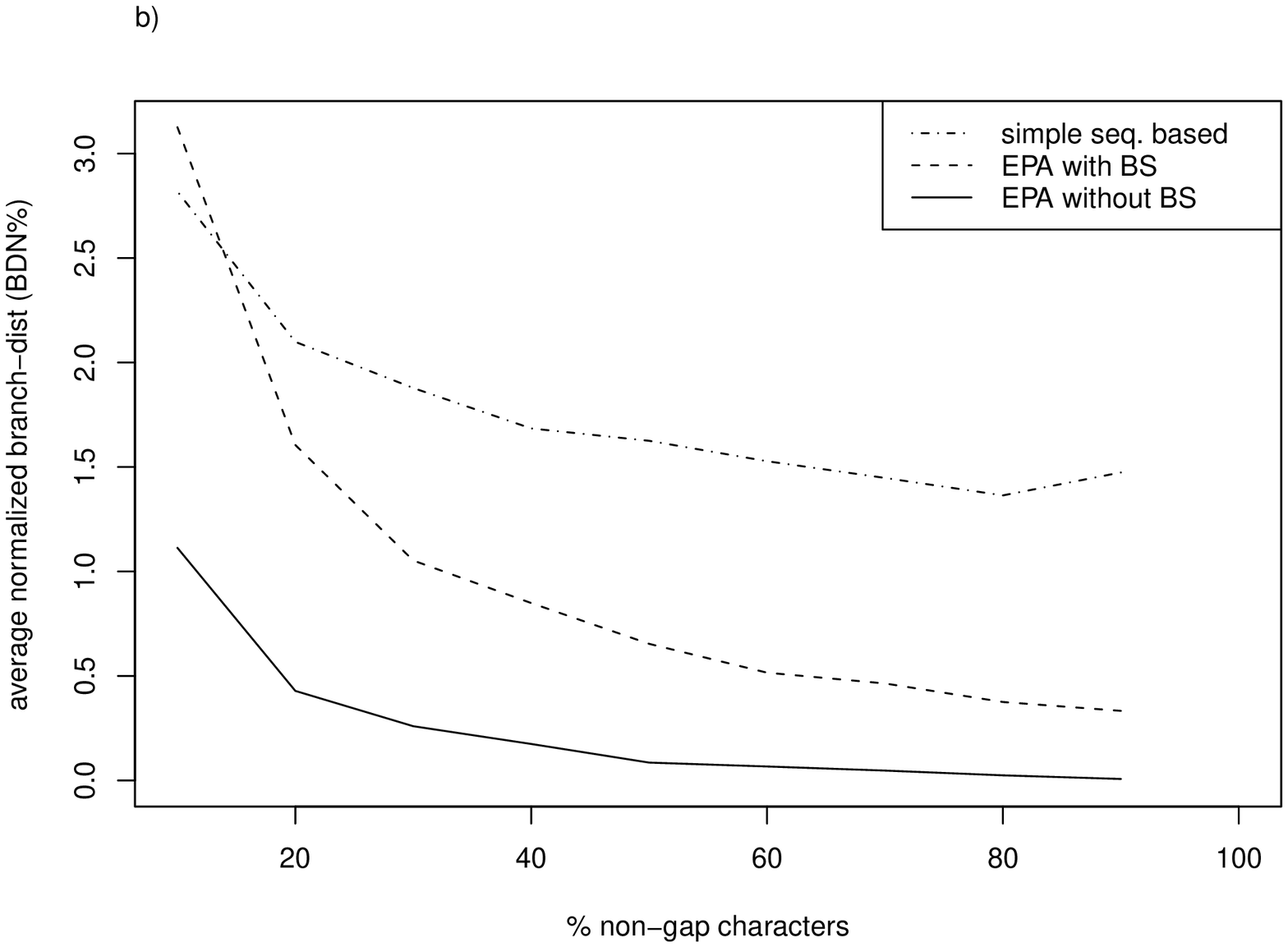}

\caption{Placement accuracy on all QS from all data sets. (a) Average node distance and (b) Normalized Branch Distance between insertion positions and real positions. \label{gappy_all}}
\end{figure}

To test the accuracy of the EPA on random-gap sequences, placement runs were carried out with bootstrapping (using 100 replicates) and without 
bootstrapping. 
For the placements from the bootstrap runs the WRMS distance from the real (original) insertion position was calculated as previously described. 
Placements were carried out on the 8 data sets for varying virtual read lengths.
Figure \ref{gappy_all} gives a detailed plot of the accuracy depending on the proportion of gaps, averaged over all candidate QS from all data sets 
(respective plots for the individual data sets can be found in the supplementary material). 
In general, SEQ-NN produces less accurate placements than the EPA. 
In cases where the QS contain less than 20\% non-gap characters (more than 80\% gaps),  the EPA with bootstrapping, produces less accurate placements
than SEQ-NN. A possible cause for this effect is discussed below. 
On the original alignment (without bootstrapping) the EPA placements are consistently approximately 1.5 times more accurate  
than SEQ-NN placements. 

\begin{figure}[ht]\centering
\includegraphics[width=0.8\columnwidth]{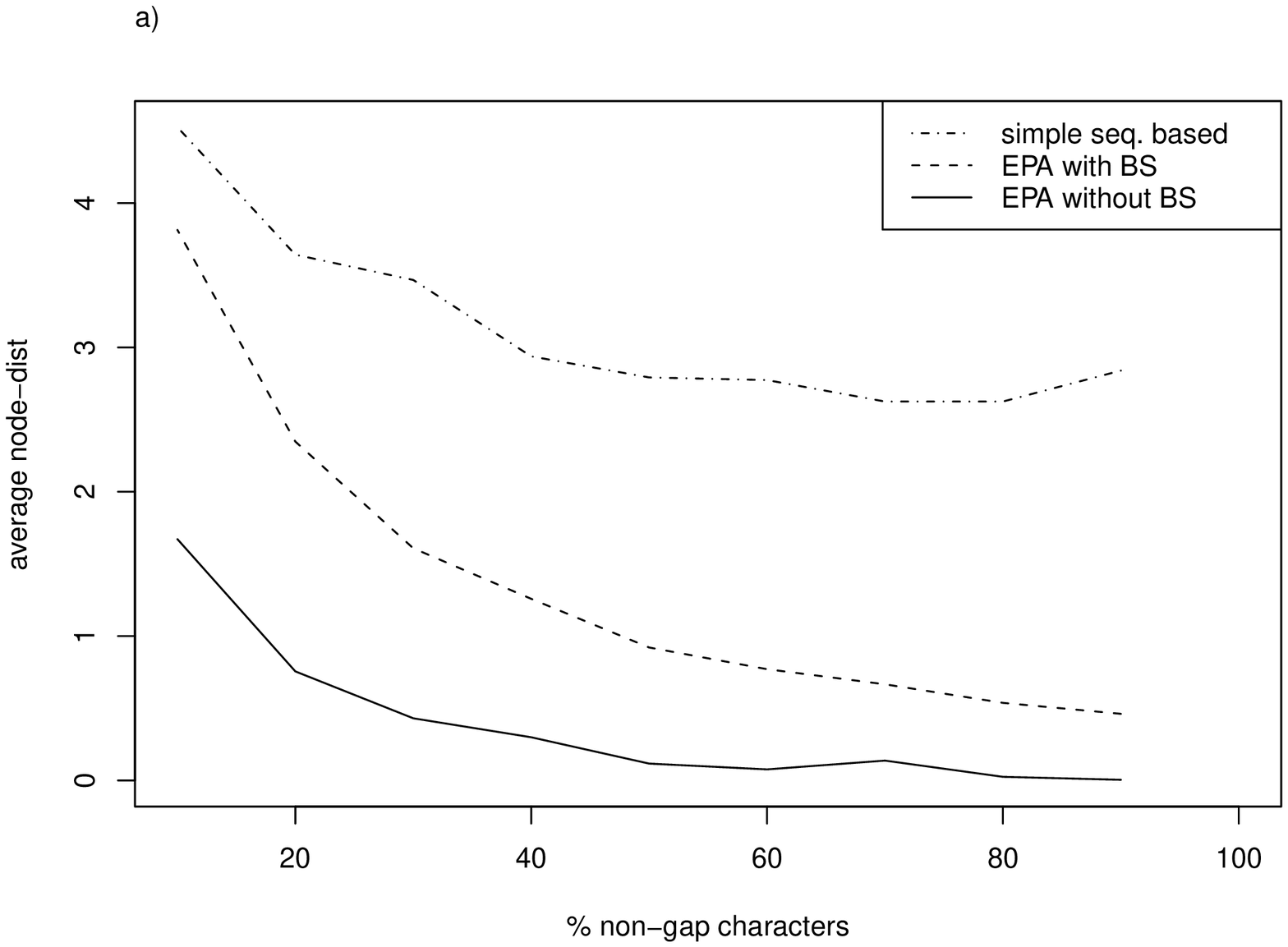}
\includegraphics[width=0.8\columnwidth]{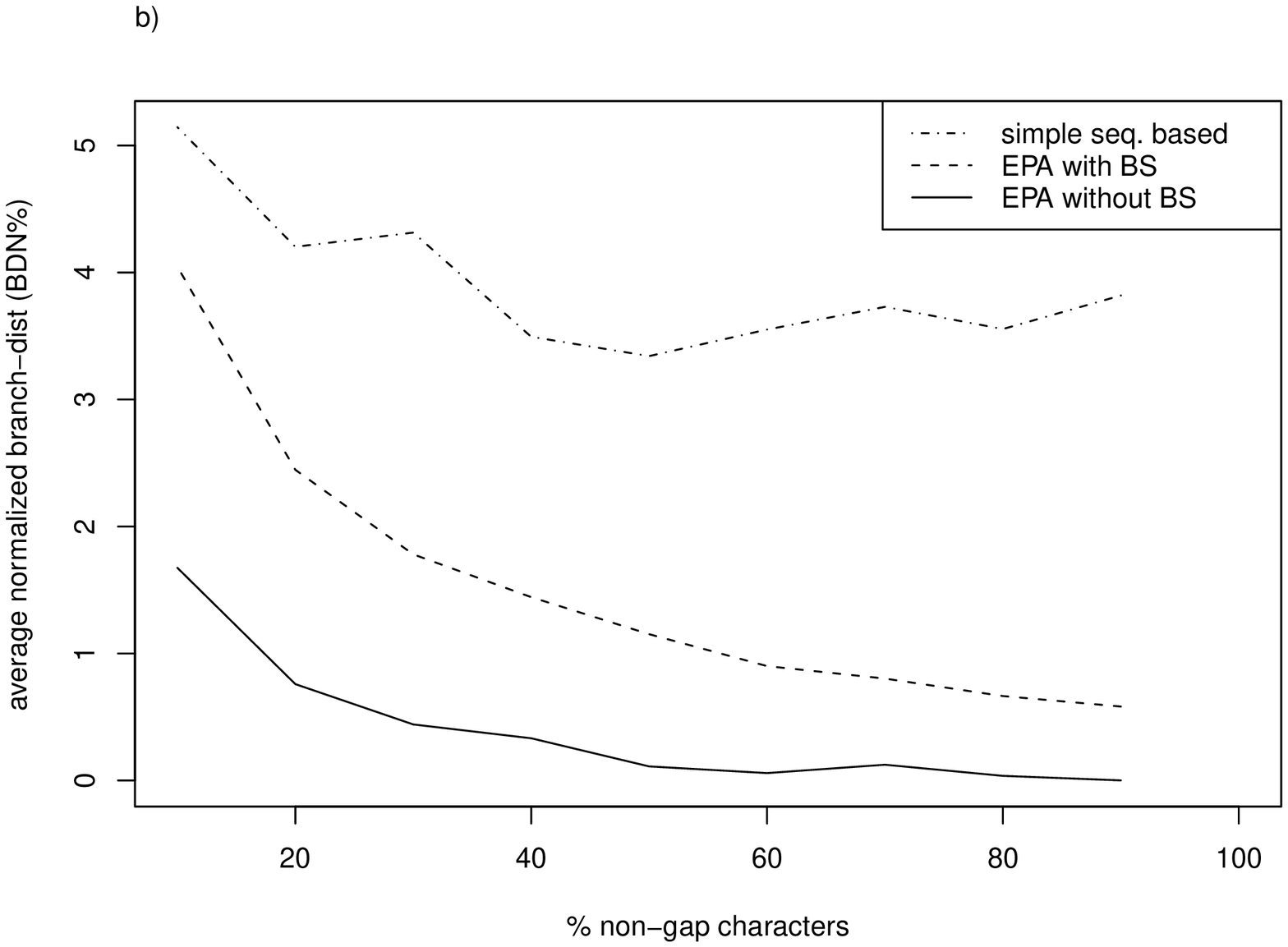}
\caption{Placement accuracy on inner QS from all data sets. (a) Average node distance and (b) normalized branch distance between insertion positions and real positions. \label{gappy_hard}}
\end{figure}

The placement accuracy improvement is even higher for for the 'hard' QS subsets that only comprise inner QS (Fig. \ref{gappy_hard}). 
For inner QS, the conceptual disadvantage of SEQ-NN becomes prevalent because the best possible placement 
that can be inferred will at least be one node away from the original insertion position (see also Figure \ref{figure_qs}). 
The difference between the EPA and SEQ-NN placements is about 3 nodes on average. For inner QS the EPA algorithm 
achieves a three-fold improvement in placement accuracy.
It is worth noting that, on average the EPA correctly places almost all QS, when the they contain less than 50\% gaps, even on this 'hard' subset of inner QS. 
This means that, for the single gene case, QS with virtual read lengths of about 400--500 base pairs are placed into the branch they 
were pruned from. This virtual read length range, that allows for sufficiently accurate placement of QS, 
is in the same length range as reads from 454 sequencing runs.
In addition, there is a trend for 454 reads to become longer as the technology matures.
The placements on the 'hard' subset are especially encouraging as they show that, in contrast to SEQ-NN, 
the good overall results of the EPA are not merely caused by the presence of a tip as direct neighbor with a high sequence similarity. 
The results on this subset are indicative for the performance on data sets with a sparse or inadequate taxon sampling.
Since it is hard to determine an adequate taxon sampling a priori for an unknown microbial community, our approach 
can actually be used to appropriately adapt the taxon sampling.

The comparison of the accuracy graphs for the EPA with and without bootstrapping helps to understand the impact of using bootstrapping for evolutionary placement.
One of the consequences of the phylogenetic bootstrapping procedure is that, for each bootstrap replicate only a fraction (less distinct alignment sites) 
of the input data is used. 
The probability for each of the $n$ alignment columns to form part of a bootstrap replicate is $1-(1-\frac{1}{n})^n \approx e^{-1} \approx 0.632$.
Thus, only 63.2\% of the available characters in each QS will be used on average. 
In practice this has a similar effect as using shorter QS with less signal for computing the insertion position, and partly explains why the 
EPA placements with bootstrapping enabled are generally worse than those without bootstrapping. 
For QS with a very low proportion of non-gap characters, the aforementioned property of the bootstrap method becomes 
particularly noticeable and results in a inferior placement accuracy compared to the simple approach used in SEQ-NN. 
In accordance with this, the accuracy of bootstrap placements improves significantly with increasing QS lengths for which it 
clearly outperforms SEQ-NN.

\subsection*{Placement Accuracy for Short Contiguous Sequence Reads}
Table \ref{table_pe_100} provides the overall results of the experiments with virtual paired-end reads of length 2x100bp
(the results for 2x50bp reads are provided in the supplementary material). 
EPA accuracy is compared against SEQ-NN with the aforementioned appropriately adapted distance function as well as against BLAST. 
We specifically report results for two sequence based methods, to assess to which extent the exclusion of gap positions from the 
original multiple alignment have a negative impact on the results derived from BLAST hits. Inversely, we also assess to which extent 
the availability of the additional information provided by the multiple sequence alignment, benefits the alignment based SEQ-NN approach. 

The placements returned by BLAST are based on a local pair-wise alignment of sequences.
In our case this means that only one half of the paired-end reads (100bp) is actually used for the placement.
As a consequence, the simple approach used in SEQ-NN (with gap penalties) consistently places the QS closer to their 
original position than BLAST. 
Nonetheless, the accuracy of the EPA placements is 1.58--3.55 times better than for SEQ-NN
that also uses the information contained in the multiple sequence alignment.

\begin{figure}[ht]
\centering
\includegraphics[width=\columnwidth]{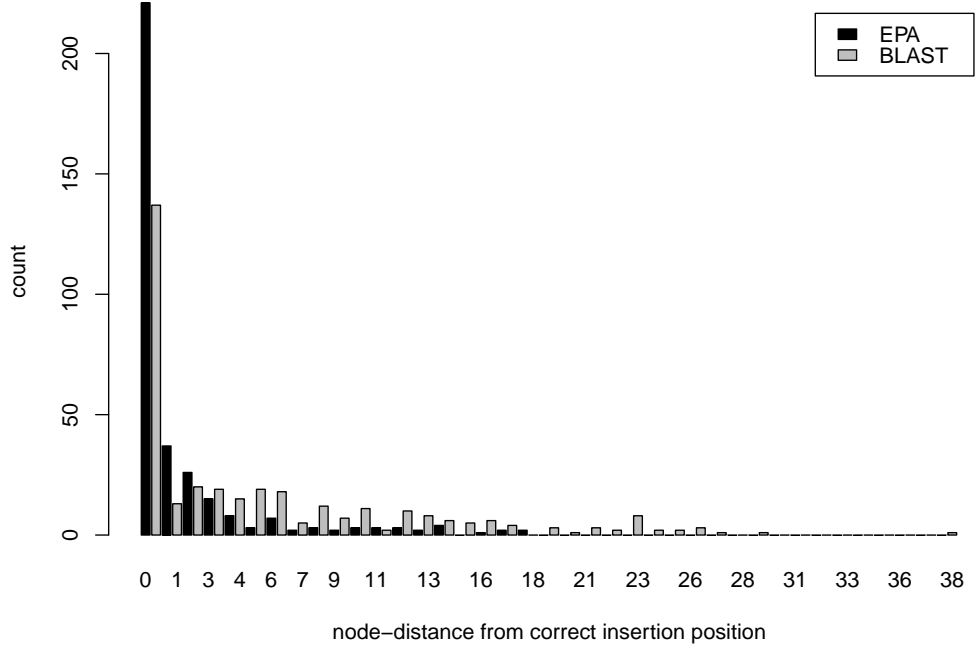}
\caption{Histogram plot of the placement accuracies (Node Distance) for the placement of 2x100 BP paired-end reads on data set D855. \label{pe_hist_nd_855_100}}
\end{figure}

Figure \ref{pe_hist_nd_855_100} provides a histogram for the distribution of individual placements computed by the EPA and BLAST for 2x100bp 
paired-end reads on data set D855. 
Respective histograms for all data sets on 2x100bp and 2x50bp reads are available in the supplementary material. 
The histograms show that placements obtained via the EPA are closer to the reference position on average
and yield smaller maximum placement errors than BLAST.

Table \ref{table_pe_100} highlights that the phylogeny-aware EPA consistently outperforms sequence comparison based methods 
and that placements are approximately twice as accurate on average.
Generally, the placement accuracy for contiguous short QS is 
consistent with the results obtained for candidate QS with random gaps.

\subsection*{Impact of Placement Algorithms and Substitution Models on Accuracy}
The preceding computational experiments have been carried out using the most {\em thorough} version of the EPA
under the GTR+$\Gamma$ and WAG+$\Gamma$ (AA) models.
In addition, we used the slow branch length optimization option for every possible insertion branch on the original alignment. 
As previously mentioned, we also devised a {\em fast} version of the EPA where the Newton-Raphson based branch length optimization 
is deactivated for QS insertions. These heuristics can speed up the EPA by one order of magnitude, when a large amount of QS is being
placed into a reference tree. 
An additional speedup of a factor of 3 to 4 can be achieved by using the GTR+CAT or PROT+CAT approximations (\cite{stamatakis2006cat})
of rate heterogeneity.

\begin{figure}[ht]\centering

\includegraphics[width=\columnwidth]{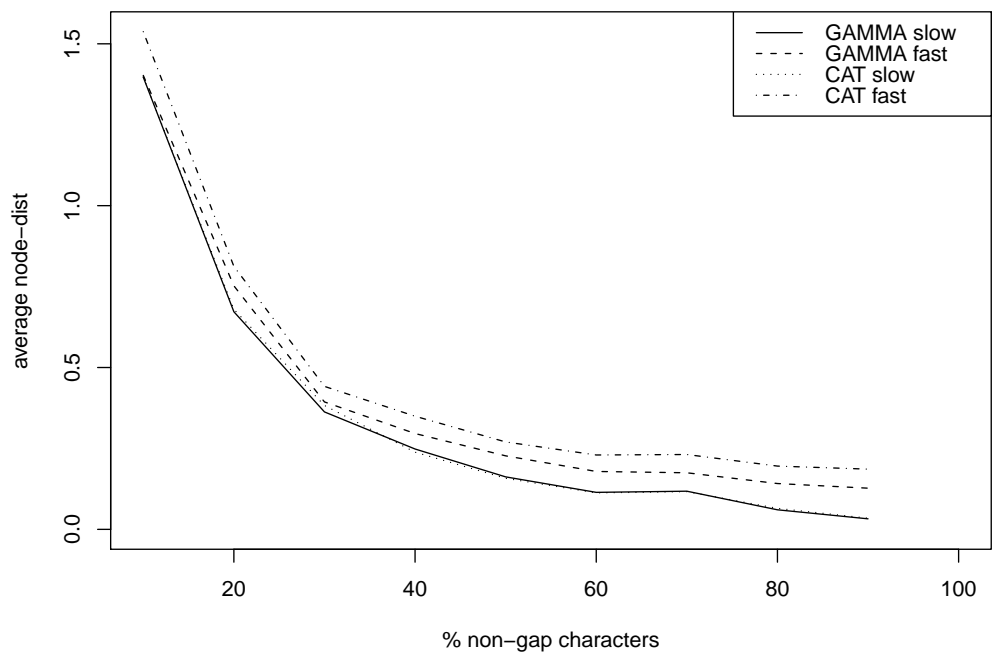}
\caption{Average Node Distance between insertion positions and real positions for different versions of the EPA (\emph{fast}/\emph{slow} insertions) algorithm and model types (GTR+$\Gamma$, GTR+CAT) on all QS from all data sets.\label{gappy_methods_all}}
\end{figure}

Figure \ref{gappy_methods_all} shows the impact of the EPA placement heuristics and rate heterogeneity model 
on the accuracy for all QS over all data sets (analogous plots for the individual data sets are available in the supplementary material). 
For the thorough insertion method there is practically no difference in placement accuracy between the $\Gamma$ model and CAT approximation.
At the same time we obtained three to four-fold run time improvements (a detailed analysis of execution times is provided in the following Section), 
which is in accordance with previous results on the CAT approximation (\cite{stamatak:IPDPS}).
For the fast insertion method, there is a noticeable decrease in placement accuracy for the CAT as well as the $\Gamma$ models. 
In particular, the {\em slow} QS insertion method performs better for long QS that contain more than 70\% of non gap characters.
However, the differences in placement accuracy between the distinct EPA models and heuristics 
are very small compared to the much larger errors returned by the sequence comparison based approaches (see Fig. \ref{gappy_all} and \ref{gappy_hard}).

\subsection*{Run Time Analysis and Heuristics for Slow Insertions}
As shown in the previous Section the loss of accuracy induced by the {\em fast} insertion method is minimal. 
Nonetheless, a slight accuracy improved can be attained by the {\em slow} insertion method. 
Using the rapid pre-scoring heuristics we have already described, it is possible to significantly accelerate the {\em slow}
insertion algorithm with little to no impact on placement accuracy. 
Here we evaluate the run time and accuracy trade offs associated with those heuristics. 
We also provide run time measurements for the standard insertion methods and BLAST.

In contrast to the previous accuracy assessments, we do not test the placement of one QS at a time into an existing RT from which the QS has been previously pruned. 
Instead, we randomly split the alignments into two subsets that each comprise 50\% of the taxa. 
The first subset is used to infer a best-known ML tree with RAxML 
into which the remaining taxa (of the second subset) are placed via the EPA. 
For run time measurements this experimental setting better corresponds to a typical application scenario of the EPA, 
where a large number of QS is placed into a reference tree. 
In contrast to the previous experiments, we can not use the position in the RT from which the candidate QS has been pruned as a reference for accuracy measurement. 
Instead, we compare the placements obtained by the various heuristics to the placements inferred by the slowest and most thorough EPA version under 
the GTR+$\Gamma$ model. This {\em slow} EPA version has shown to be the most accurate placement algorithm 
in the previous experiments. 
Here, we assume the {\em slow} EPA placements to be the true placements. 
In this test we reduce the length of the QS to 50\% non-gap characters. 
The non-gap characters are a contiguous sequence fragment that starts at the beginning of the respective sequence, i.e., the QS
represent roughly the first half of the gene. 

All performance tests were carried out on a typical current desktop computer with a Intel Core2 Quad CPU Q9550 running at 2.83GHz with 8GB 
of main memory and Ubuntu Linux 8.10. 
All programs were compiled as optimized 64bit binaries with the gcc compiler (version 4.3.2), and only one core of the CPU was used. 
The EPA uses SSE3 instructions to accelerate the likelihood computations (introduced with RAxML version 7.2.0). 
Running a BLAST search of all QS against a database comprising the remaining sequences takes 216 seconds for the largest data set in this test (D1604). 
We use BLAST with the default settings without affine gap penalties. 
As already mentioned, affine gap penalties did not improve the accuracy, but resulted in much higher run times, therefore we kept them disabled. 
On D1605 the run time for the {\em slow} insertion method is 7409 seconds under GTR+$\Gamma$ and 1846 seconds under GTR+CAT. 
With fast insertions the run time amounts to 251 seconds under GTR+$\Gamma$ and only 172 seconds under GTR+CAT.
Thus, the EPA with the {\em fast} insertion method under CAT is faster than a simple BLAST search.

\begin{figure}[ht]
\centering
\includegraphics[width=0.8\columnwidth]{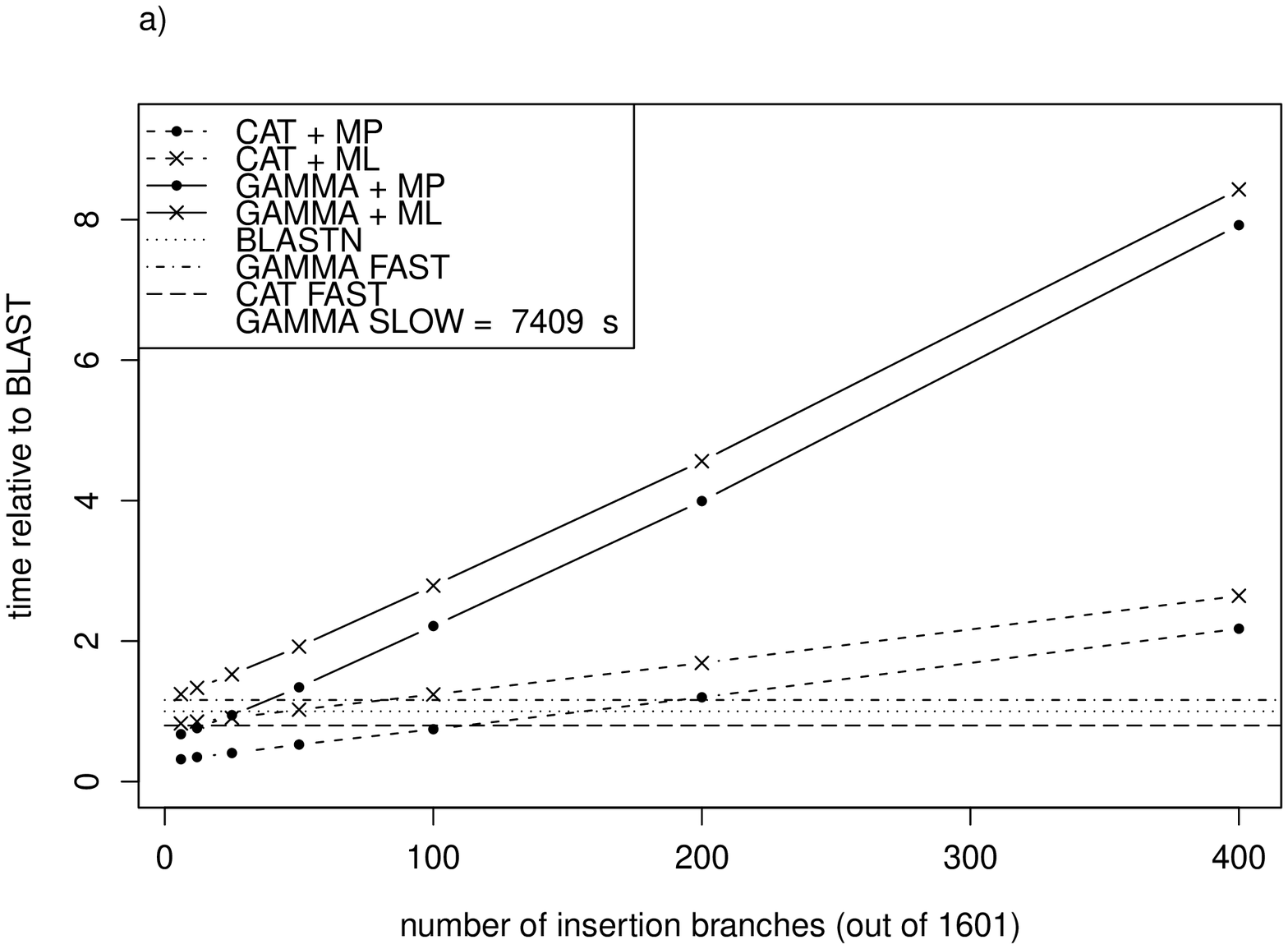}

\includegraphics[width=0.8\columnwidth]{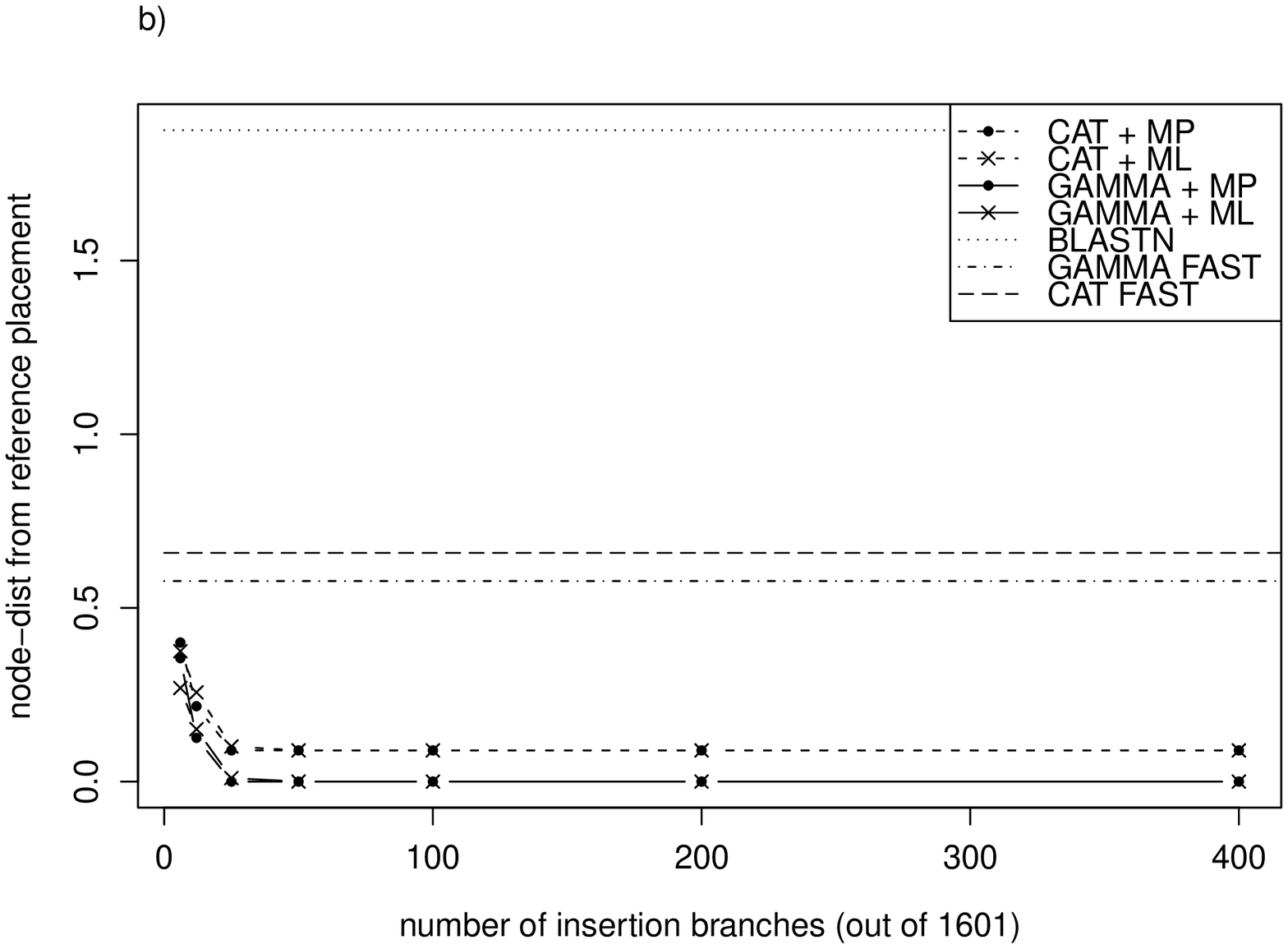}
\caption{\label{plot_heuristics}Accuracy and Runtime of the as a function of the number of branches considered for \emph{slow} insertion after heuristic filtering. (a) Runtime relative to BLAST (216 s). The plots also contain accuracy and time measurements for \emph{fast} insertions (no filter heuristics) and BLAST as horizontal lines. (b) Node Dist from optimum placement (\emph{slow} EPA under GTR+$\Gamma$). }
\end{figure}

For the pre-scoring heuristics, the run times depend on the parameter $fh$ that determines the fraction of pre-scored insertion branches that will 
subsequently be scored using the {\em slow} insertion method. 
In Figure \ref{plot_heuristics}a the run times of the different pre-scoring heuristics as well as fast insertions without heuristics relative to 
BLAST are shown for data set D1604.
 
The behavior of the heuristics and the parameter $fh$ is as expected.
It produces a constant initial overhead and scales linearly with the fraction of branches selected for {\em slow} insertions. 
The initial overhead is smaller for the MP heuristics, while the run time of the thorough insertion phase only depends on $fh$.
Therefore,  the run time graphs of the ML- and MP-based heuristics are parallel in the plots. 

Figure \ref{plot_heuristics}b shows the accuracy on the largest data set D1604 (placement of 802 QS into a reference tree with 802 RS). 
The fraction of insertion branches considered for the slow insertion phase is controlled by the parameter $fh$. 
In the plot the accuracy of the heuristics for values of $fh=1/n, n={4,8,16,32,64,128,256}$ are shown. 
The results suggest that on this data set it is sufficient to more thoroughly analyze only 50 out of 1601 ($fh = \frac{1}{32}$) 
candidate insertion branches proposed by the heuristics to gain the best possible accuracy 
(even for $fh = \frac{1}{64}$ there is only a very small deviation from the reference results). 
Another important result is that the MP heuristics produces equally accurate placements as the ML heuristics, 
on all except the smallest values of $fh$. 
This is particularly promising since the MP implementation in RAxML can be significantly accelerated by SSE3 vectorization and other 
low-level code optimizations.
We conclude that the MP heuristics with a parameter setting of $fh:=1/32$ (using the $\Gamma$ model for {\em slow} insertions) 
is sufficient for achieving placement accuracy comparable to the reference placement, but with computational requirements 
(290 seconds) that are in the same order of magnitude as a simple and significantly less accurate BLAST search. 

Even lower run times (113 seconds) can be achieved by using the CAT model for {\em slow} insertions, at the expense 
of a slight loss in accuracy.
Based on the results in the previous Section, we expect the accuracy difference 
between the CAT approximation and the $\Gamma$ model to be negligible in a real world scenario. 

In the execution time tests the differences in accuracy between the {\em fast} and {\em slow} insertion methods as well as between the $\Gamma$ and CAT 
models are generally larger than in the previous Sections. 
This is not surprising, given the setup of this experiment that was not designed to measure the insertion accuracy relative to 
an assumed correct position, but the deviation between our best, yet slowest, method and less accurate, accelerated methods. 
Here, we do not constrain the experiment to QS with high support values in the reference tree, but chose QS at random
which may introduce a certain degree of imprecision to this evaluation. 
In addition, the RT (comprising 50\% of the taxa in the original RA) is smaller than in the previous evaluations and thus more sparsely sampled. 
Nonetheless, the deviation between the {\em fast} and {\em slow} EPA versions amounts to less than half a node on average
and the general finding that {\em slow} insertions under CAT are more accurate than {\em fast} insertions 
under $\Gamma$ is consistent with previous experiments.

\section*{Conclusion}

We have presented an accurate and scalable approach for phylogeny-aware sequence comparison 
and compared its accuracy and run times to alignment-based as well as alignment-free 
sequence comparison based methods. A phylogeny-aware approach has methodological advantages over standard 
sequence based approaches and the Evolutionary Placement Algorithm is freely available for 
download as open source code. 
We demonstrate that our approach, that can, e.g.,
be used for analyses of microbial communities, is at least twice as accurate than standard techniques.
More importantly, we demonstrate that achieving significantly better accuracy does not require longer inference times
and that our approach is as fast as a simple BLAST based search when using additional heuristics.

The algorithm is also relatively straight-forward to parallelize (the parallelization of the EPA will be covered
elsewhere) by applying a multi-grain parallelization technique. On a multi-core system with 32 cores and 64GB 
of main memory we were able to classify 100,000 QS in parallel into a reference tree with 4,000 taxa within 1.5 hours.

A major challenge that remains to be solved consists in aligning the QS to a given reference alignment.
Throughout this paper we have assumed that such an alignment was given. Ideally, one would like to simultaneously
place and align the QS to the respective insertion branch. We have already implemented a simplistic version 
of such an alignment method under ML in the EPA. Our alignment procedure still lacks an appropriate indel model,
since gaps are treated as undetermined characters in most standard ML implementations. Nonetheless, our method works
surprisingly well on QS with approximately 50\% gaps and is more accurate than BLAST with an average placement distance of one node 
(as in Fig. \ref{plot_heuristics}b), but less accurate and significantly slower than the alignment-based EPA insertions.
Therefore, future work will focus on the development of rapid methods for simultaneous QS placement and alignment.

\newpage 
\section*{Acknowledgments}
The authors would like to thank Rob Knight, Steven Kembel, Micah Hamady, Christian von Mehring and 
Manuel Stark for useful discussions on algorithm design and for providing test data sets.

\bibliographystyle{sysbio}
\bibliography{references}


\newpage
\begin{table}[h]
\centering

\caption{\label{table_datasets}The data sets used for evaluation of the placement algorithm. Column type indicates the sequence type (either Nucleotide (N) or Amino Acid (AA)), length the number of alignment columns, \#taxa the number of sequences, \# QS the overall number of candidate query sequences and \# inner QS the number of inner QS  }

\begin{tabular}{ccccccc}

Data & type & length & \# taxa & \# QS & \# inner QS\\
\hline
D140 & AA & 1104 & 140 & 95 & 9\\
D150 & N & 1269 & 150 & 66 & 10\\
D218 & N & 2294 & 218 & 80 & 14 \\
D500 & N & 1398 & 500 & 205 & 29\\
D628 & N & 1228 & 628 & 210 & 20\\
D714 & N & 1241 & 714 & 293 & 61\\
D855 & N & 1436 & 855 & 344 & 48\\
D1604 & N & 1276 & 1604 & 541 & 83\\

\hline

\end{tabular}

\end{table}

\newpage
\begin{table}[h]
\begin{footnotesize}
\centering

\caption{\label{table_pe_100}Accuracy on 2x100 BP paired-end reads. The given values are the distances between the real and the proposed insertion position as node-distance (ND) and the normalized branch-distance in percent (BDN \%). The methods used are the Evolutionary Placement Algorithm (EPA) (\emph{slow} insertions under the $GTR+\Gamma$ model), pairwise distance based nearest neighbor with affine gap penalties (SEQ NN) and BLAST based nearest neighbor. The rows denoted by (A) represent all candidate QS, while for the rows denoted with (I) only inner QS are used.}
\begin{tabular}{ccccccc}
& \multicolumn{3}{c}{ND} & \multicolumn{3}{c}{BDN \%}\\
Data & EPA & SEQ NN & BLAST & EPA & SEQ NN & BLAST\\
\hline

 150 (A) & 1.26 & 2.61 (2.07) & 3.67 (2.91) & 0.85 & 1.79 (2.11) & 2.29 (2.69)\\
 218 (A) & 1.96 & 4.41 (2.25) & 4.85 (2.47) & 3.96 & 7.14 (1.80) & 7.76 (1.96)\\
 500 (A) & 0.94 & 3.34 (3.55) & 5.52 (5.87) & 1.76 & 5.01 (2.85) & 7.28 (4.14)\\
 628 (A) & 0.59 & 1.29 (2.19) & 1.93 (3.27) & 0.63 & 1.32 (2.10) &  1.4 (2.22)\\
 714 (A) & 1.39 & 2.74 (1.97) & 3.86 (2.78) & 1.81 & 3.27 (1.81) & 4.86 (2.69)\\
 855 (A) & 2.12 & 3.35 (1.58) &  6.2 (2.92) & 1.18 & 2.01 (1.70) & 3.37 (2.86)\\
1604 (A) & 1.57 & 2.76 (1.76) & 3.92 (2.50) & 0.9 & 1.42 (1.58) & 2.08 (2.31) \\
 
\hline

 150 (I) &  1.9 &  2.4 (1.26) &  5.3 (2.79) & 2.17 & 4.99 (2.30) &  5.49 (2.53)\\
 218 (I) & 4.43 & 6.36 (1.44) & 5.21 (1.18) & 7.35 & 9.49 (1.29) &  8.44 (1.15)\\
 500 (I) & 1.59 &    5 (3.14) & 9.41 (5.92) & 2.85 & 7.16 (2.51) & 11.08 (3.89)\\
 628 (I) & 0.85 & 2.35 (2.76) & 2.95 (3.47) & 0.44 & 1.24 (2.82) &   0.9 (2.05)\\
 714 (I) & 1.66 & 4.28 (2.58) &  4.9 (2.95) & 2.61 & 5.48 (2.10) &  6.99 (2.68)\\
 855 (I) &  2.9 &  4.5 (1.55) & 7.54 (2.60) & 1.95 & 3.07 (1.57) &   4.4 (2.26)\\
1604 (I) & 1.55 & 3.29 (2.12) & 5.25 (3.39) & 1.62 & 3.09 (1.91) &   4.7 (2.90)\\


\hline

\end{tabular}
\end{footnotesize}
\end{table}

\newpage
\listoffigures
\newpage



\newpage

\newpage

\newpage

\newpage 

\newpage 

\newpage 

\newpage 

\newpage 

\newpage 

\newpage 
\newpage

\end{document}


\begin{table}[h]
\centering

\caption{Accuracy on 2x50 BP paired-end reads. The given values are the distances between the real and the proposed insertion position as node-distance (ND) and the normalized branch-distance in percent (BDN \%). The methods used are the EPA (\emph{slow} insertions under the $GTR+\Gamma$ model), pairwise distance based nearest neighbor with affine gap penalities (SEQ NN) and BLAST based nearest neighbor. The rows denoted by (A) represent all QS, while for the rows denoted with (I) only inner QS are used.}
\begin{tabular}{ccccccc}
& \multicolumn{3}{c}{ND} & \multicolumn{3}{c}{BDN \%}\\
Data & EPA & SEQ NN & BLAST & EPA & SEQ NN & BLAST\\
\hline

 \hline
  150 (A) & 3.58 & 4.18 (1.17) &  6.67 (1.86) & 2.36 & 3.41 (1.44) &  5.48 (2.32)\\
  218 (A) & 2.92 & 5.25 (1.80) &  6.86 (2.35) & 5.36 &  8.7 (1.62) & 12.72 (2.37)\\
  500 (A) & 3.74 &  5.6 (1.50) & 12.06 (3.22) & 5.67 & 8.35 (1.47) & 16.95 (2.99)\\
  628 (A) & 1.31 &  2.3 (1.76) &  3.56 (2.72) & 1.13 & 2.07 (1.83) &  2.74 (2.42)\\
  714 (A) & 2.13 & 3.27 (1.54) &  4.18 (1.96) & 2.88 & 4.07 (1.41) &  5.36 (1.86)\\
  855 (A) & 3.82 & 5.27 (1.38) &  9.52 (2.49) & 2.47 &  3.4 (1.38) &  7.13 (2.89)\\
 1604 (A) & 2.49 & 3.62 (1.45) &  5.79 (2.33) & 1.28 & 2.01 (1.57) &  3.84 (3.00)\\
\hline
  150 (I) &  2.1 &  4.4 (2.10) &   9.2 (4.38) & 5.22 & 6.29 (1.20) & 10.79 (2.07)\\
  218 (I) &    4 &    6 (1.50) &  9.29 (2.32) & 7.56 & 9.19 (1.22) & 15.95 (2.11)\\
  500 (I) & 3.86 & 5.59 (1.45) & 11.03 (2.86) & 6.63 & 8.87 (1.34) & 14.18 (2.14)\\
  628 (I) &  1.8 & 2.45 (1.36) &  4.55 (2.53) & 1.19 & 1.19 (1.00) &   1.5 (1.26)\\
  714 (I) & 2.61 & 4.87 (1.87) &  4.48 (1.72) & 3.89 & 6.26 (1.61) &  6.24 (1.60)\\
  855 (I) & 3.79 & 5.31 (1.40) & 10.69 (2.82) & 2.66 & 3.55 (1.33) &  8.94 (3.36)\\
 1604 (I) & 2.24 & 3.93 (1.75) &  6.18 (2.76) & 1.82 & 3.46 (1.90) &  5.72 (3.14)\\
                  
\hline                  
                        
\end{tabular}           
\end{table}             

\newpage

\begin{figure}[ht]\centering

\includegraphics[width=0.49\columnwidth]{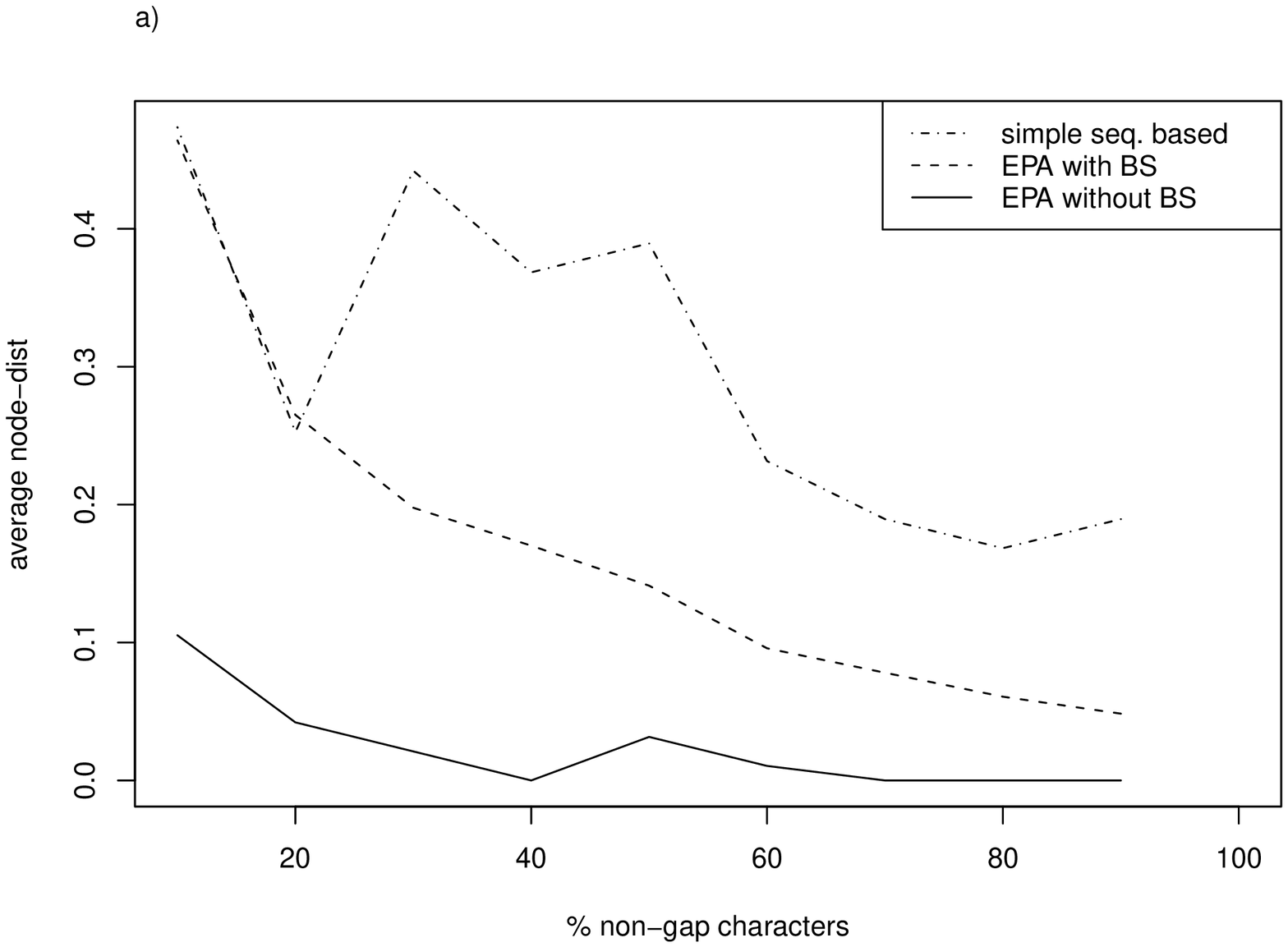}
\includegraphics[width=0.49\columnwidth]{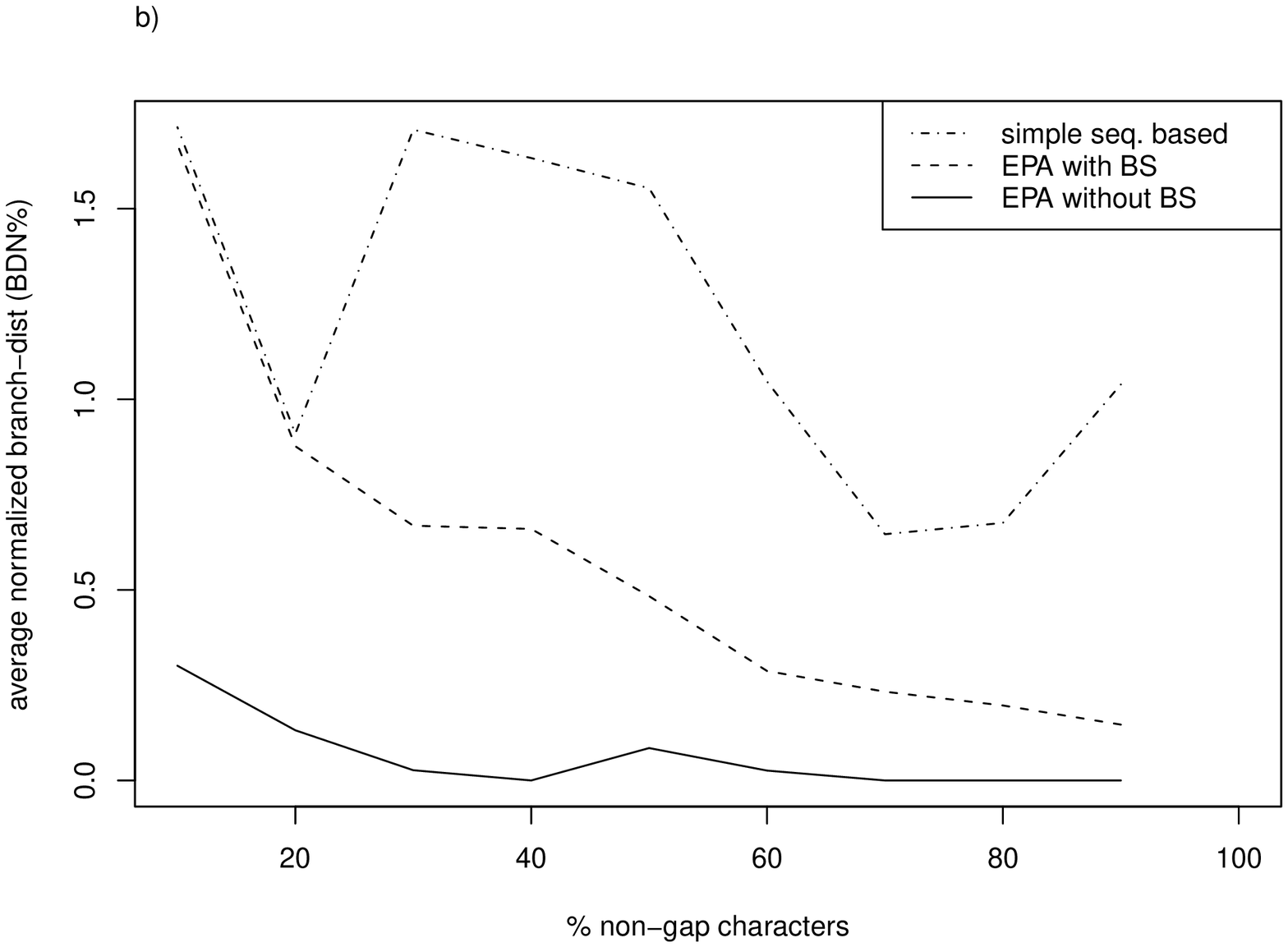}

\caption{Prediction accuracy on all QS from data set D140. (a) Average node distance and (b) Normalized Branch Distance between insertion positions and real positions. \label{gappy_all}}
\end{figure}

\begin{figure}[ht]\centering
\includegraphics[width=0.49\columnwidth]{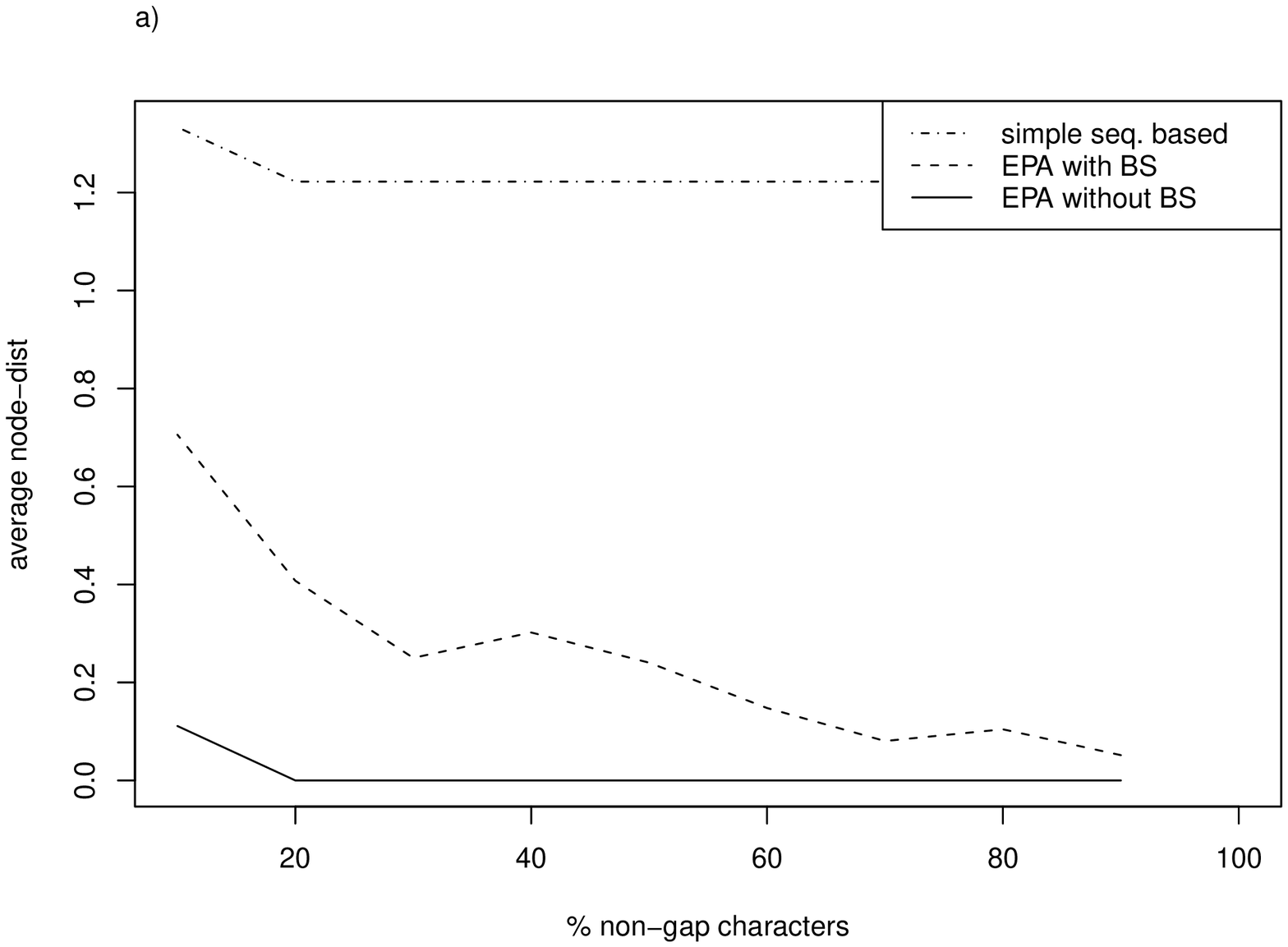}
\includegraphics[width=0.49\columnwidth]{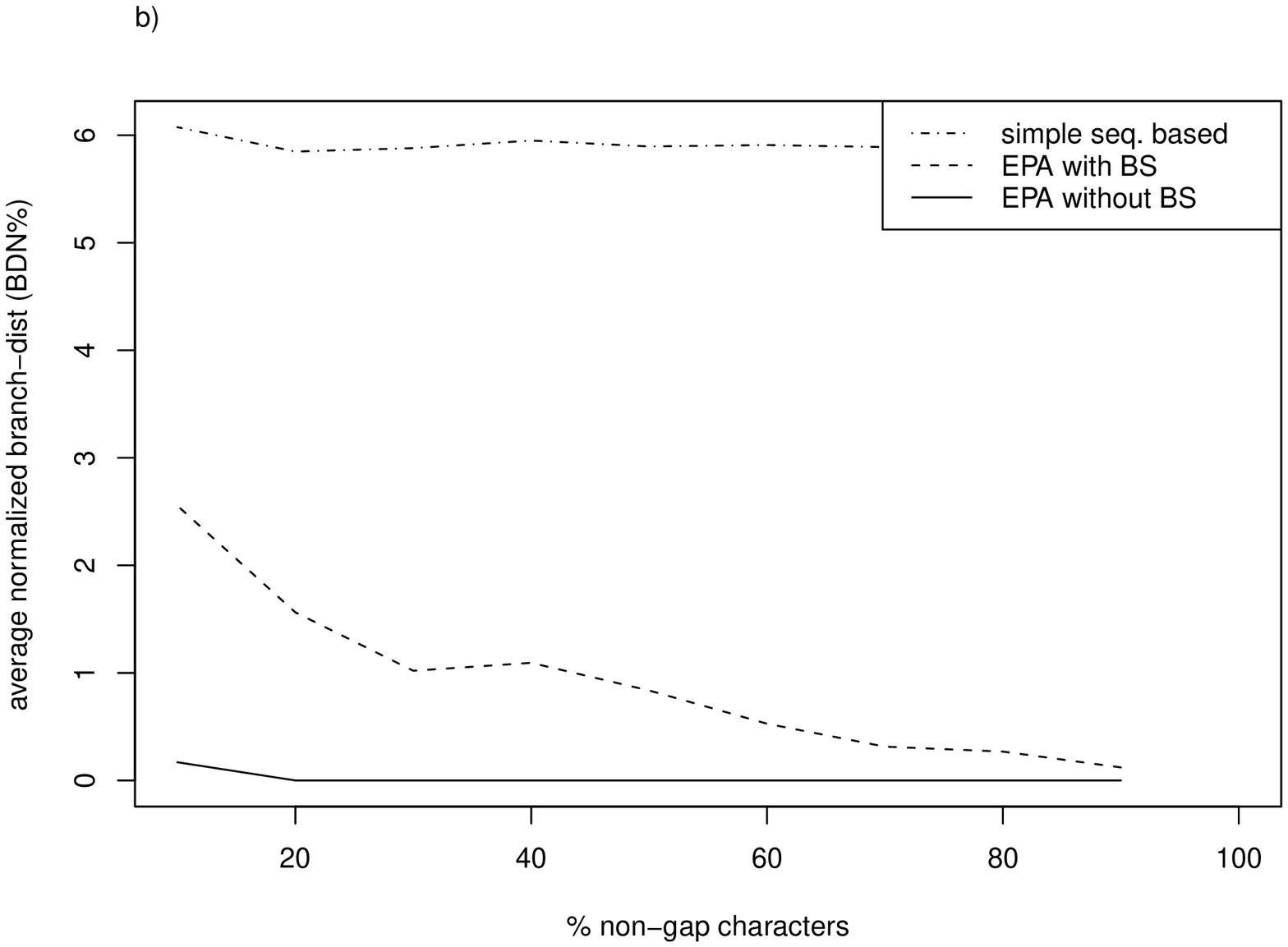}
\caption{Prediction accuracy on inner QS from data set D140. (a) Average node distance and (b) normalized branch distance between insertion positions and real positions.}
\end{figure}
\newpage 


\begin{figure}[ht]\centering

\includegraphics[width=0.49\columnwidth]{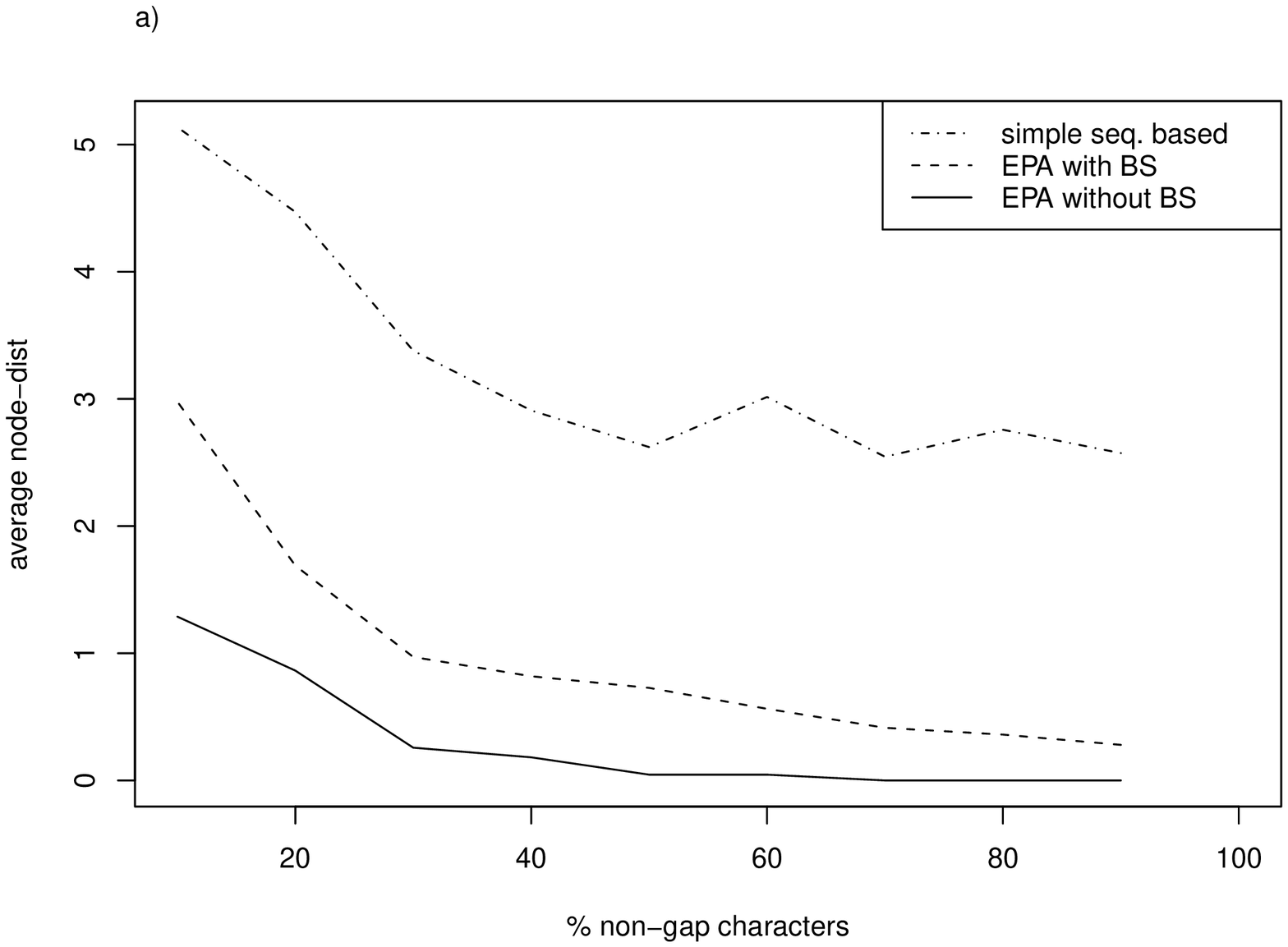}
\includegraphics[width=0.49\columnwidth]{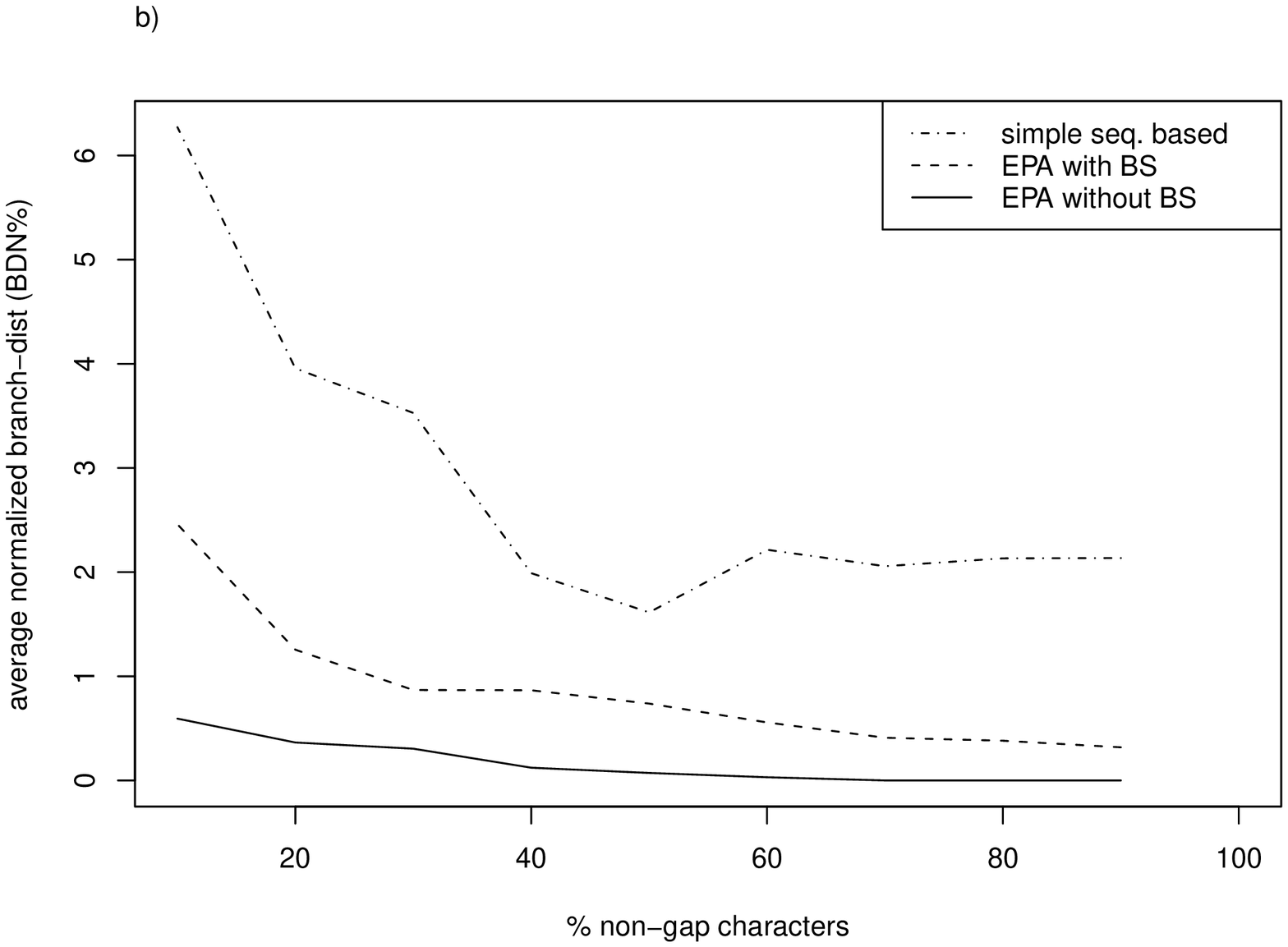}

\caption{Prediction accuracy on all QS from data set D150. (a) Average node distance and (b) Normalized Branch Distance between insertion positions and real positions.. \label{gappy_all}}
\end{figure}

\begin{figure}[ht]\centering
\includegraphics[width=0.49\columnwidth]{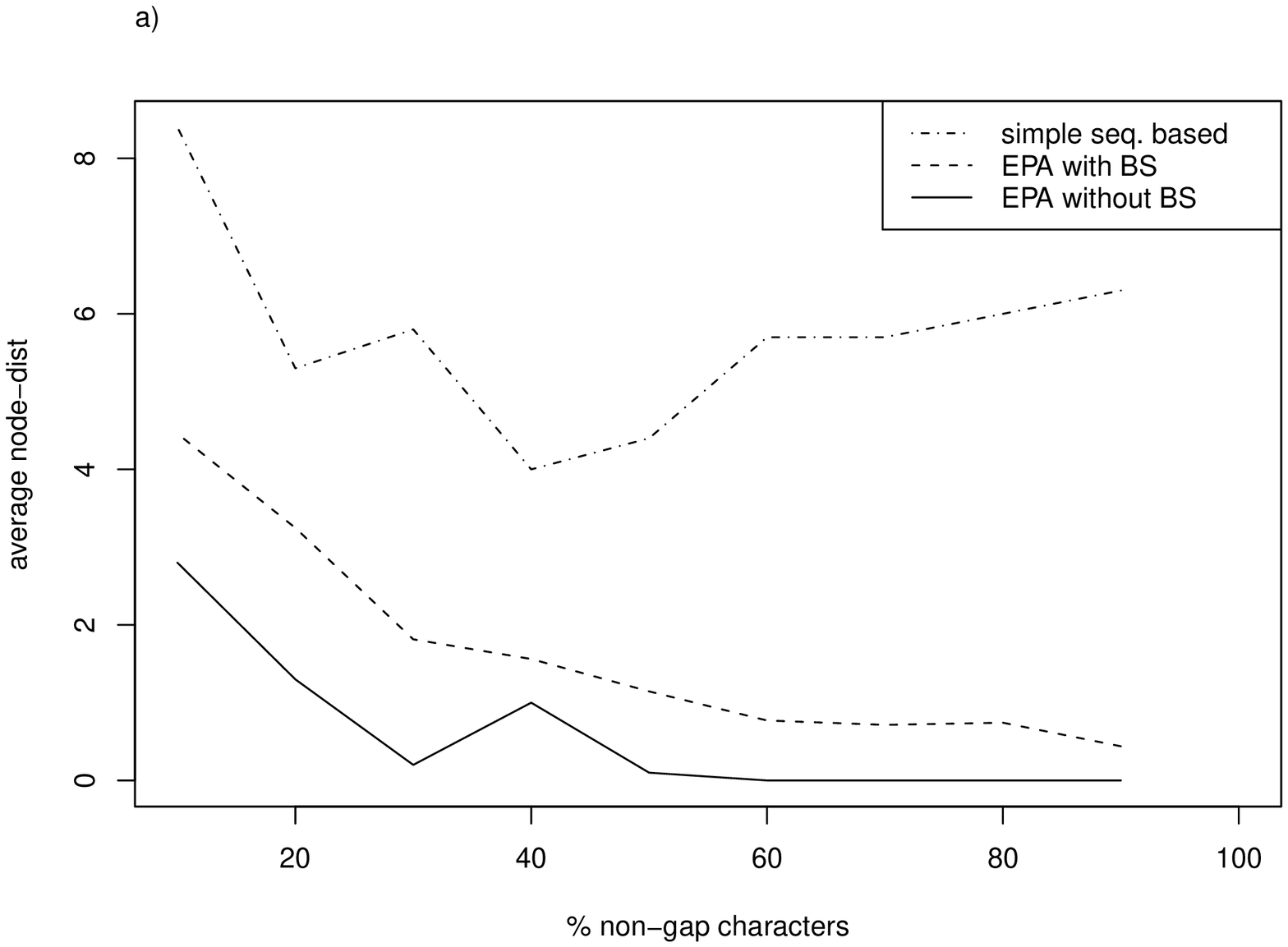}
\includegraphics[width=0.49\columnwidth]{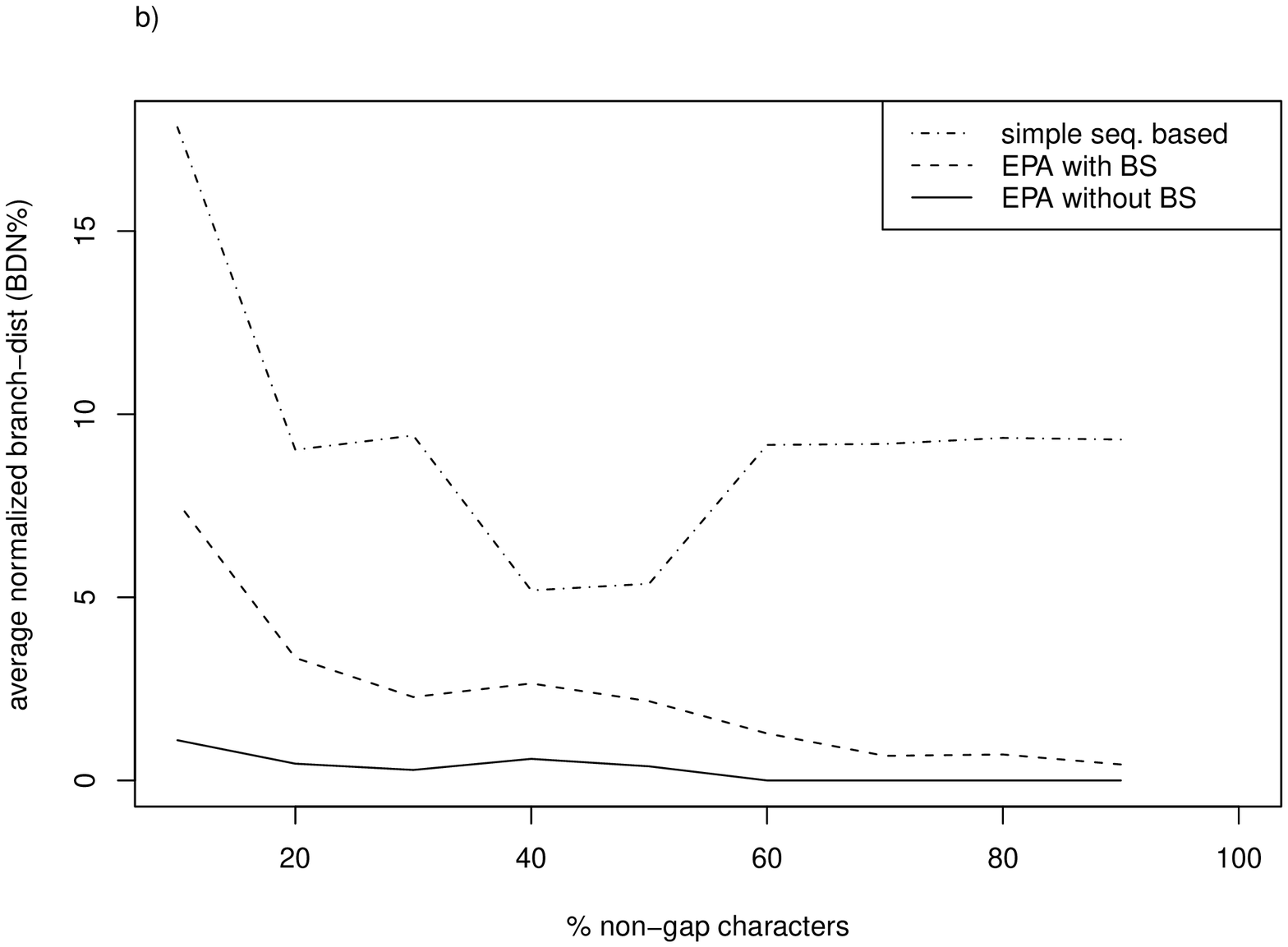}
\caption{Prediction accuracy on inner QS from data set D150. (a) Average node distance and (b) normalized branch distance between insertion positions and real positions. \label{gappy_hard}}
\end{figure}

\newpage 


\begin{figure}[ht]\centering

\includegraphics[width=0.49\columnwidth]{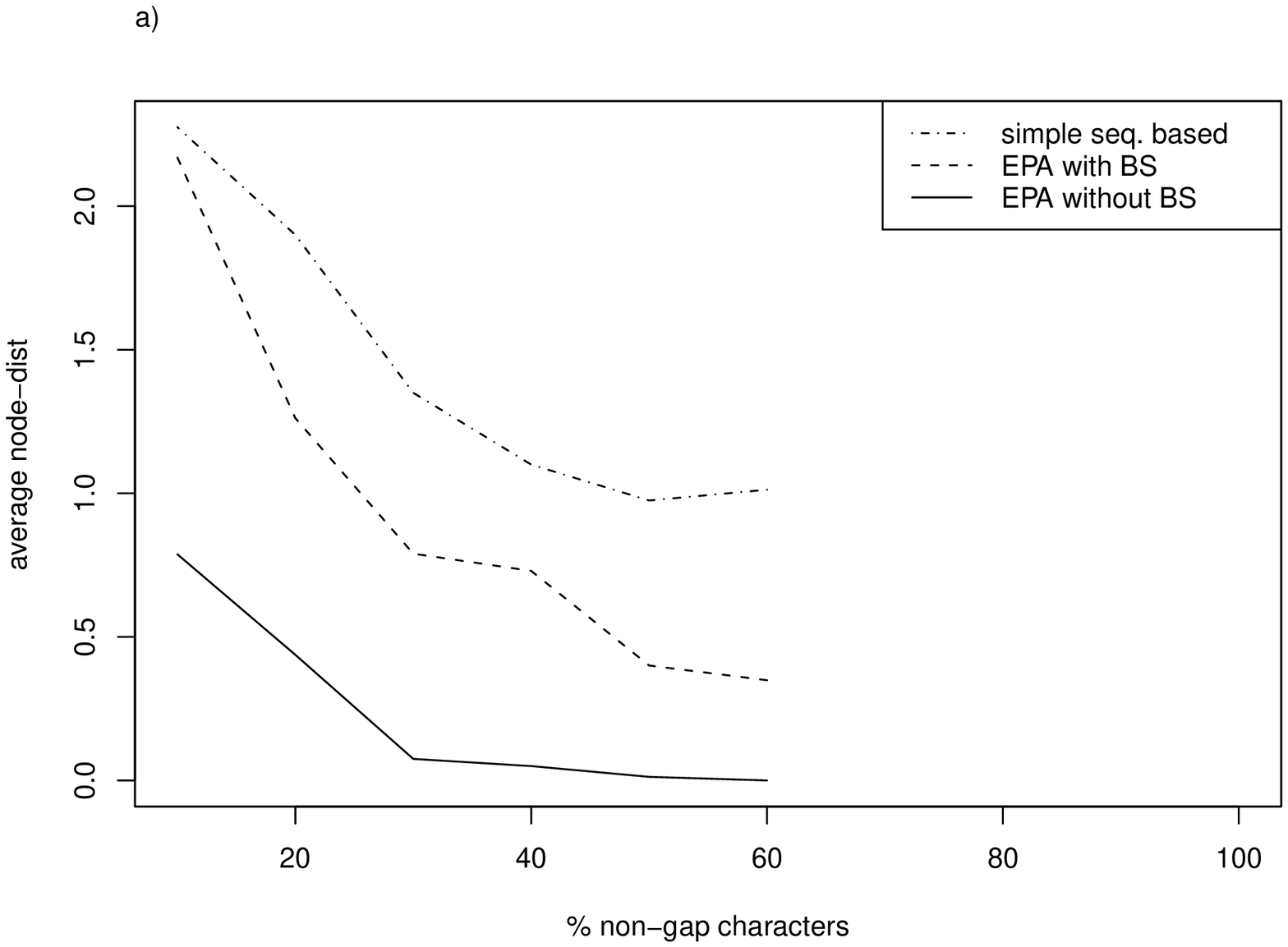}
\includegraphics[width=0.49\columnwidth]{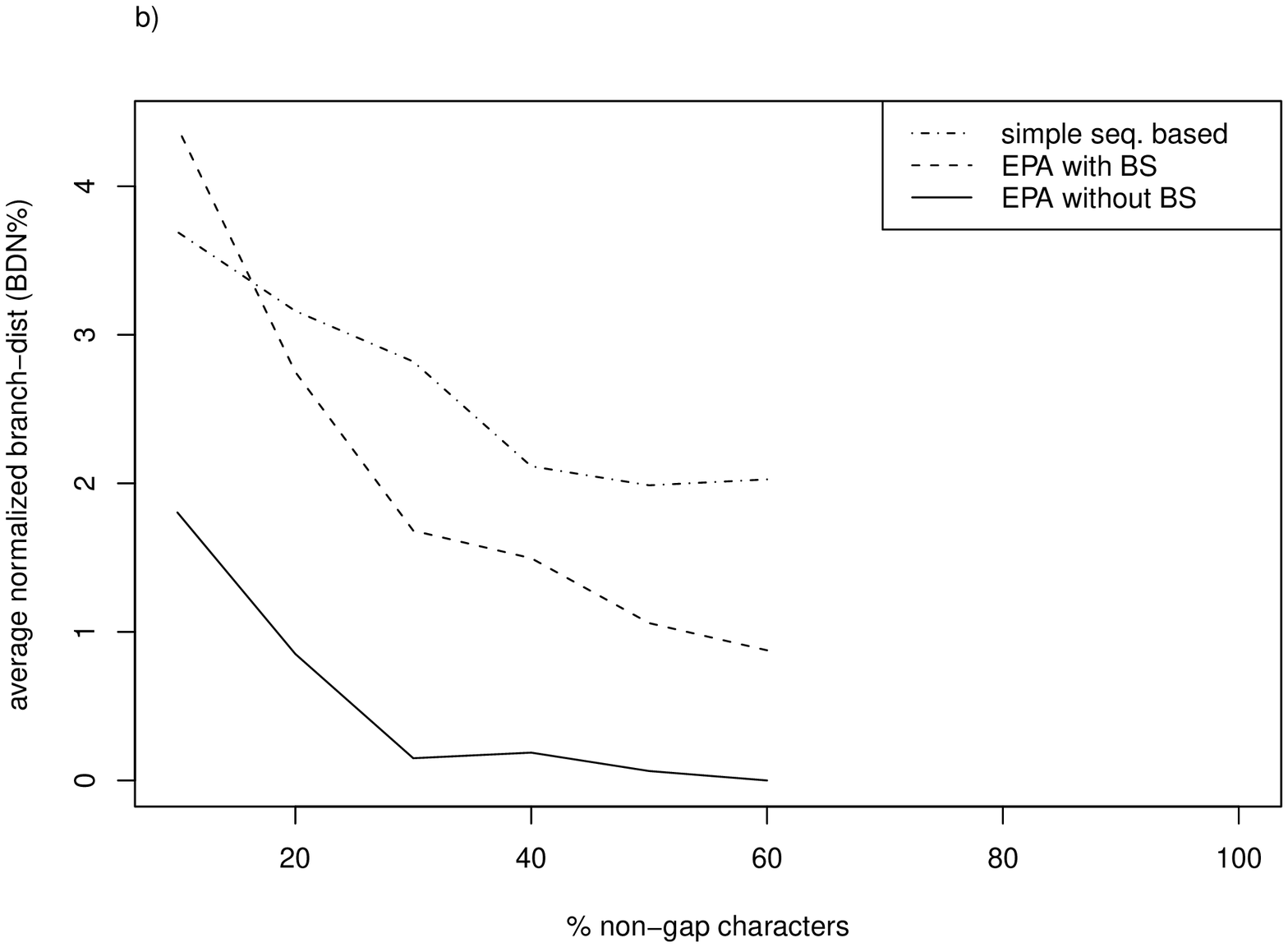}

\caption{Prediction accuracy on all QS from data set D218. (a) Average node distance and (b) Normalized Branch Distance between insertion positions and real positions.. \label{gappy_all}}
\end{figure}

\begin{figure}[ht]\centering
\includegraphics[width=0.49\columnwidth]{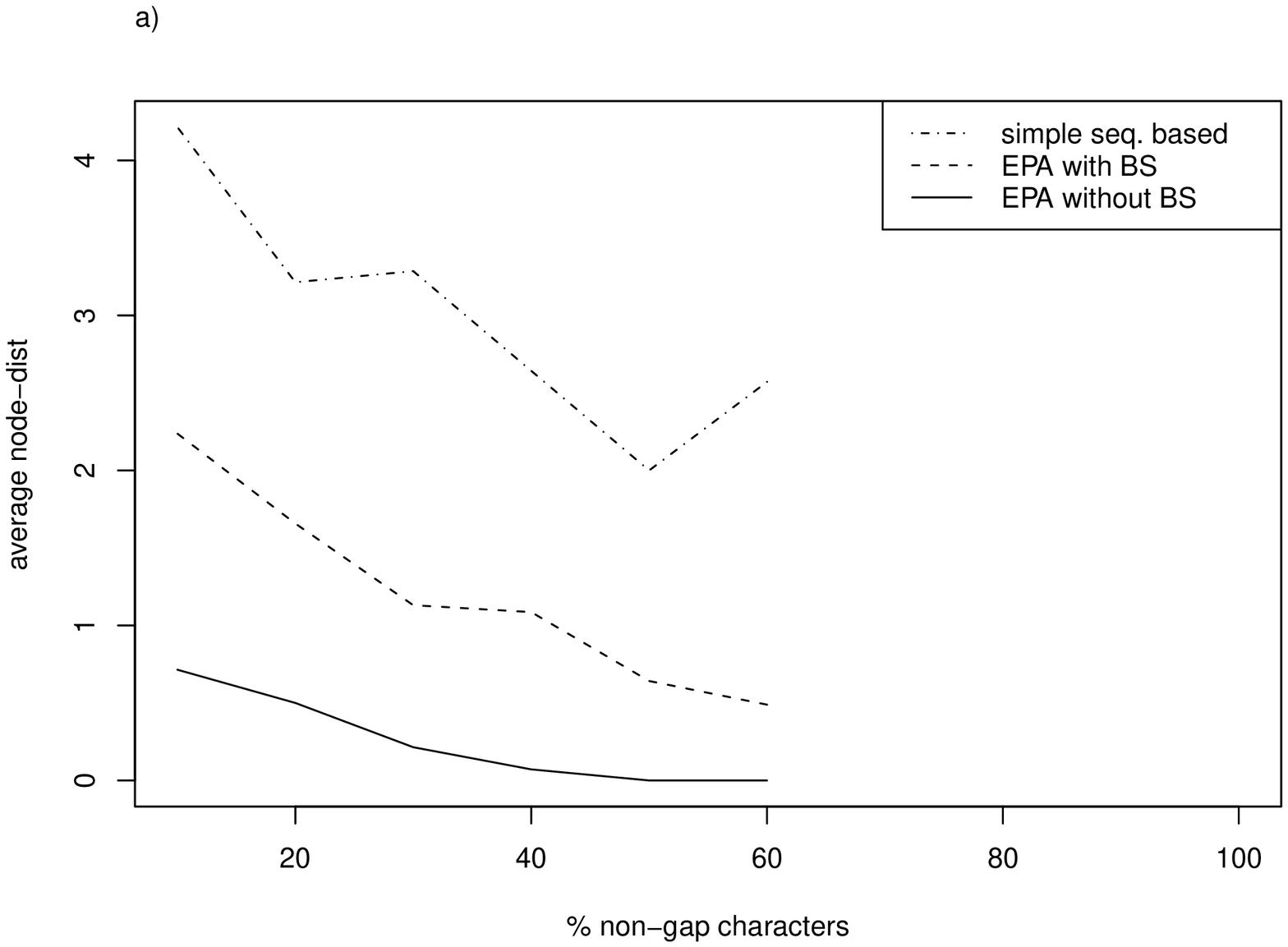}
\includegraphics[width=0.49\columnwidth]{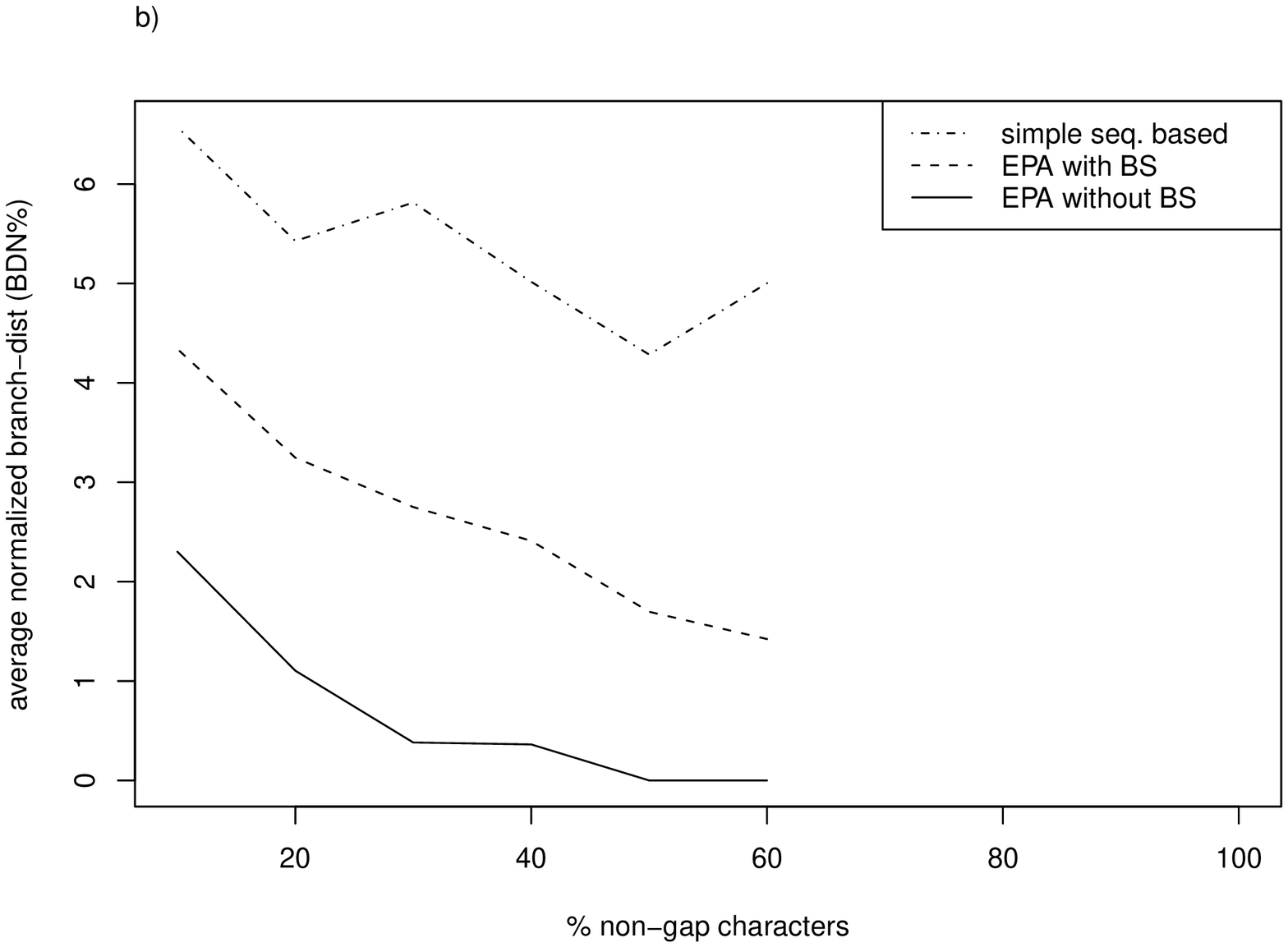}
\caption{Prediction accuracy on inner QS from data set D218. (a) Average node distance and (b) normalized branch distance between insertion positions and real positions. \label{gappy_hard}}
\end{figure}

\newpage 


\begin{figure}[ht]\centering

\includegraphics[width=0.49\columnwidth]{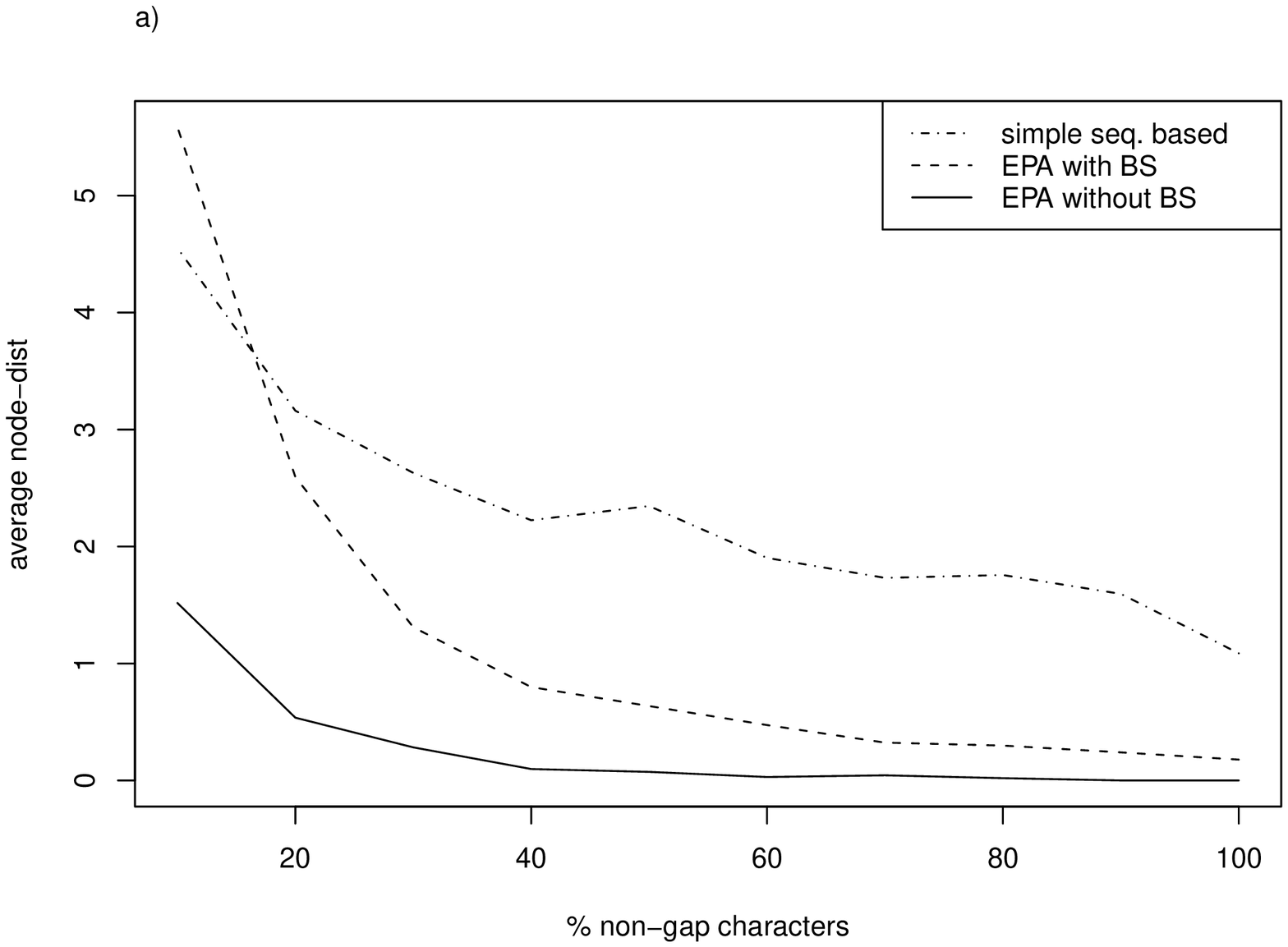}
\includegraphics[width=0.49\columnwidth]{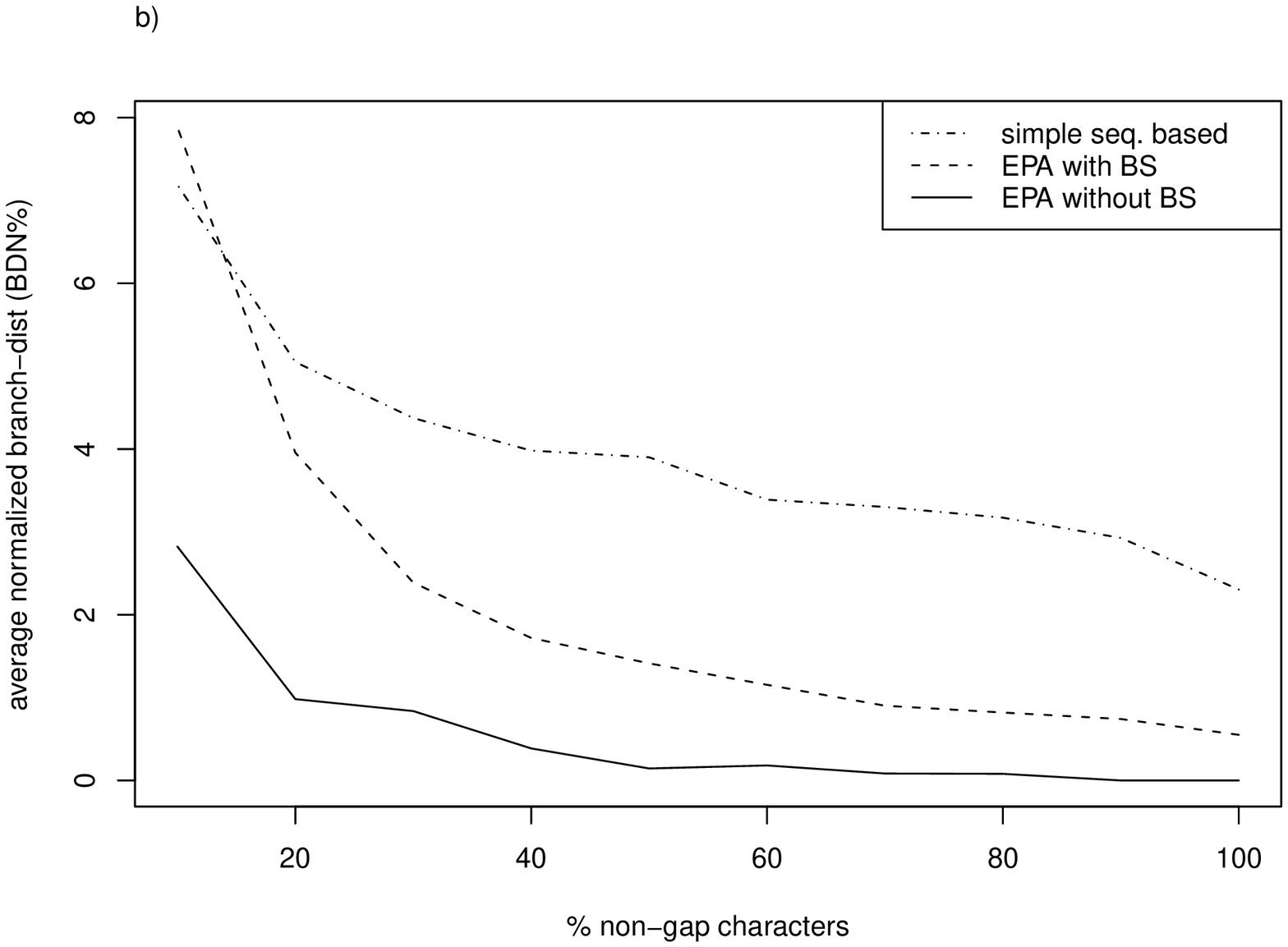}

\caption{Prediction accuracy on all QS from data set D500. (a) Average node distance and (b) Normalized Branch Distance between insertion positions and real positions.. \label{gappy_all}}
\end{figure}

\begin{figure}[ht]\centering
\includegraphics[width=0.49\columnwidth]{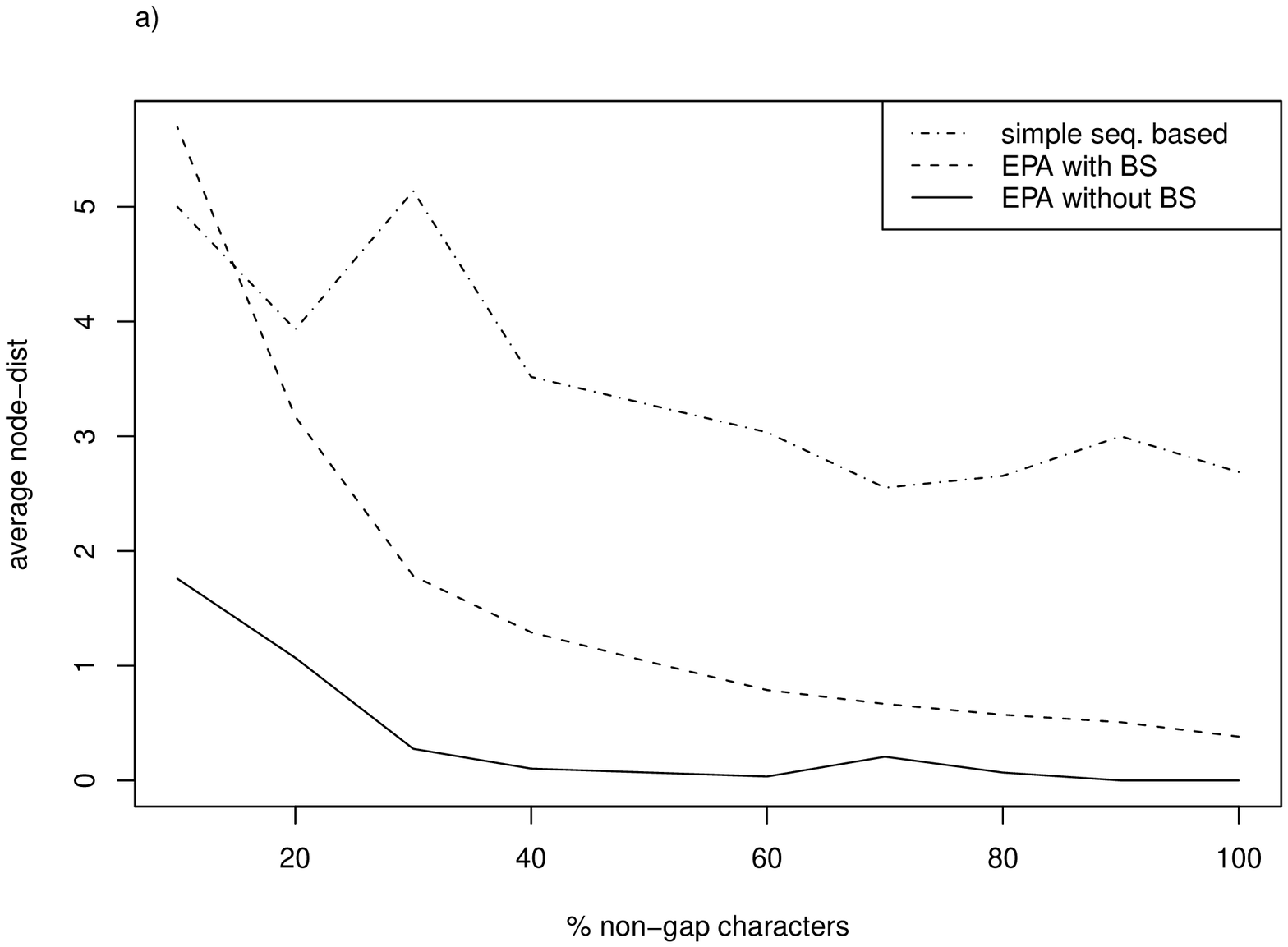}
\includegraphics[width=0.49\columnwidth]{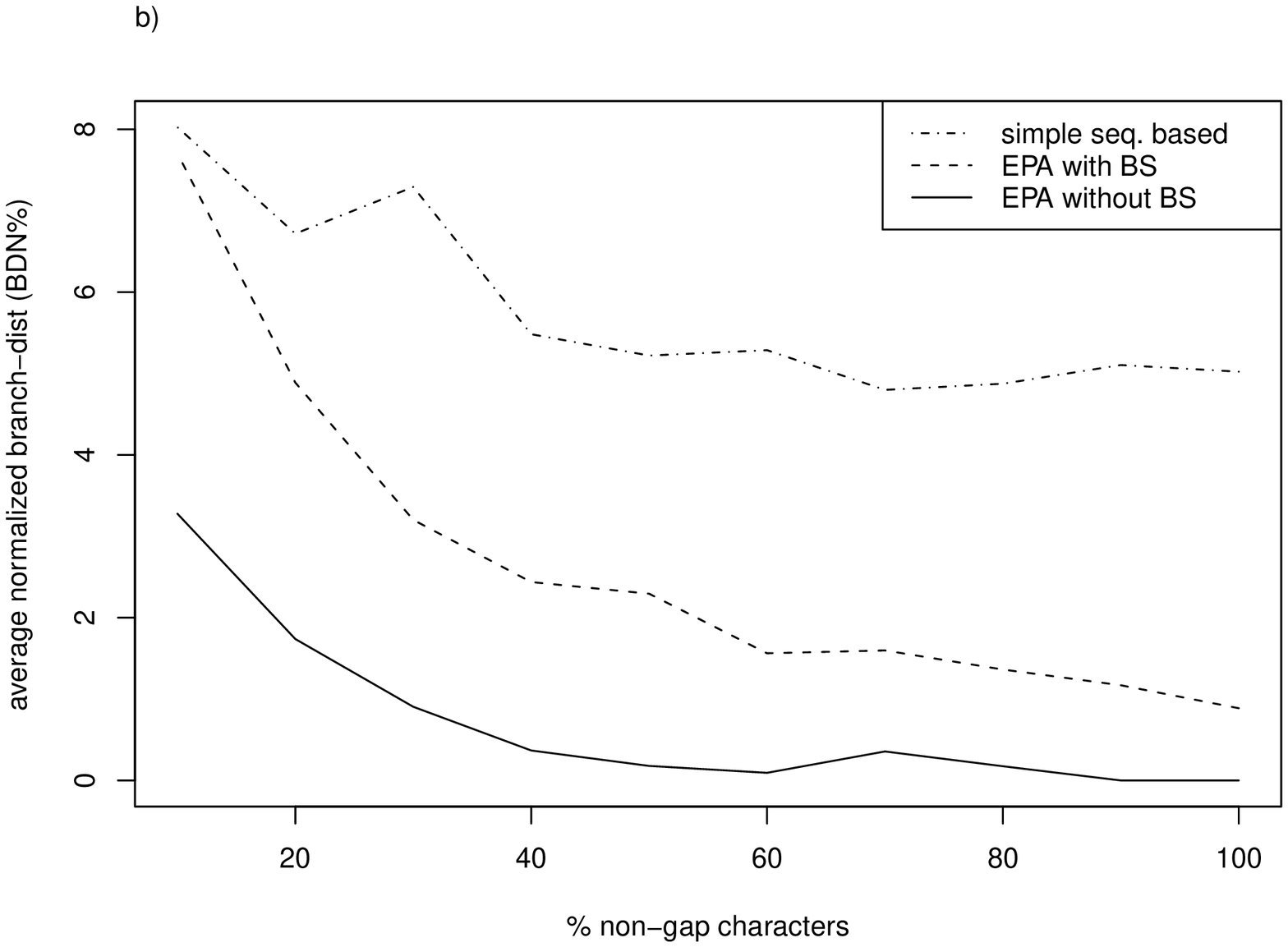}
\caption{Prediction accuracy on inner QS from data set D500. (a) Average node distance and (b) normalized branch distance between insertion positions and real positions. \label{gappy_hard}}
\end{figure}

\newpage 


\begin{figure}[ht]\centering

\includegraphics[width=0.49\columnwidth]{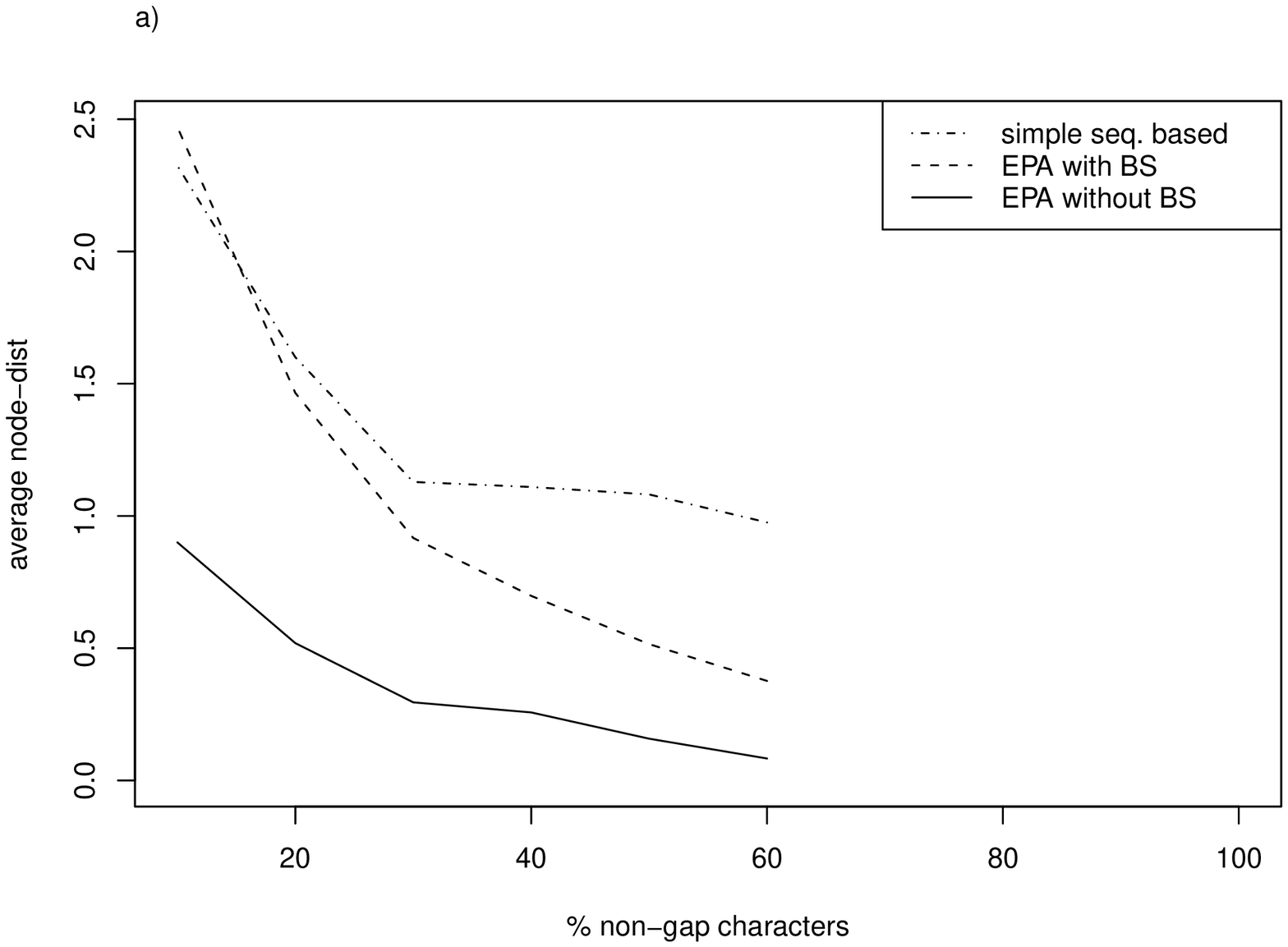}
\includegraphics[width=0.49\columnwidth]{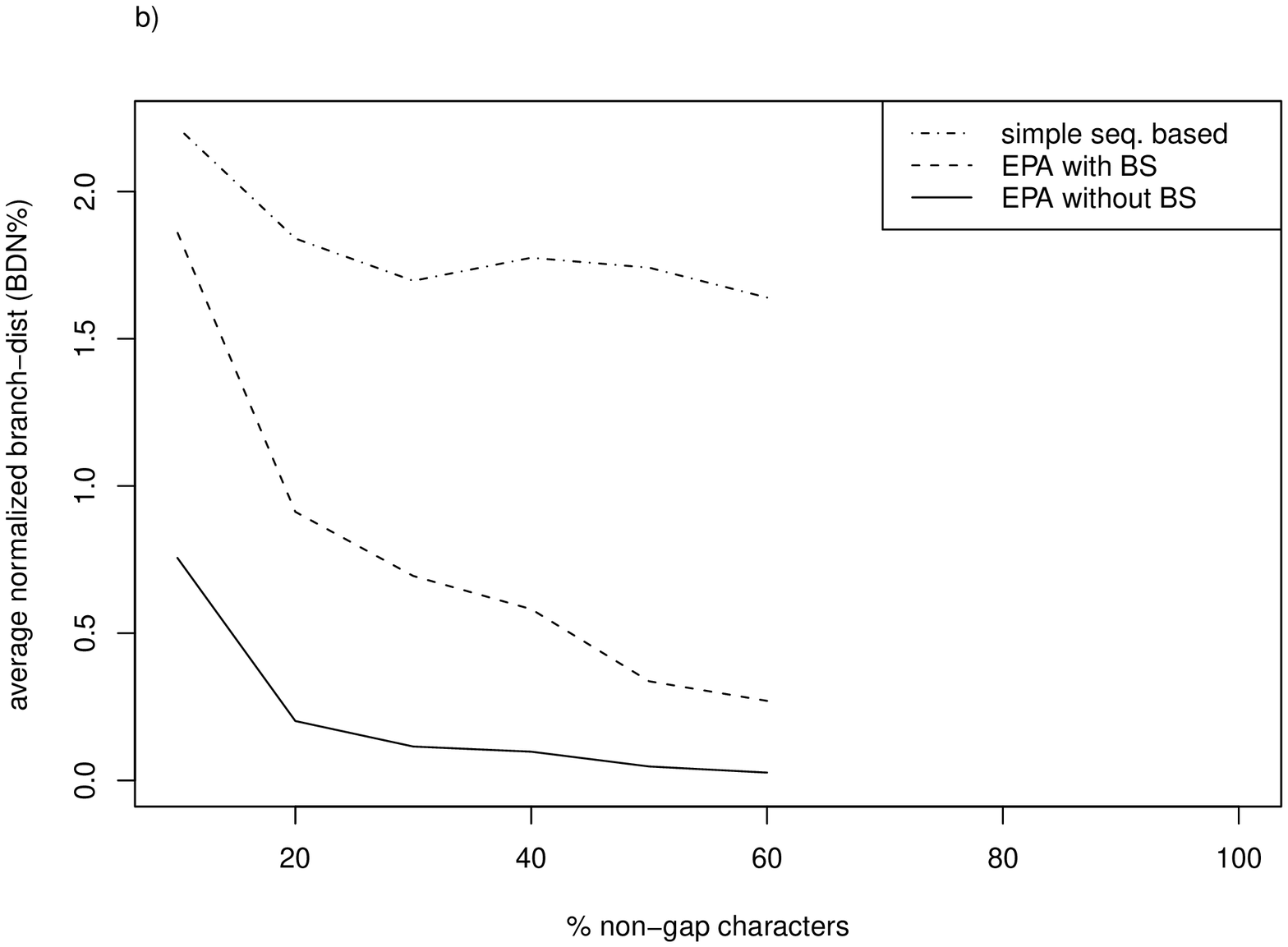}

\caption{Prediction accuracy on all QS from data set D628. (a) Average node distance and (b) Normalized Branch Distance between insertion positions and real positions.. \label{gappy_all}}
\end{figure}

\begin{figure}[ht]\centering
\includegraphics[width=0.49\columnwidth]{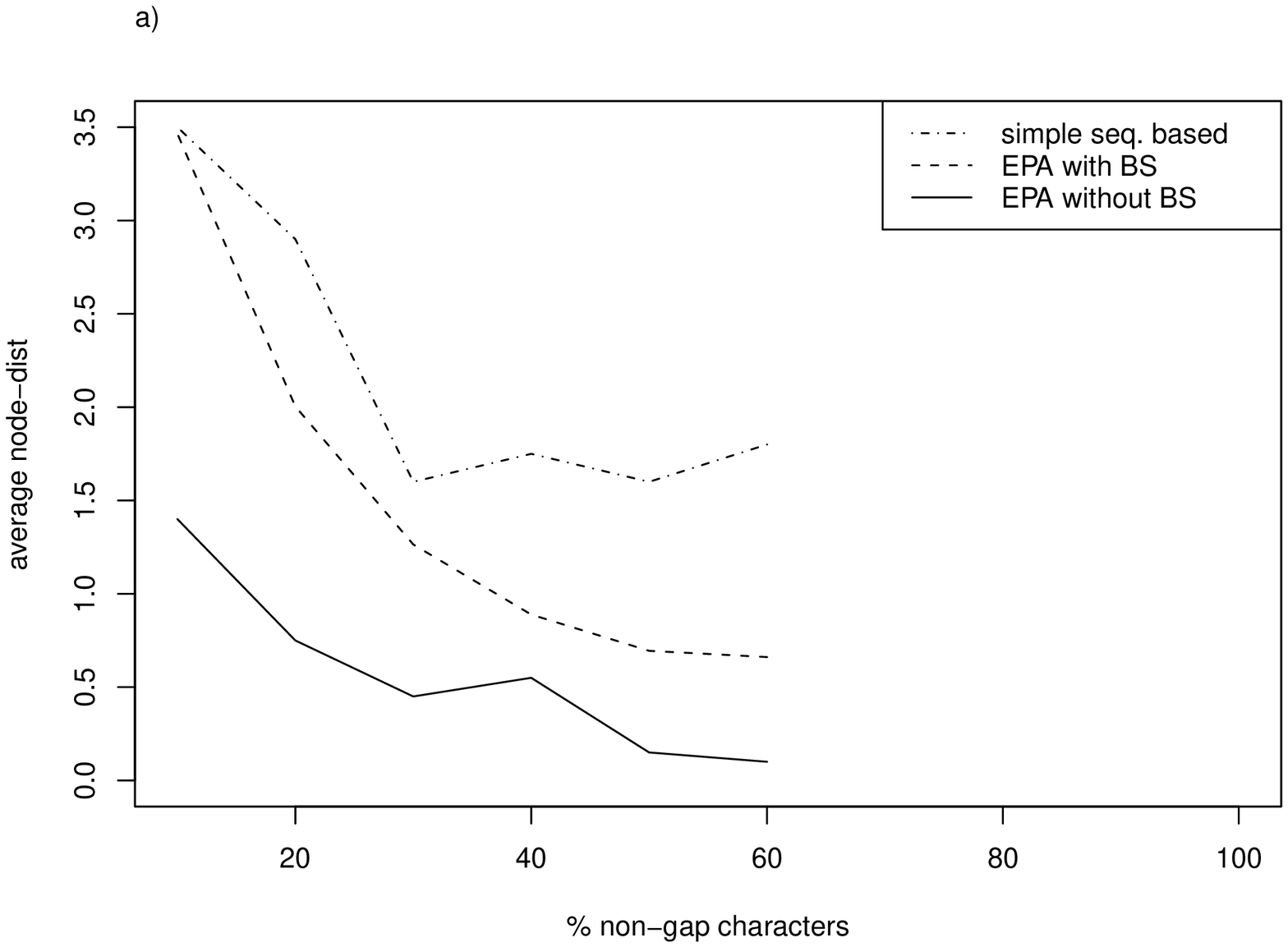}
\includegraphics[width=0.49\columnwidth]{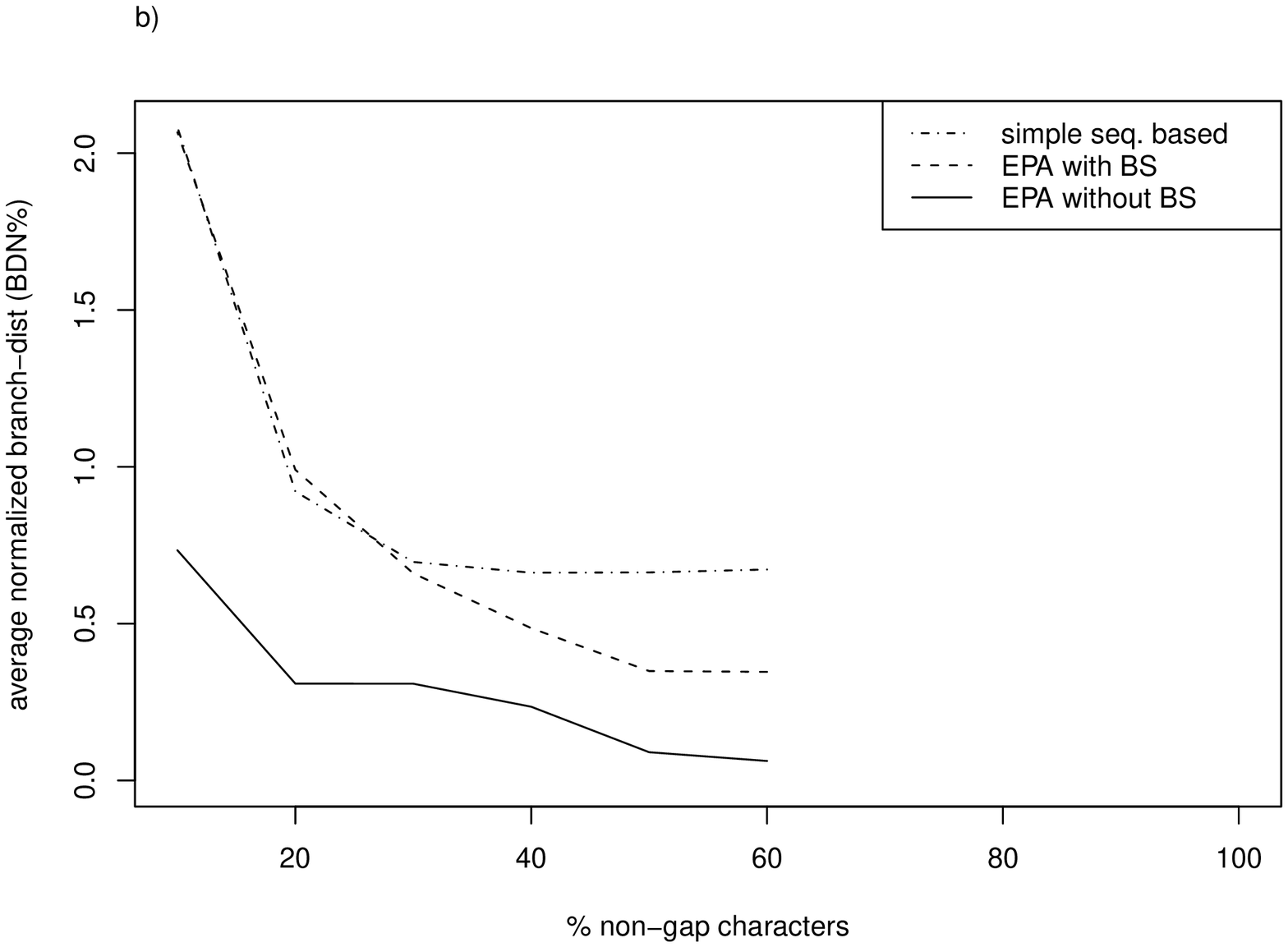}
\caption{Prediction accuracy on inner QS from data set D628. (a) Average node distance and (b) normalized branch distance between insertion positions and real positions. \label{gappy_hard}}
\end{figure}

\newpage 


\begin{figure}[ht]\centering

\includegraphics[width=0.49\columnwidth]{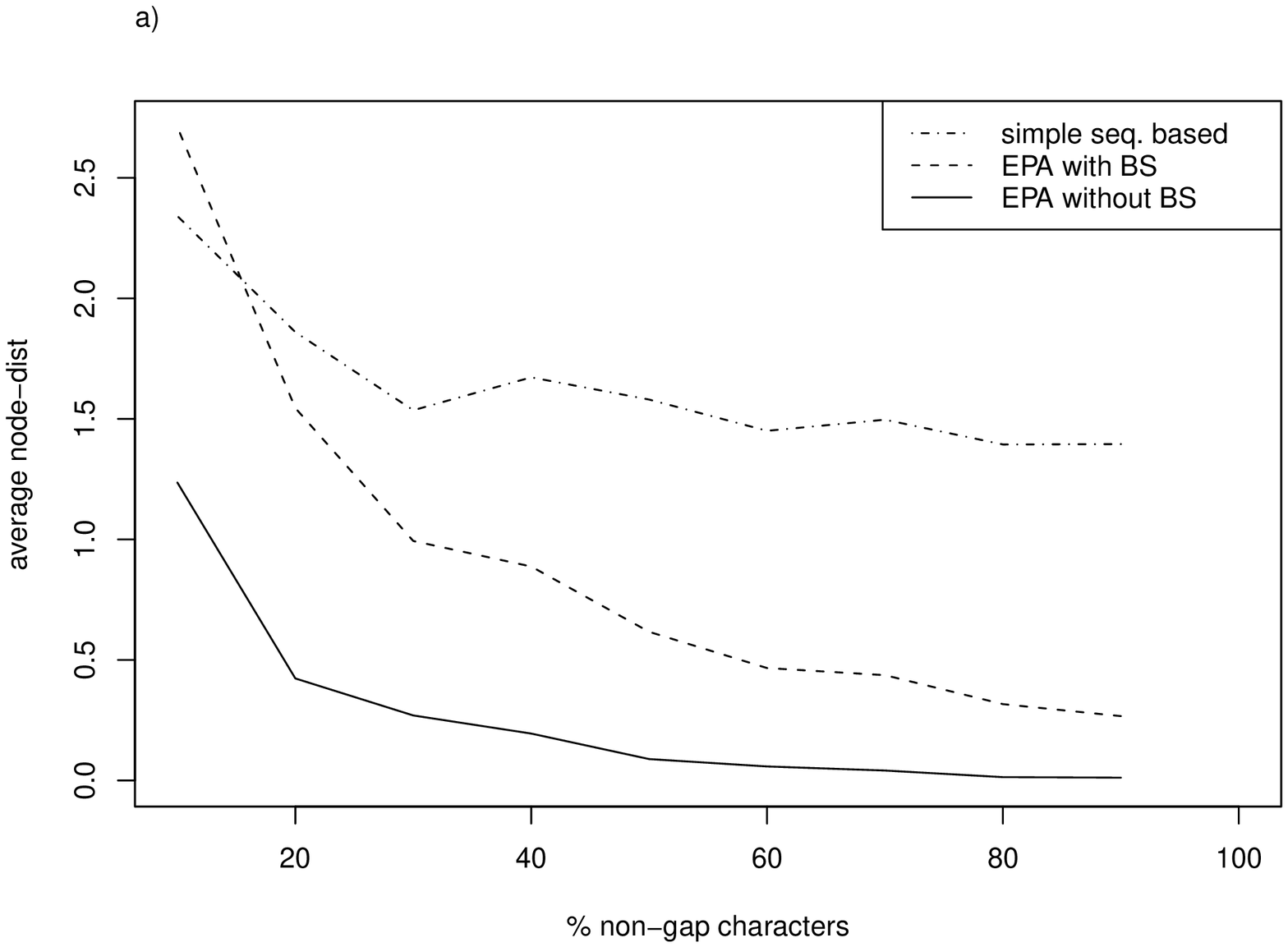}
\includegraphics[width=0.49\columnwidth]{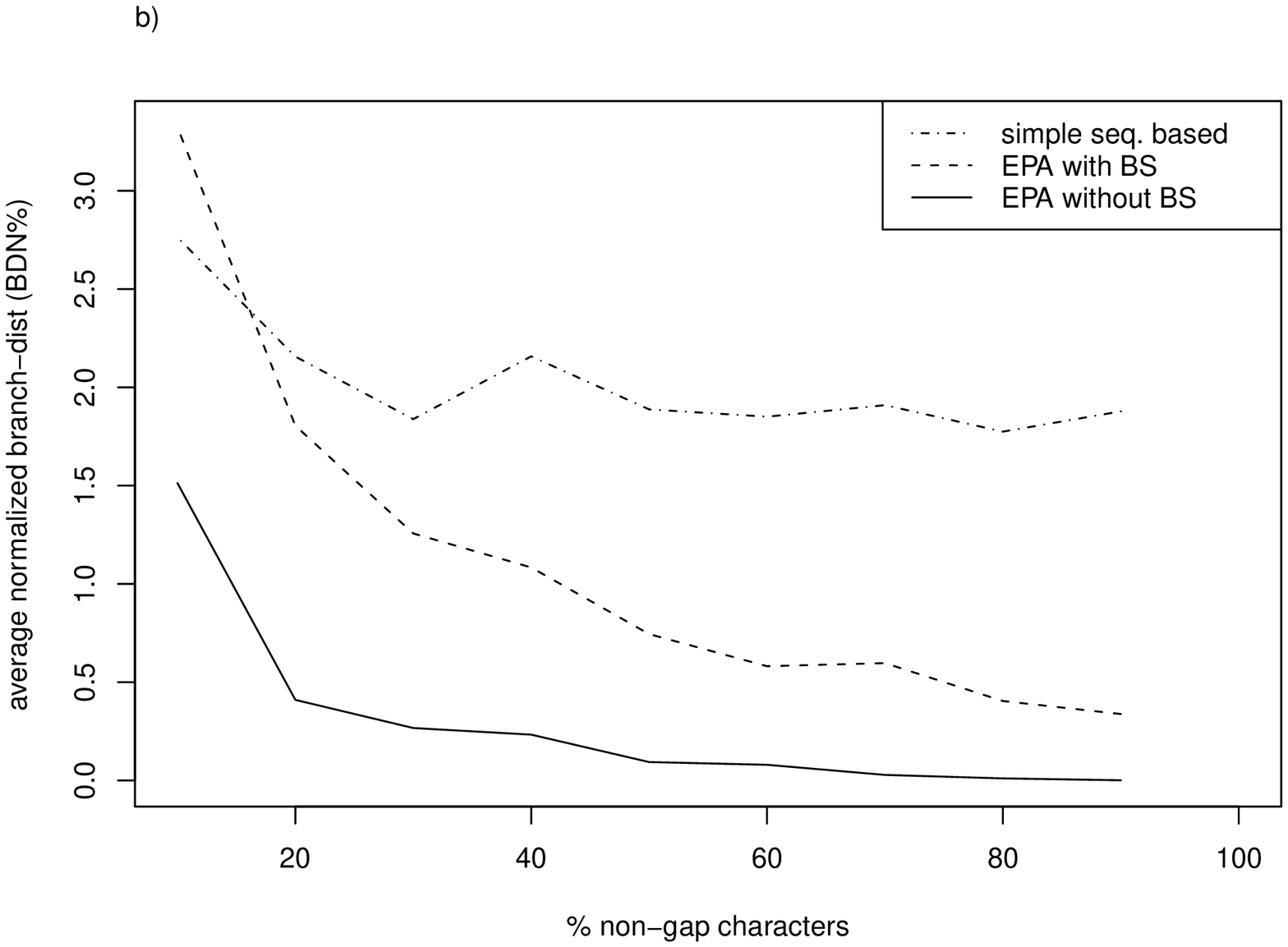}

\caption{Prediction accuracy on all QS from data set D714. (a) Average node distance and (b) Normalized Branch Distance between insertion positions and real positions.. \label{gappy_all}}
\end{figure}

\begin{figure}[ht]\centering
\includegraphics[width=0.49\columnwidth]{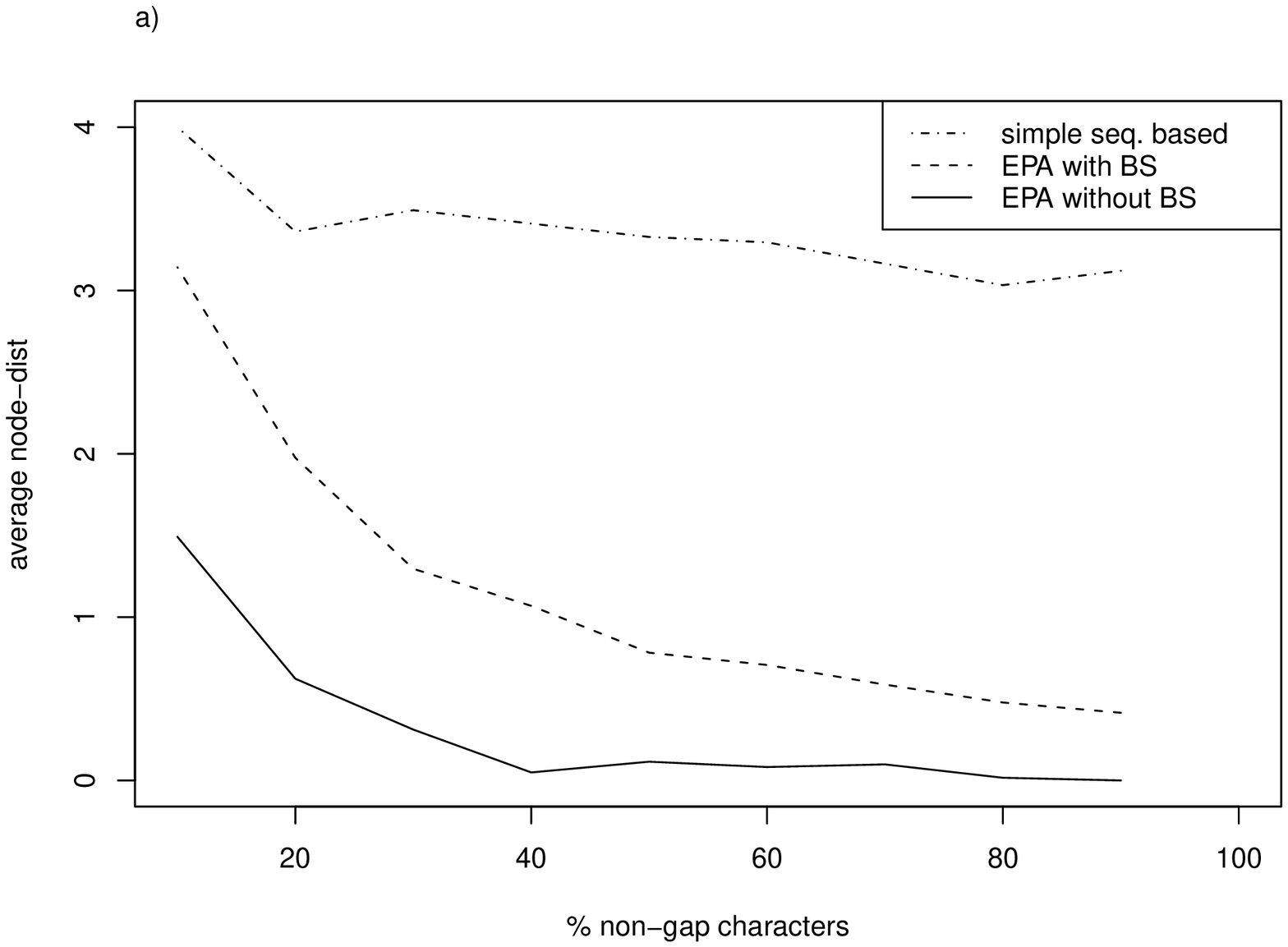}
\includegraphics[width=0.49\columnwidth]{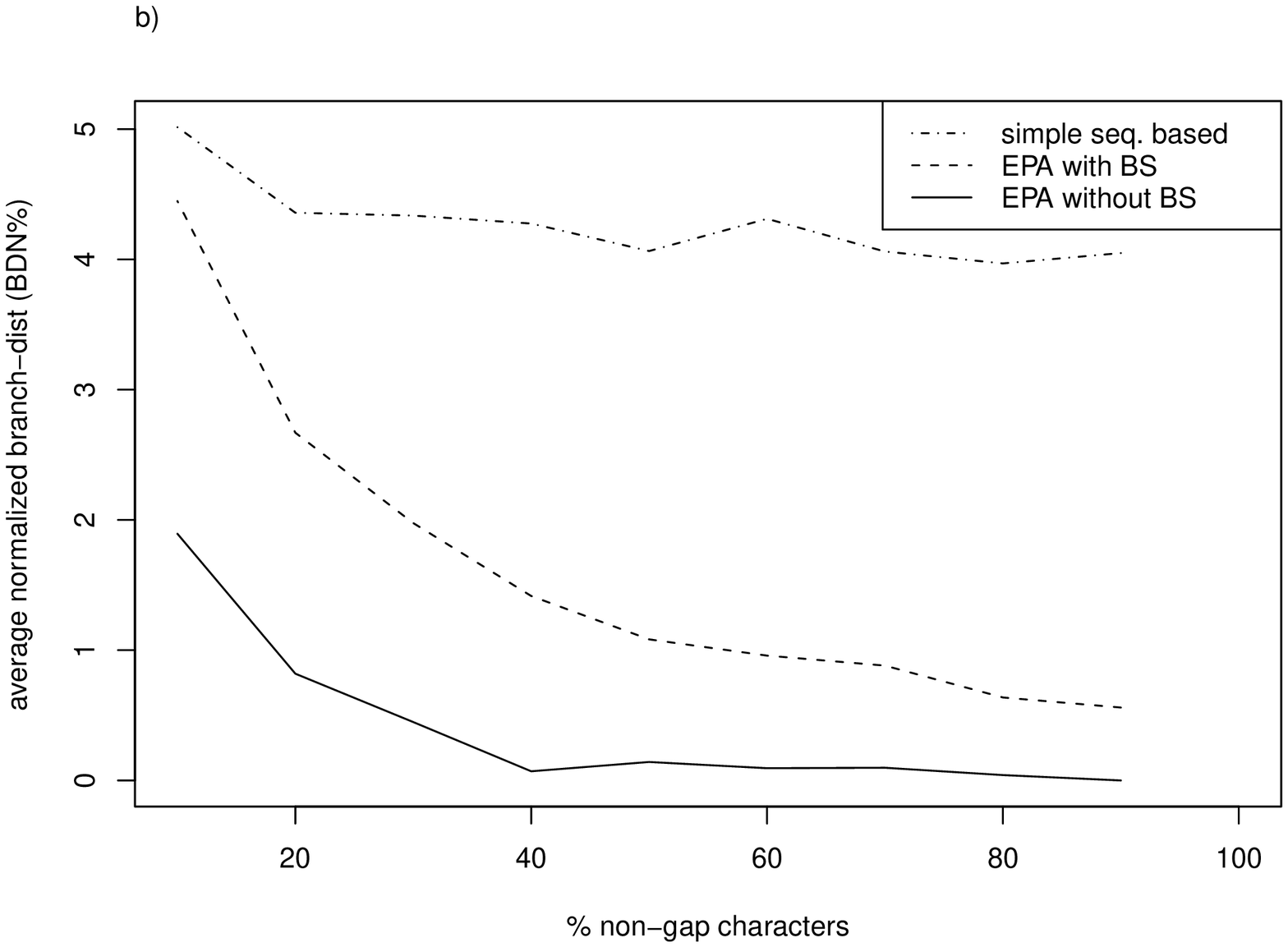}
\caption{Prediction accuracy on inner QS from data set D714. (a) Average node distance and (b) normalized branch distance between insertion positions and real positions. \label{gappy_hard}}
\end{figure}

\newpage 


\begin{figure}[ht]\centering

\includegraphics[width=0.49\columnwidth]{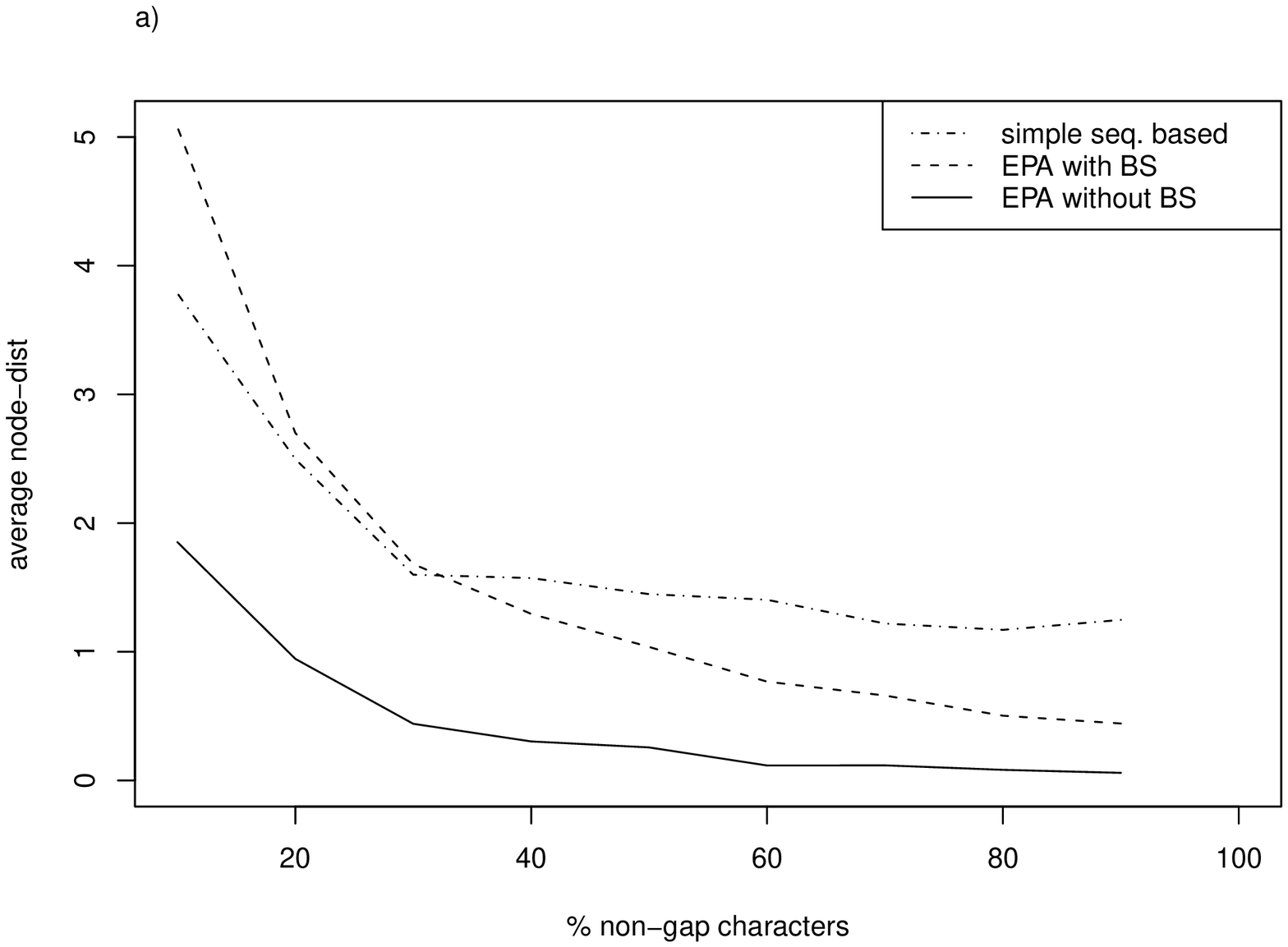}
\includegraphics[width=0.49\columnwidth]{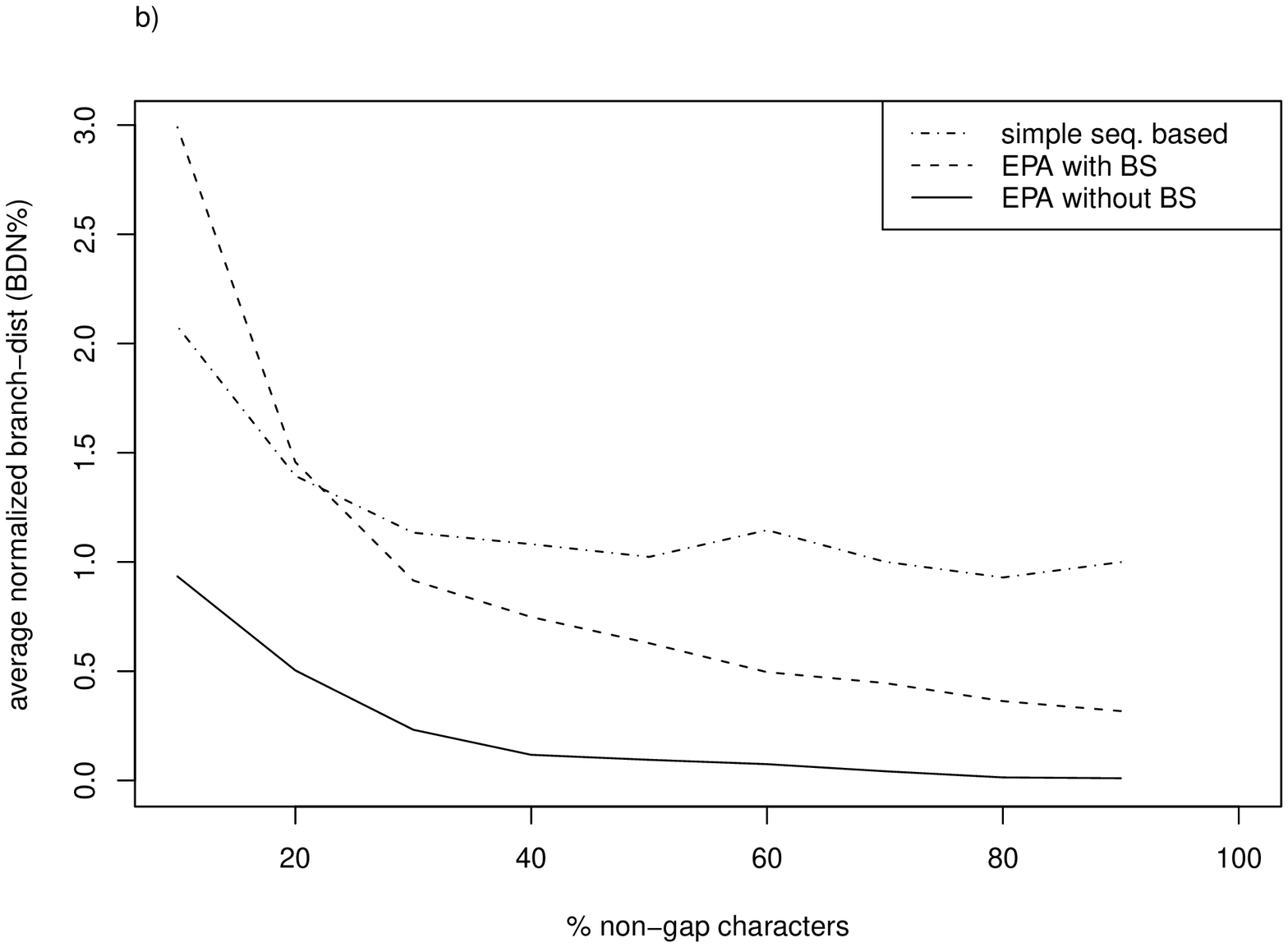}

\caption{Prediction accuracy on all QS from data set D855. (a) Average node distance and (b) Normalized Branch Distance between insertion positions and real positions.. \label{gappy_all}}
\end{figure}

\begin{figure}[ht]\centering
\includegraphics[width=0.49\columnwidth]{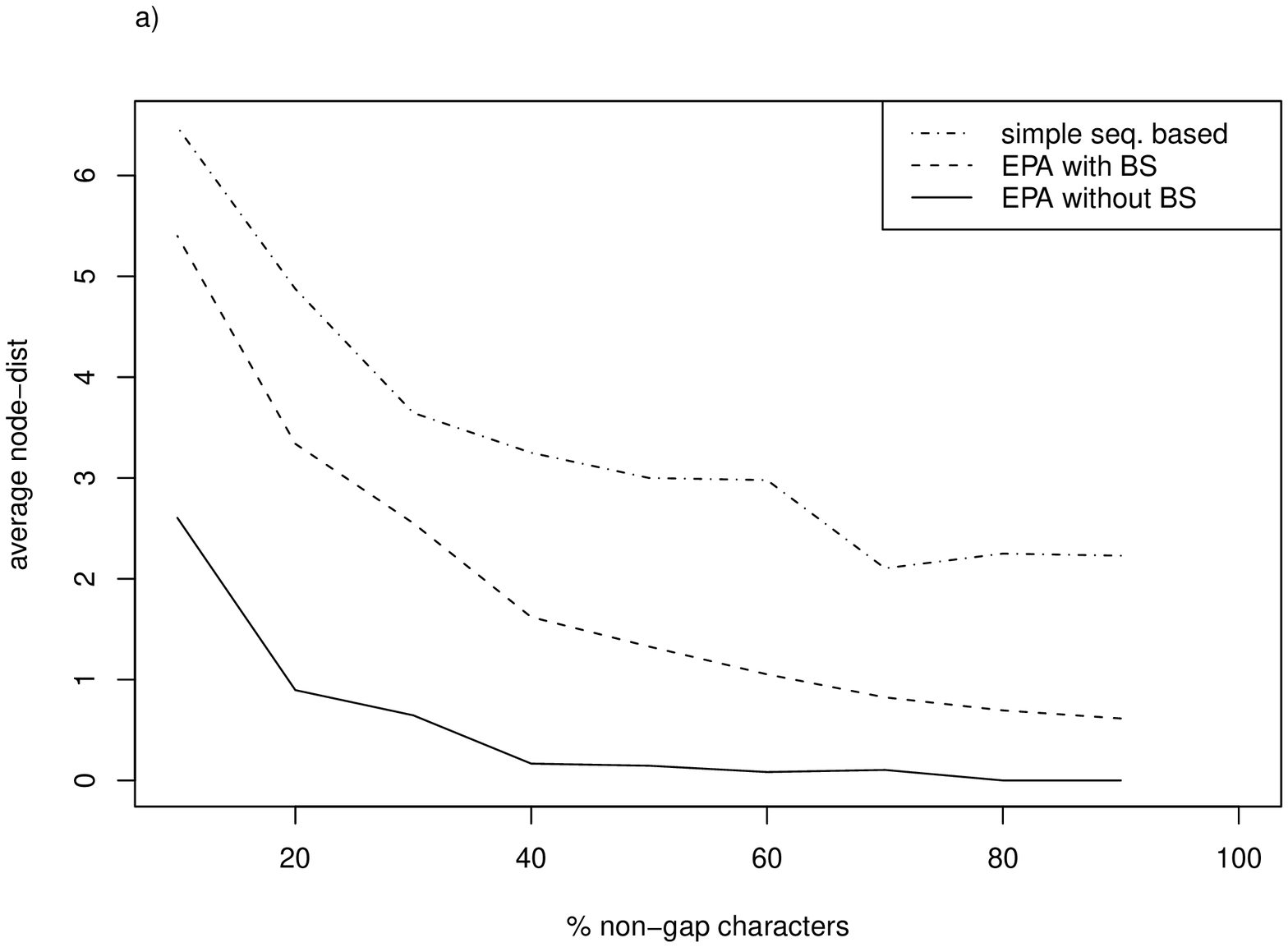}
\includegraphics[width=0.49\columnwidth]{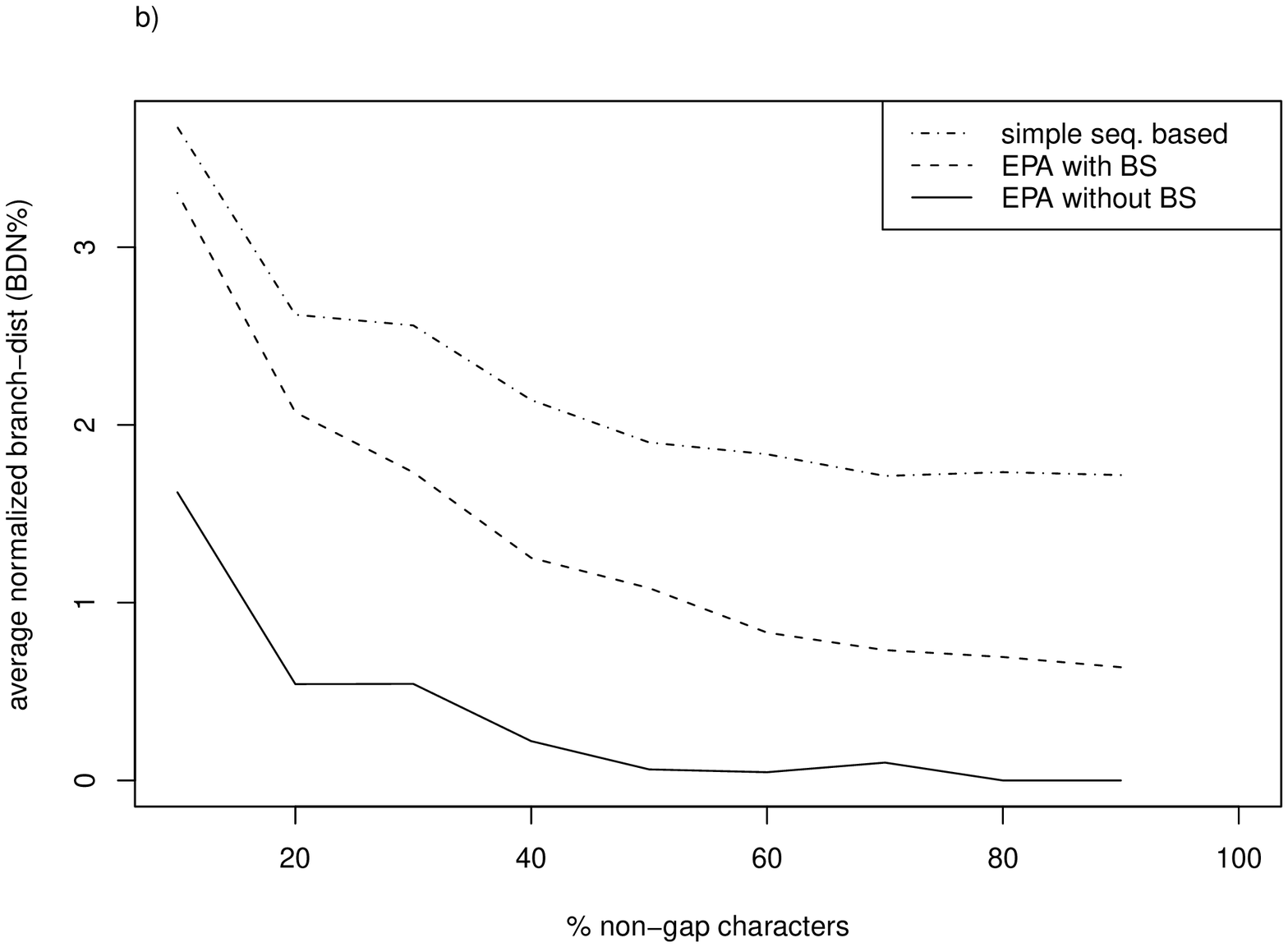}
\caption{Prediction accuracy on inner QS from data set D855. (a) Average node distance and (b) normalized branch distance between insertion positions and real positions. \label{gappy_hard}}
\end{figure}

\newpage 


\begin{figure}[ht]\centering

\includegraphics[width=0.49\columnwidth]{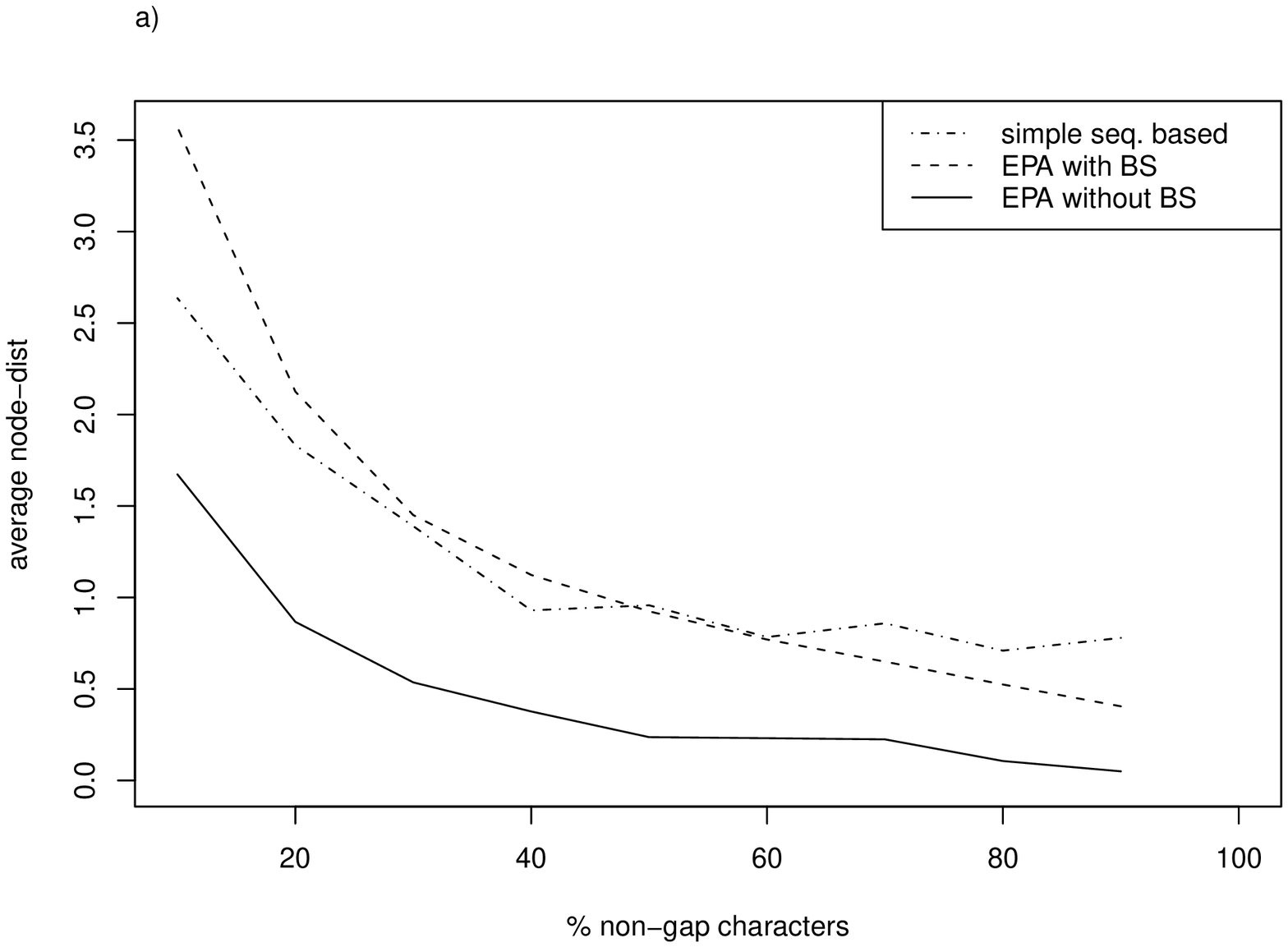}
\includegraphics[width=0.49\columnwidth]{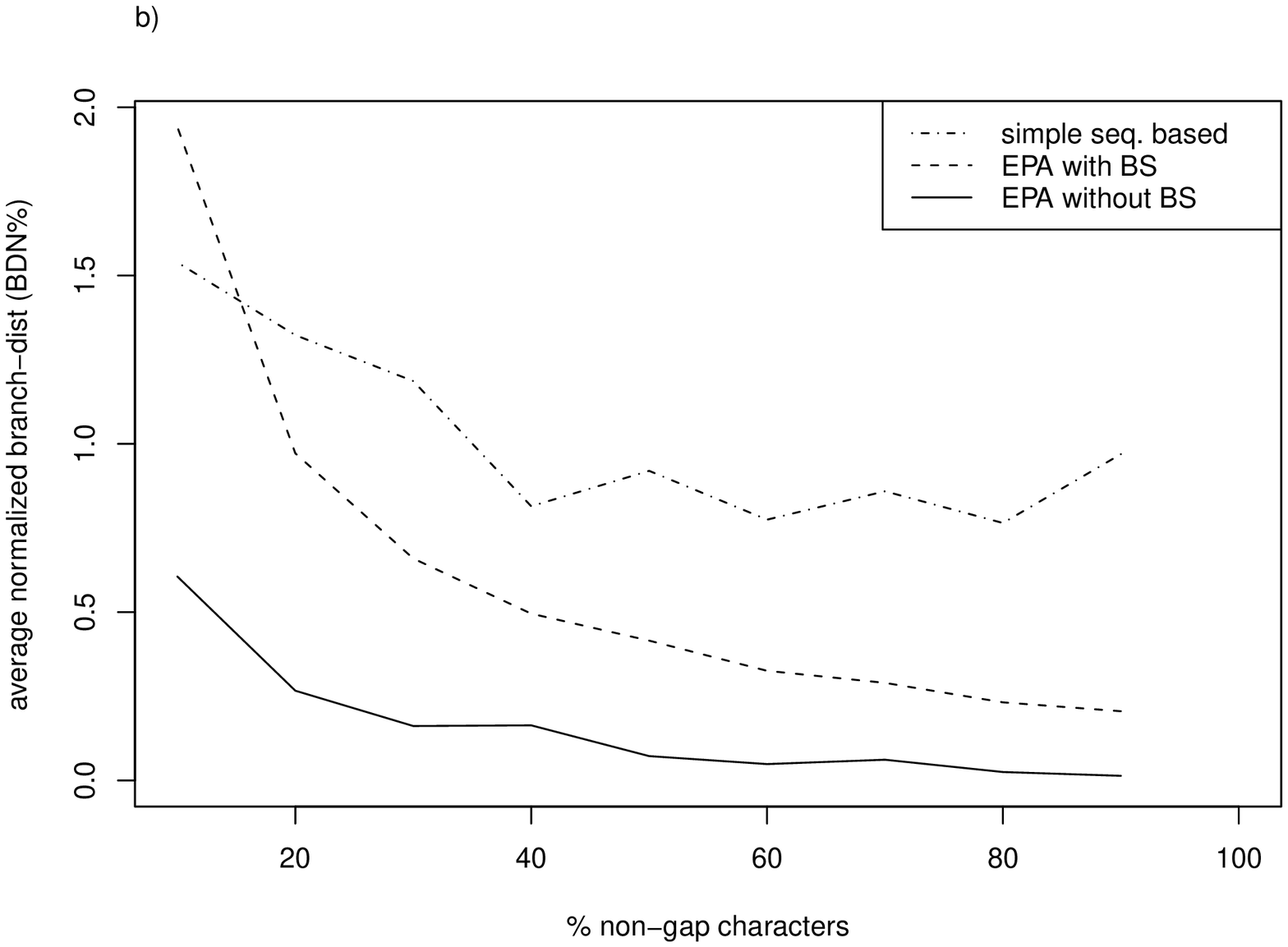}

\caption{Prediction accuracy on all QS from data set D1604. (a) Average node distance and (b) Normalized Branch Distance between insertion positions and real positions.. \label{gappy_all}}
\end{figure}

\begin{figure}[ht]\centering
\includegraphics[width=0.49\columnwidth]{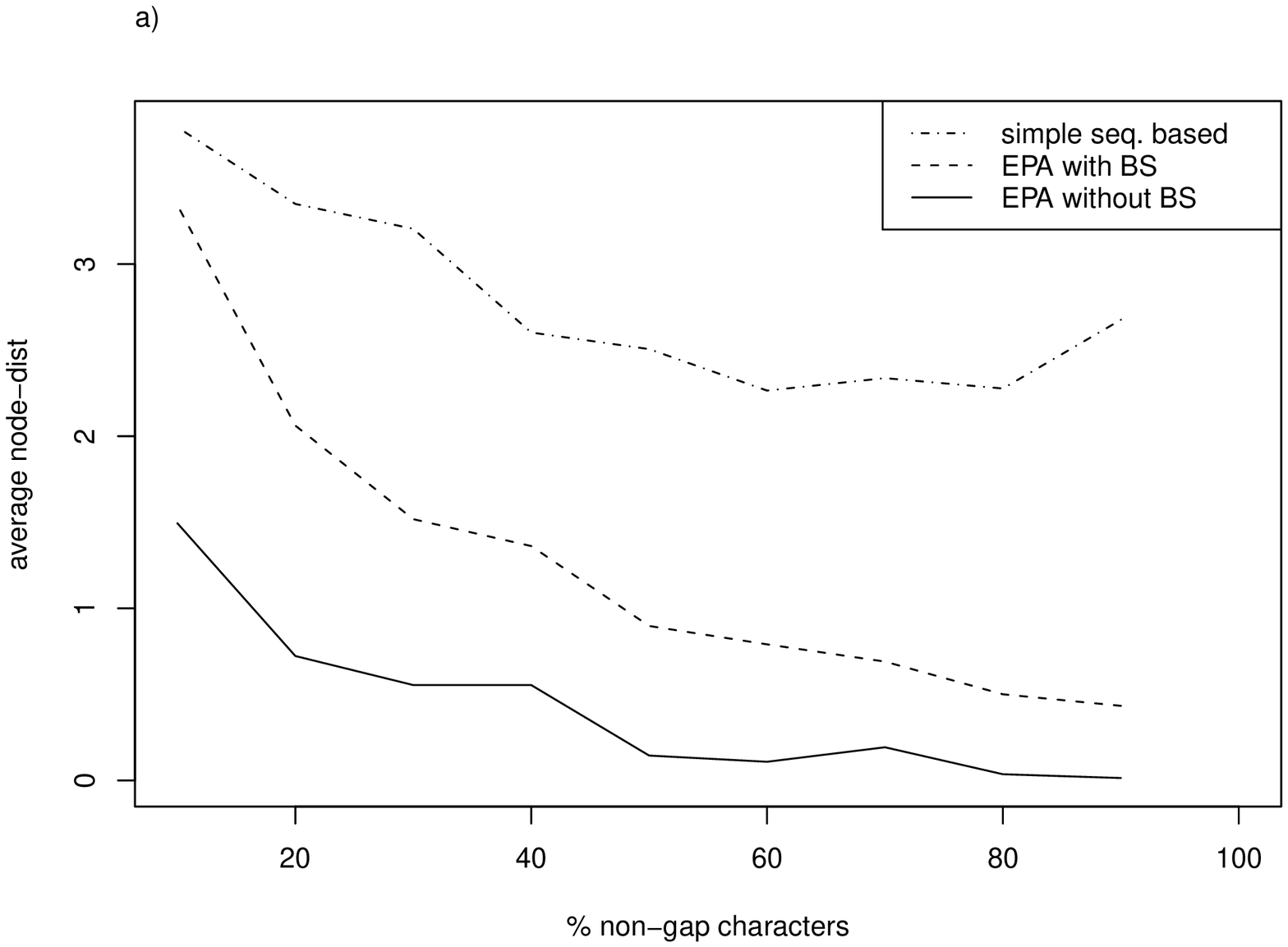}
\includegraphics[width=0.49\columnwidth]{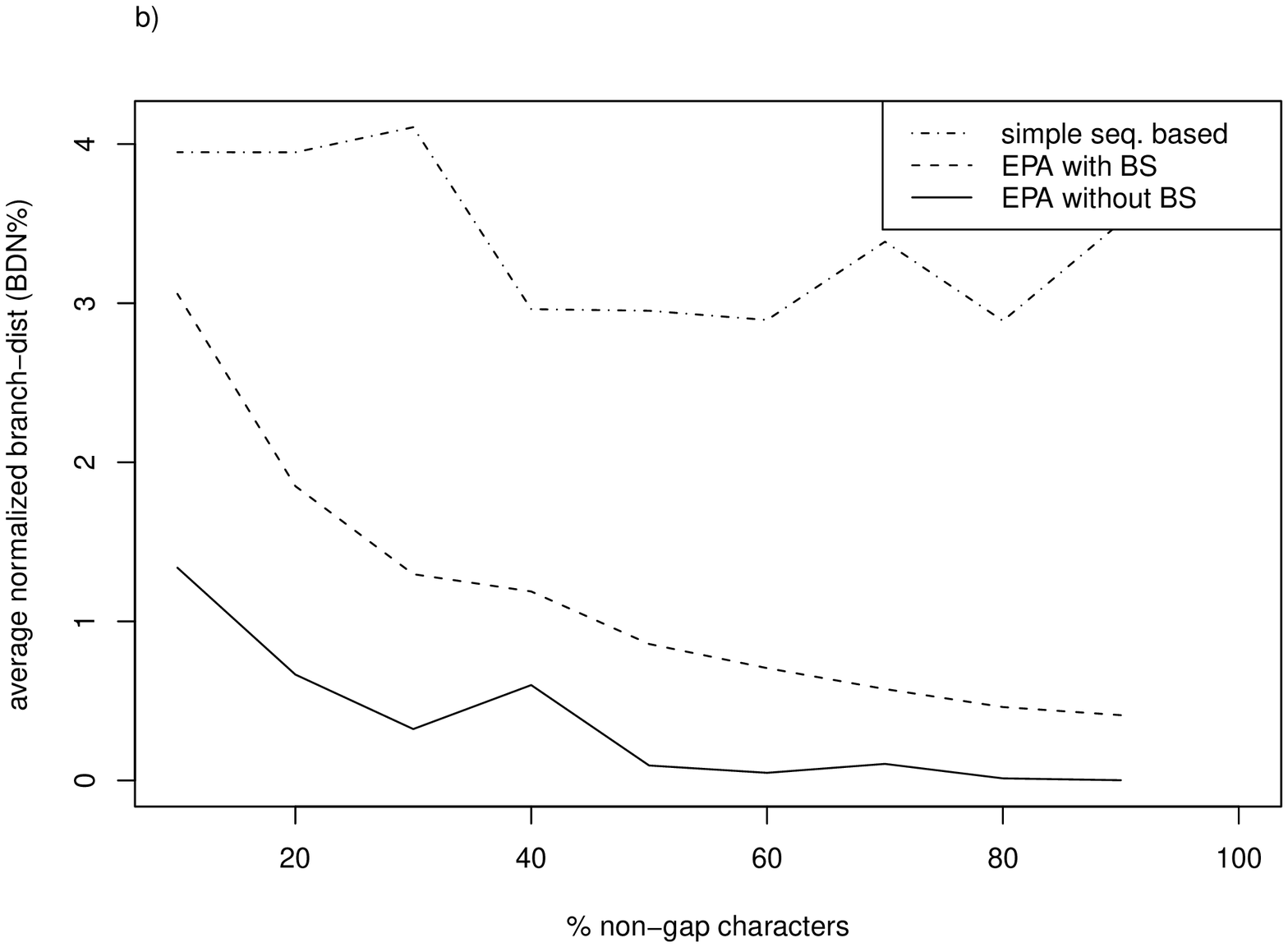}
\caption{Prediction accuracy on inner QS from data set D1604. (a) Average node distance and (b) normalized branch distance between insertion positions and real positions. \label{gappy_hard}}
\end{figure}

\newpage 


\clearpage 
\begin{figure}[ht]
\centering
\includegraphics[width=0.49\columnwidth]{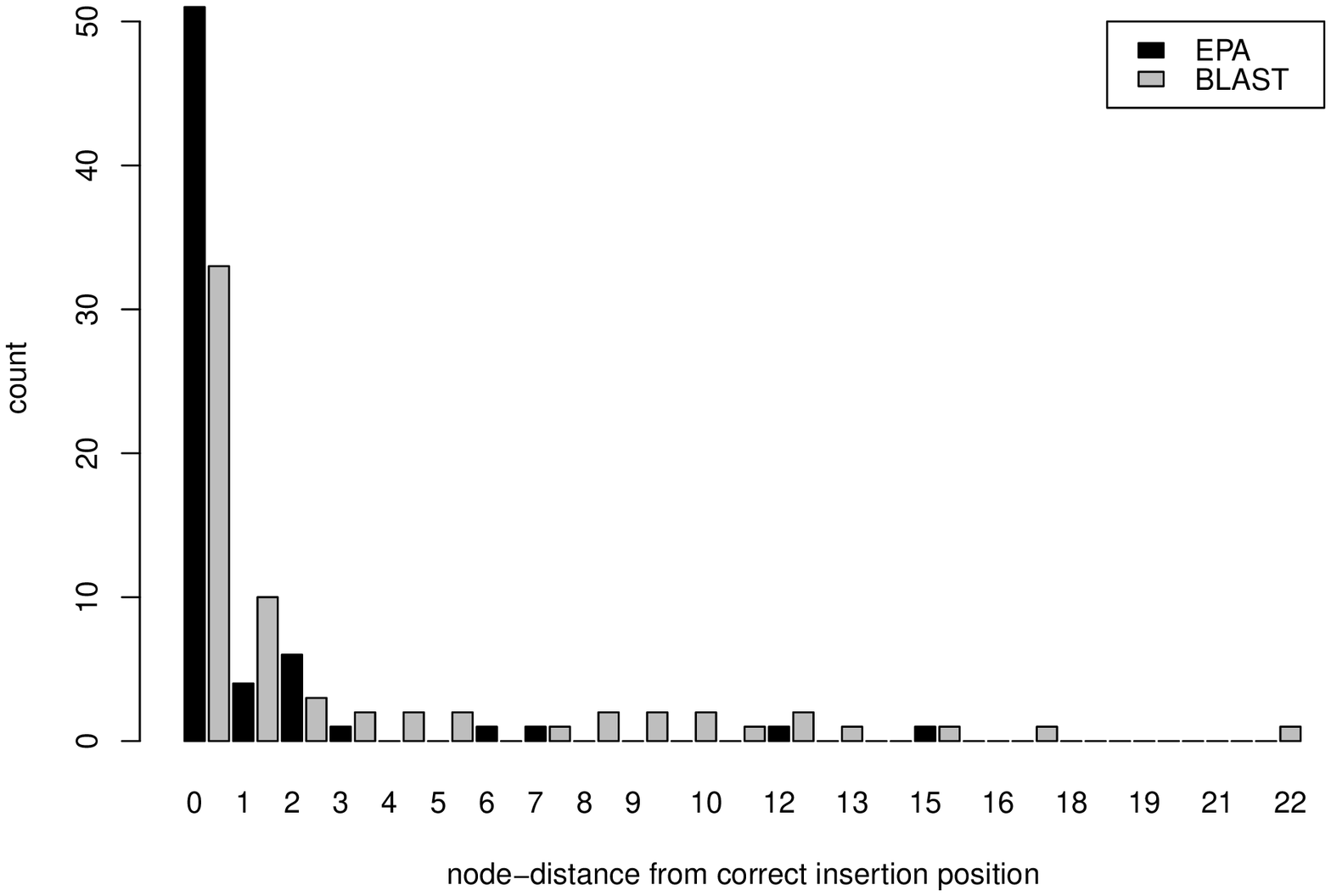}
\includegraphics[width=0.49\columnwidth]{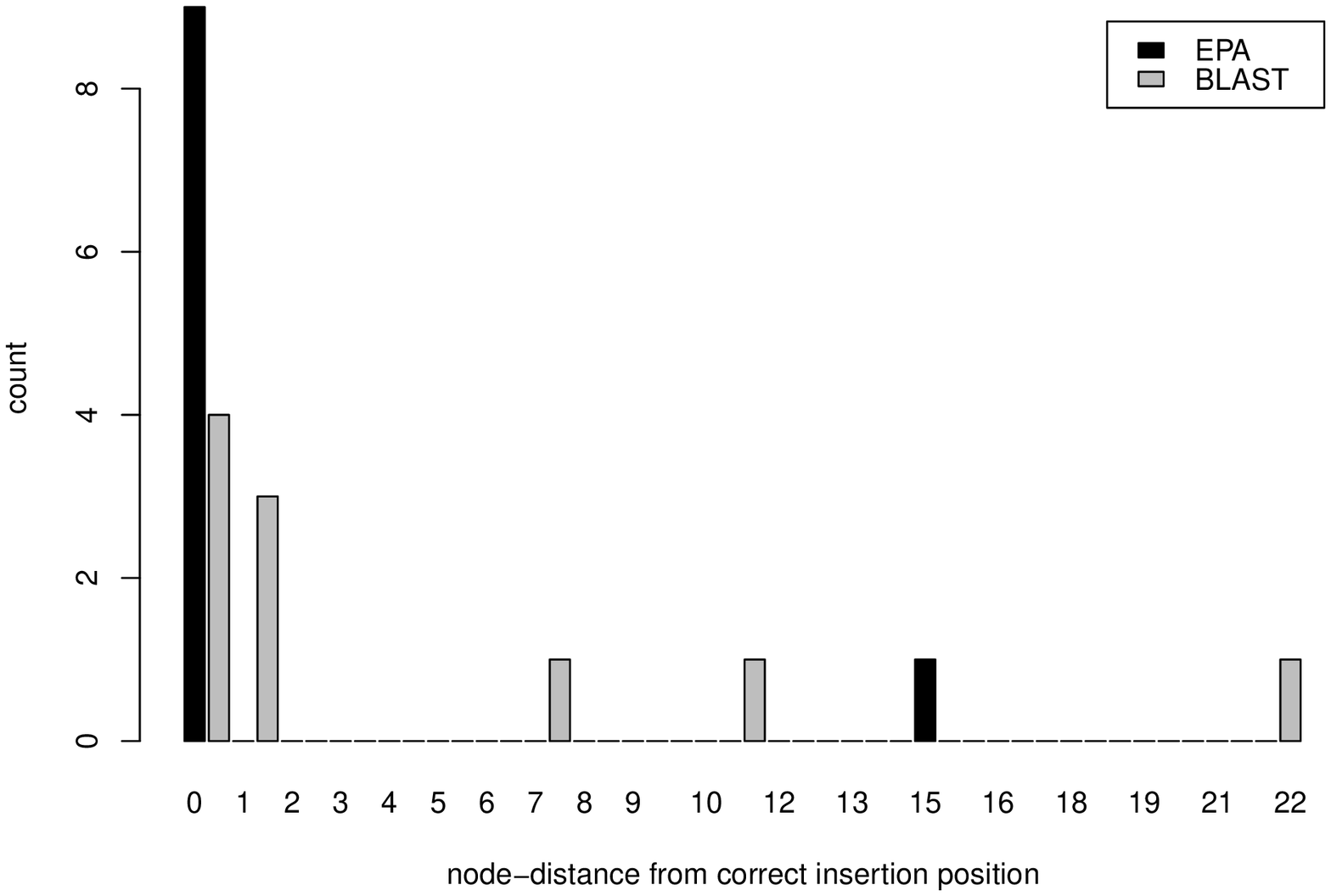}
\caption{Histogram plot of the prediction accuracies (Node Distance) for the placement of 2x100 BP paired-end reads on data set D150.The left plot comprises all QS, the right plot only inner QS.}
\end{figure}

\begin{figure}[ht]
\centering
\includegraphics[width=0.49\columnwidth]{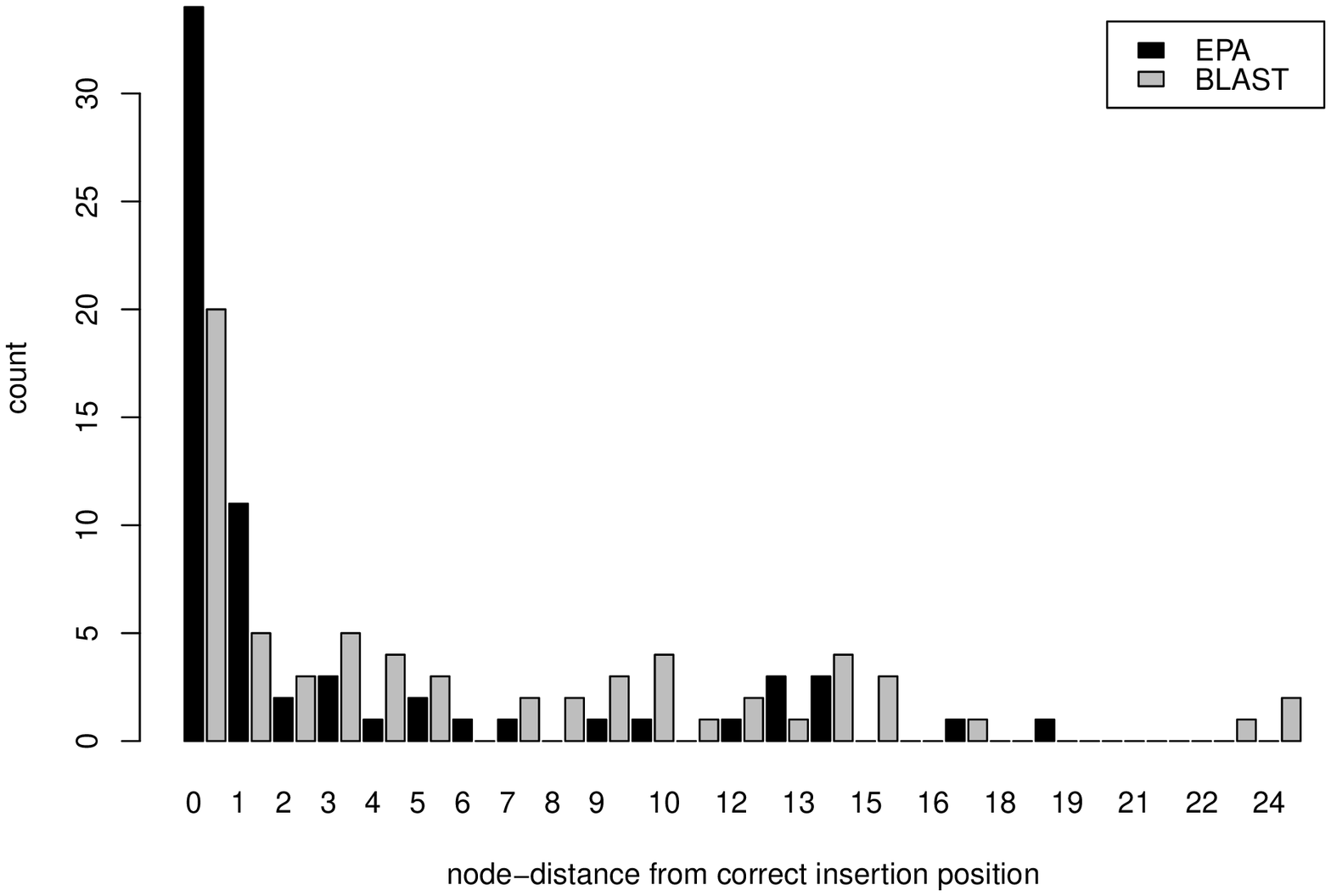}
\includegraphics[width=0.49\columnwidth]{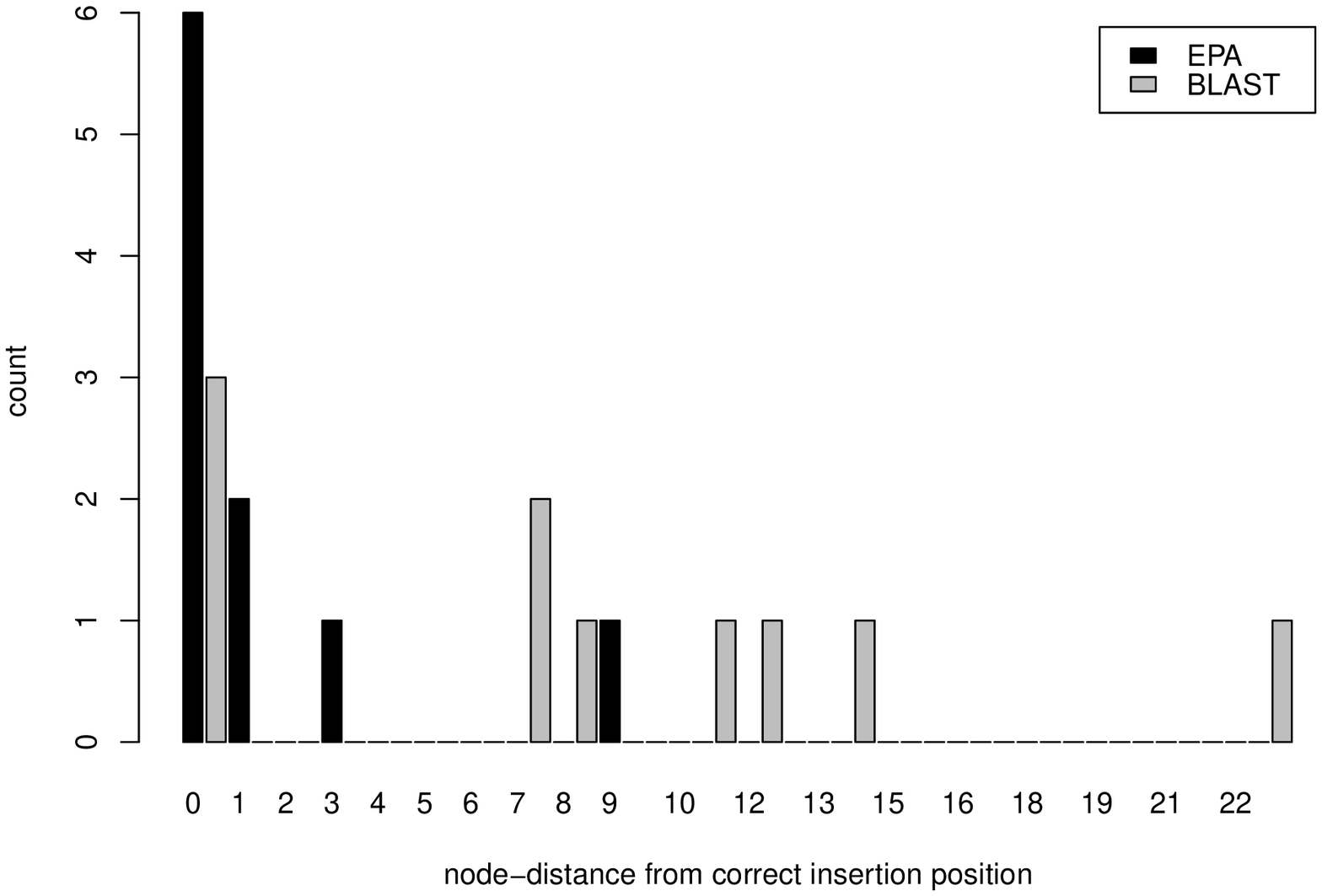}
\caption{Histogram plot of the prediction accuracies (Node Distance) for the placement of 2x50 BP paired-end reads on data set D150. The left plot comprises all QS, the right plot only inner QS.}
\end{figure}


\begin{figure}[ht]
\centering
\includegraphics[width=0.49\columnwidth]{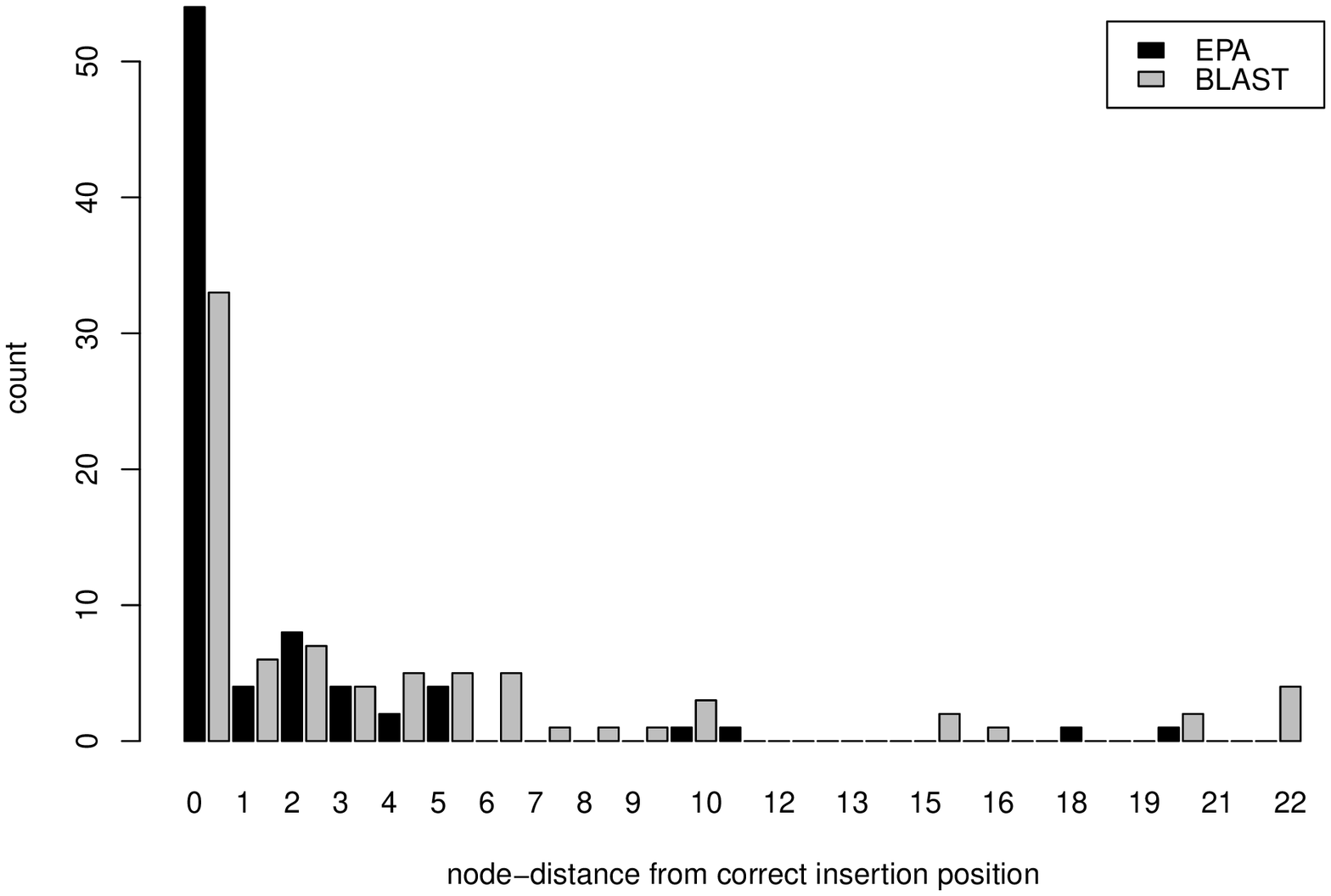}
\includegraphics[width=0.49\columnwidth]{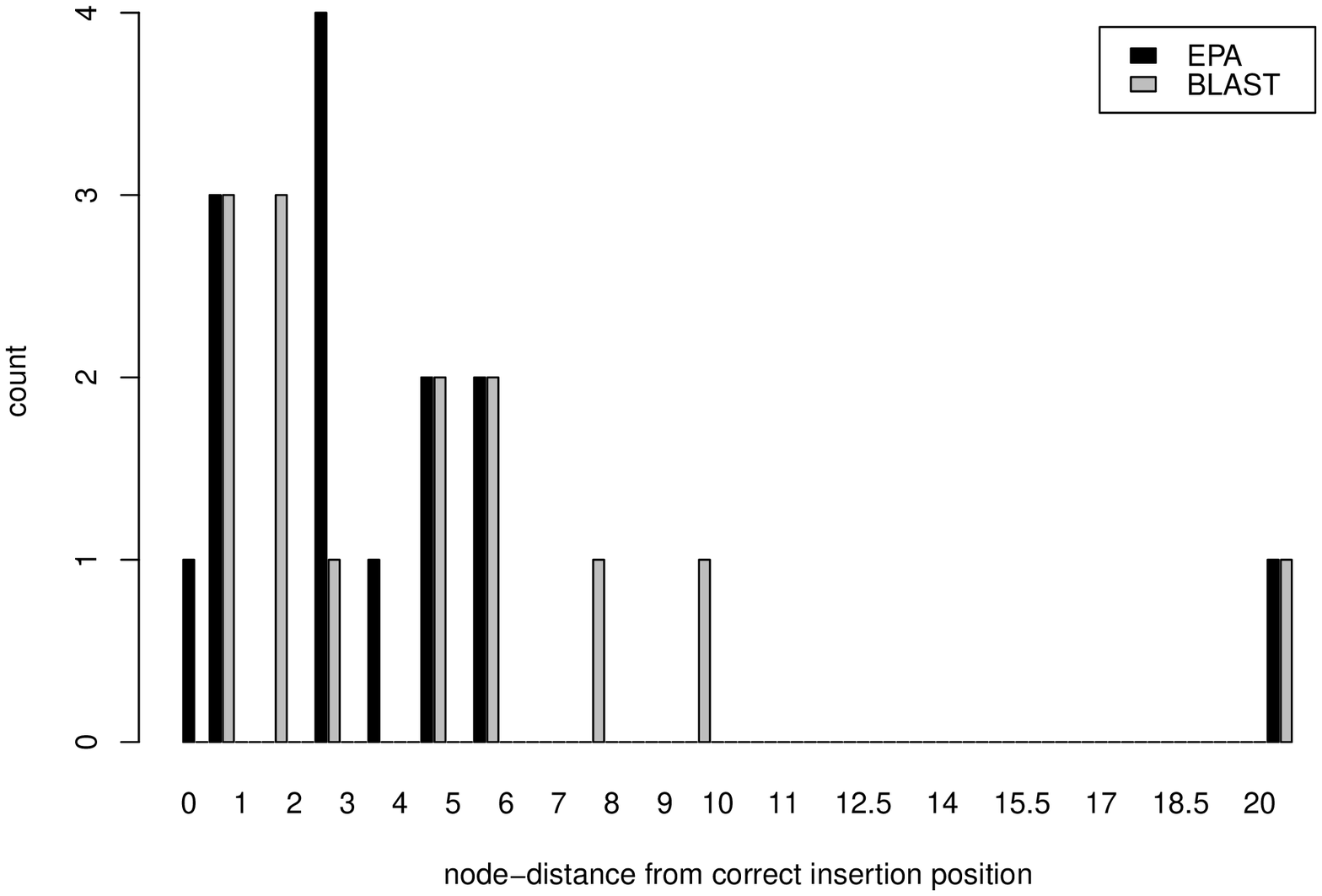}
\caption{Histogram plot of the prediction accuracies (Node Distance) for the placement of 2x100 BP paired-end reads on data set D218.The left plot comprises all QS, the right plot only inner QS.}
\end{figure}

\begin{figure}[ht]
\centering
\includegraphics[width=0.49\columnwidth]{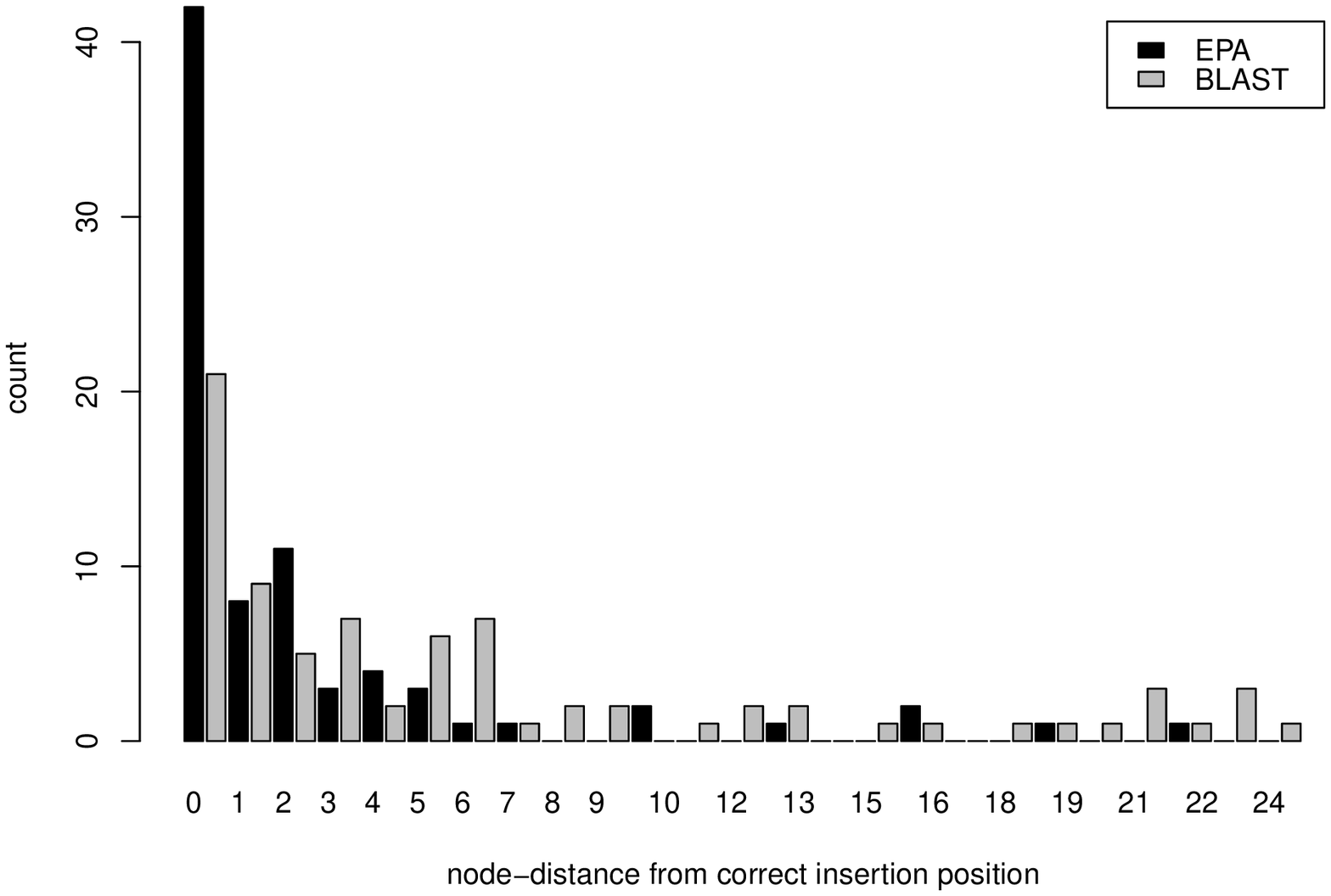}
\includegraphics[width=0.49\columnwidth]{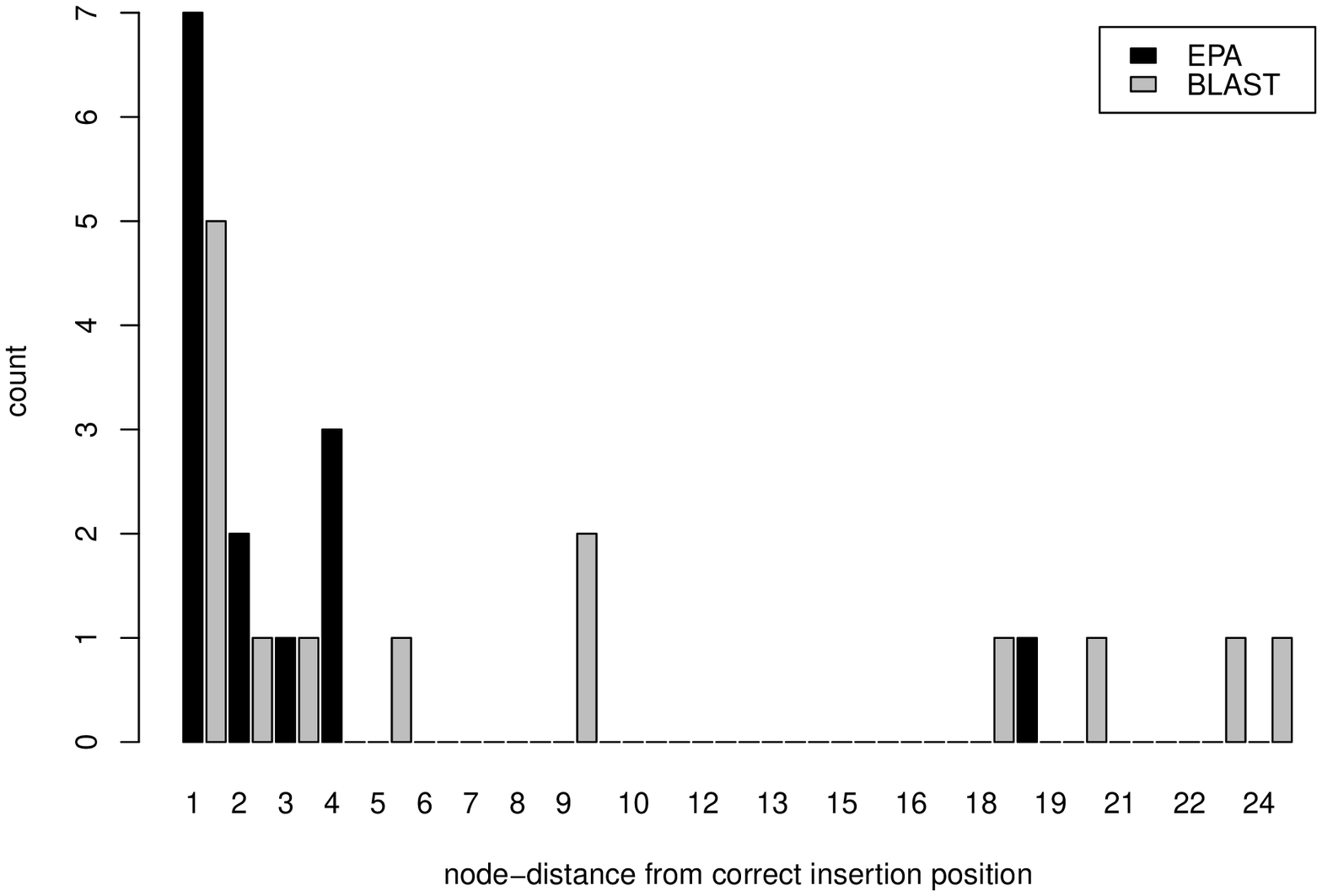}
\caption{Histogram plot of the prediction accuracies (Node Distance) for the placement of 2x50 BP paired-end reads on data set D218. The left plot comprises all QS, the right plot only inner QS.}
\end{figure}


\begin{figure}[ht]
\centering
\includegraphics[width=0.49\columnwidth]{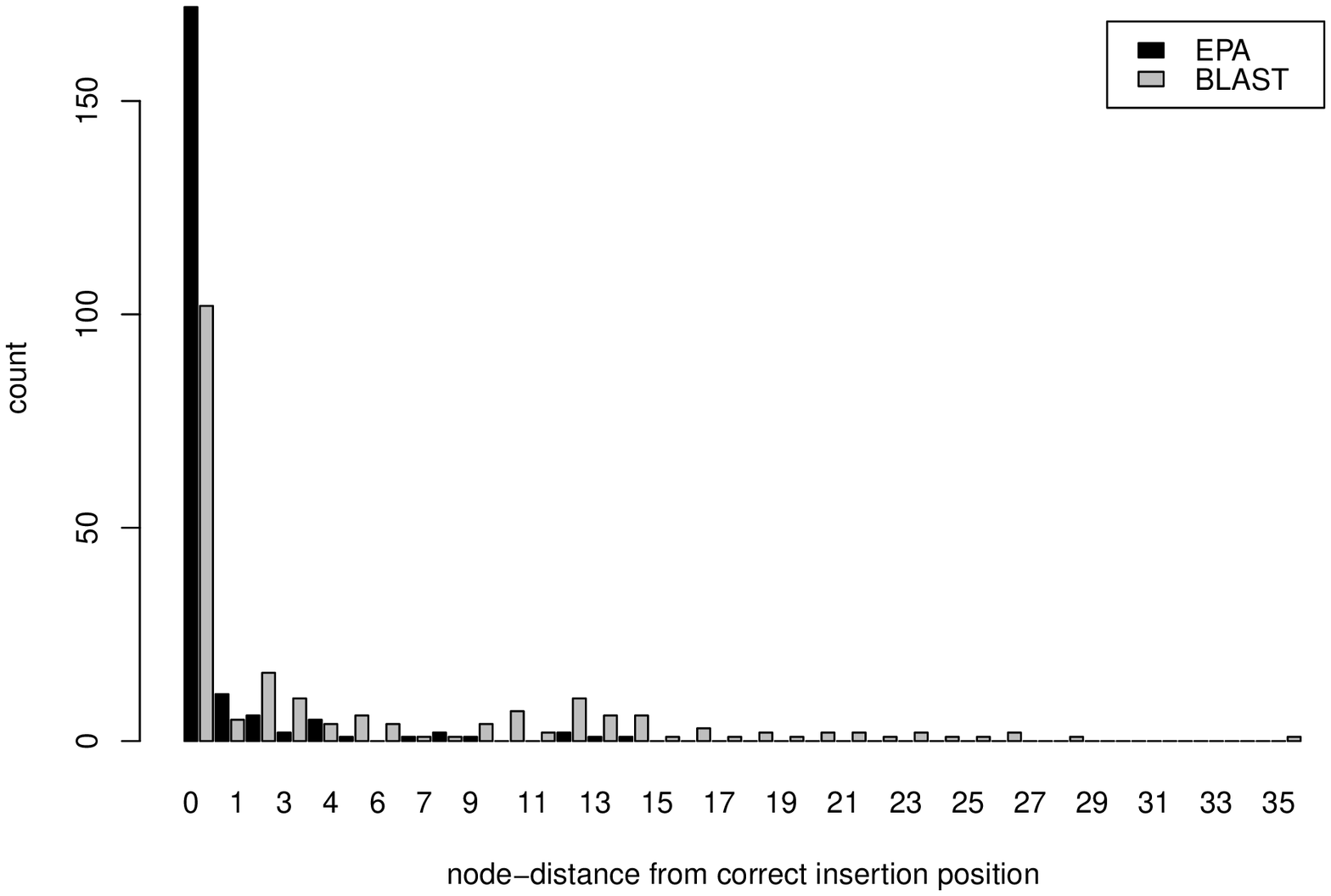}
\includegraphics[width=0.49\columnwidth]{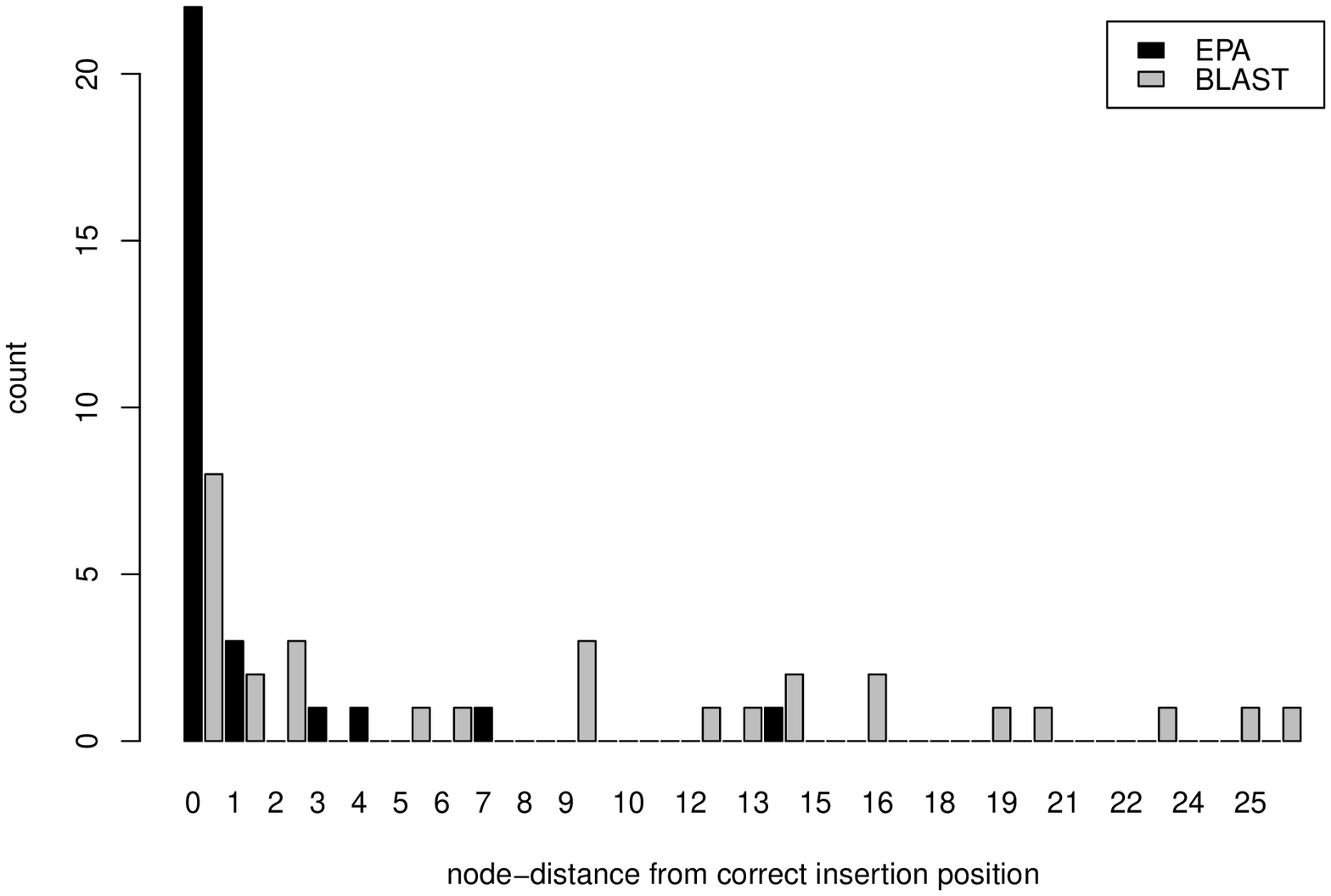}
\caption{Histogram plot of the prediction accuracies (Node Distance) for the placement of 2x100 BP paired-end reads on data set D500.The left plot comprises all QS, the right plot only inner QS.}
\end{figure}

\begin{figure}[ht]
\centering
\includegraphics[width=0.49\columnwidth]{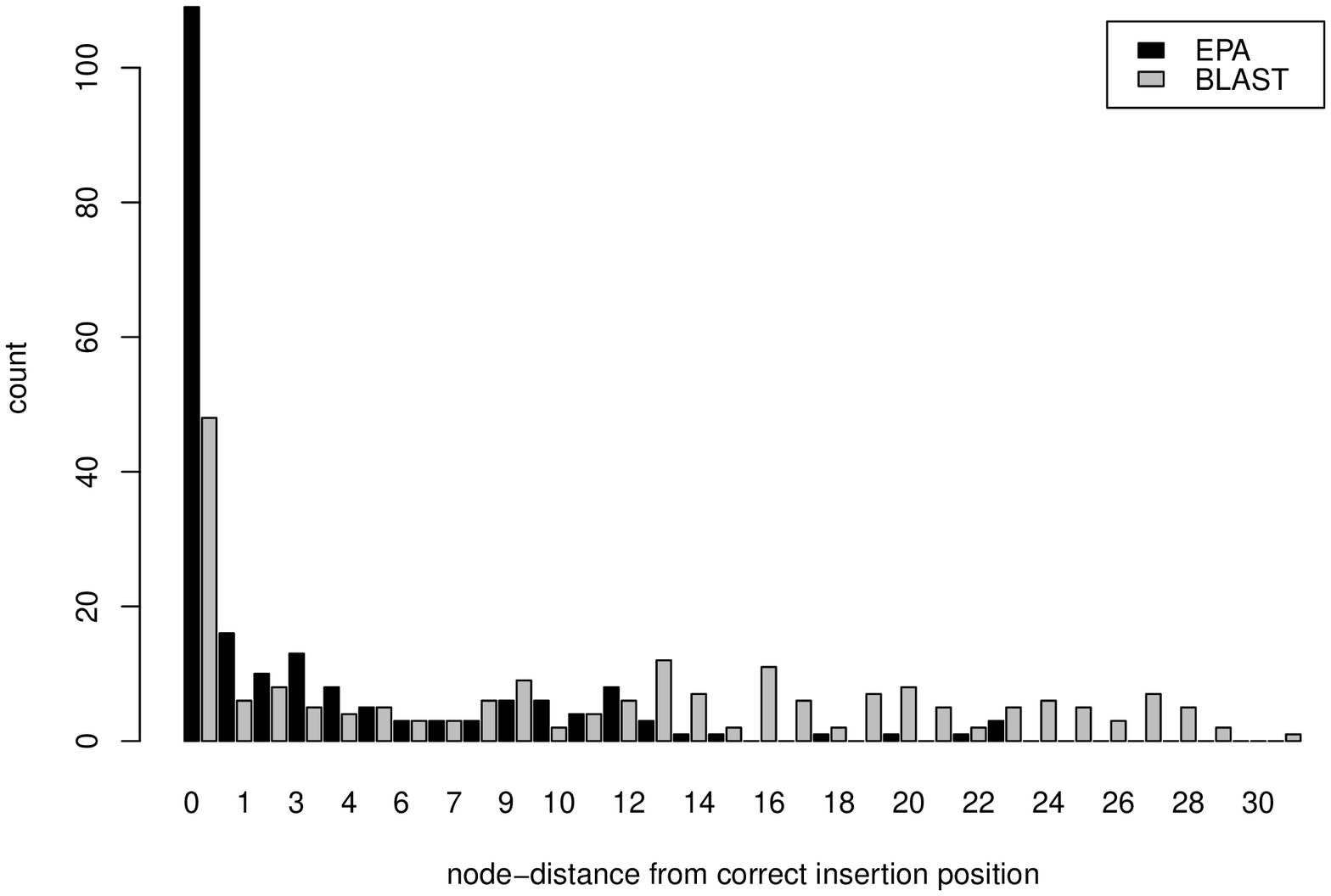}
\includegraphics[width=0.49\columnwidth]{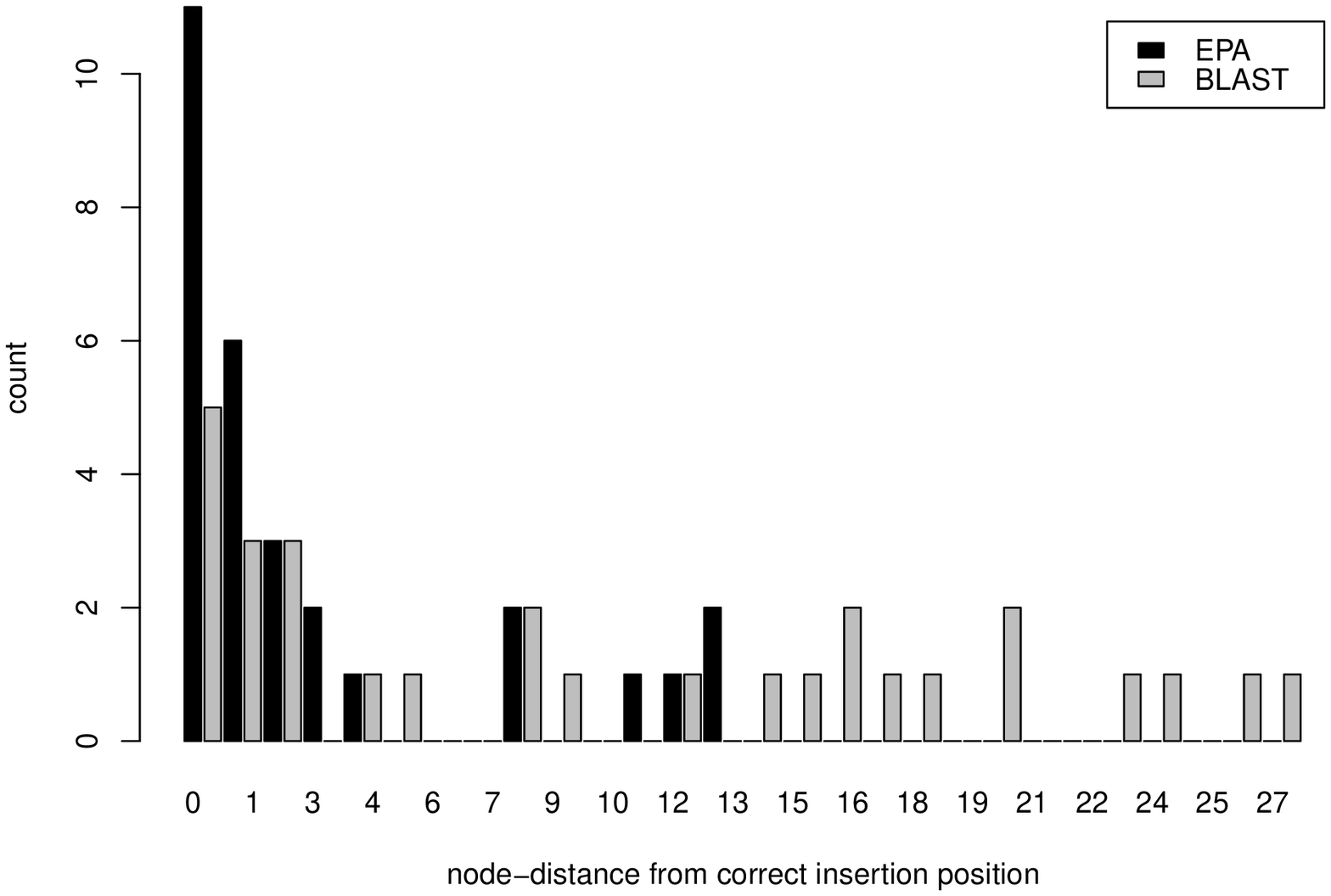}
\caption{Histogram plot of the prediction accuracies (Node Distance) for the placement of 2x50 BP paired-end reads on data set D500. The left plot comprises all QS, the right plot only inner QS.}
\end{figure}


\begin{figure}[ht]
\centering
\includegraphics[width=0.49\columnwidth]{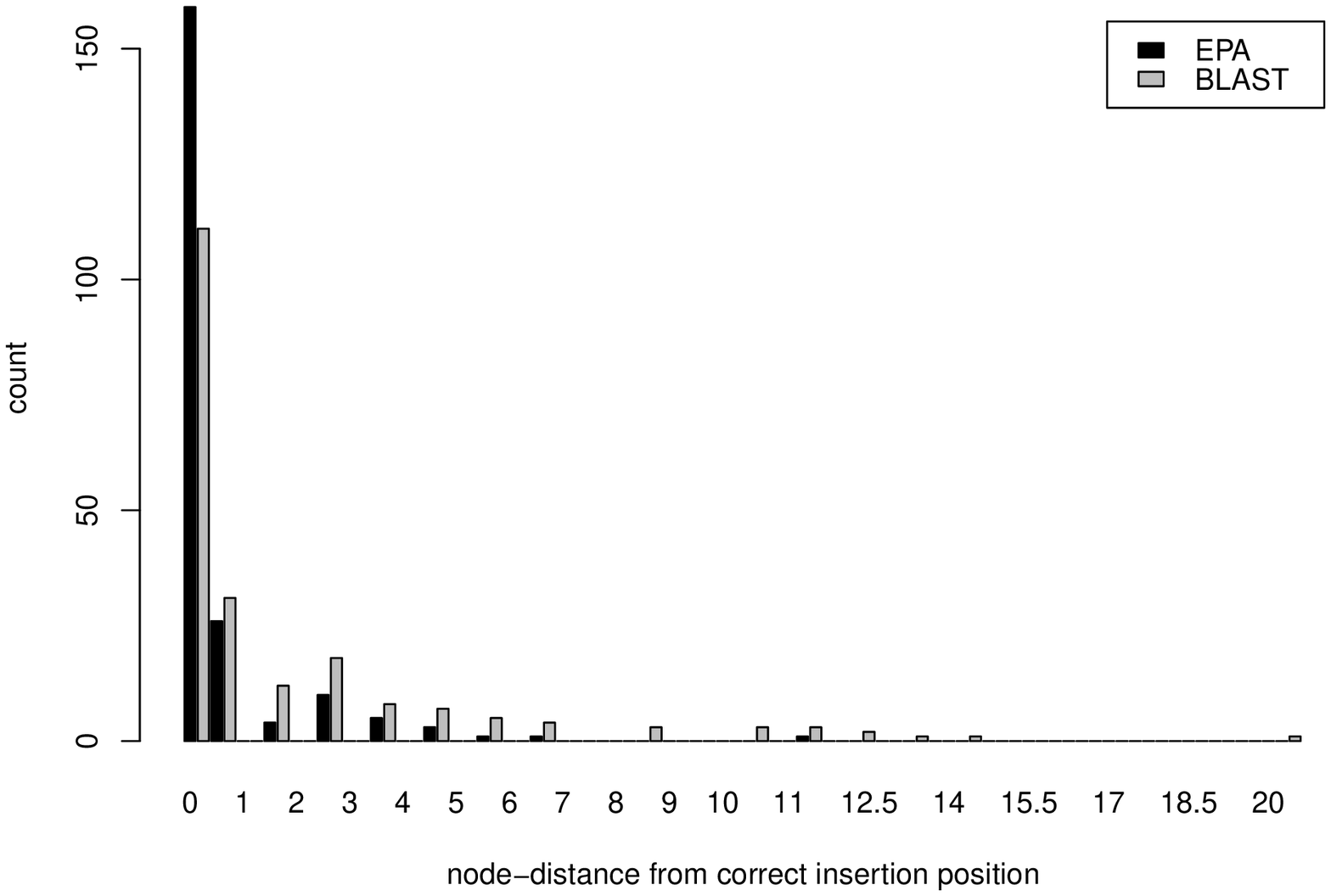}
\includegraphics[width=0.49\columnwidth]{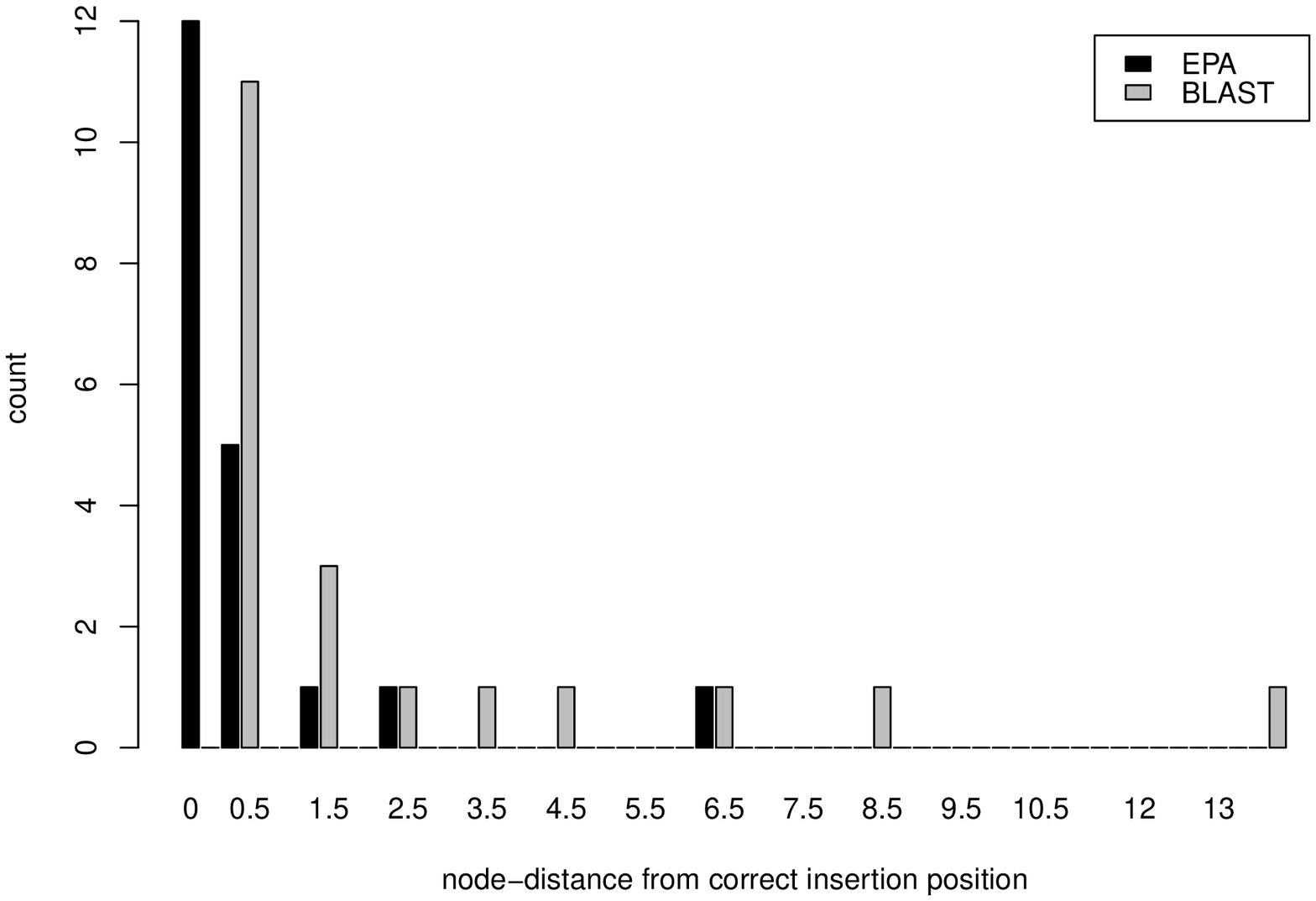}
\caption{Histogram plot of the prediction accuracies (Node Distance) for the placement of 2x100 BP paired-end reads on data set D628.The left plot comprises all QS, the right plot only inner QS.}
\end{figure}

\begin{figure}[ht]
\centering
\includegraphics[width=0.49\columnwidth]{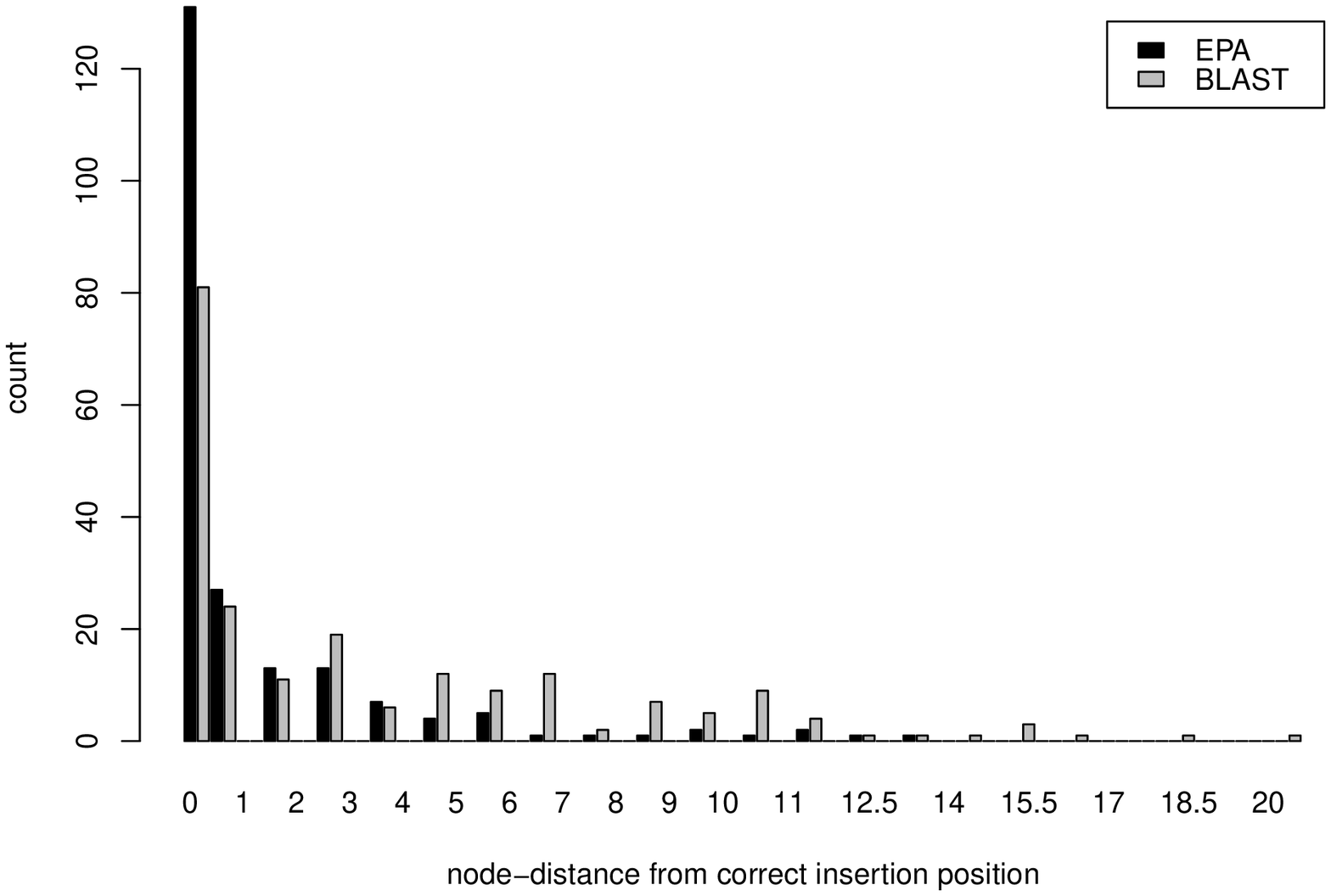}
\includegraphics[width=0.49\columnwidth]{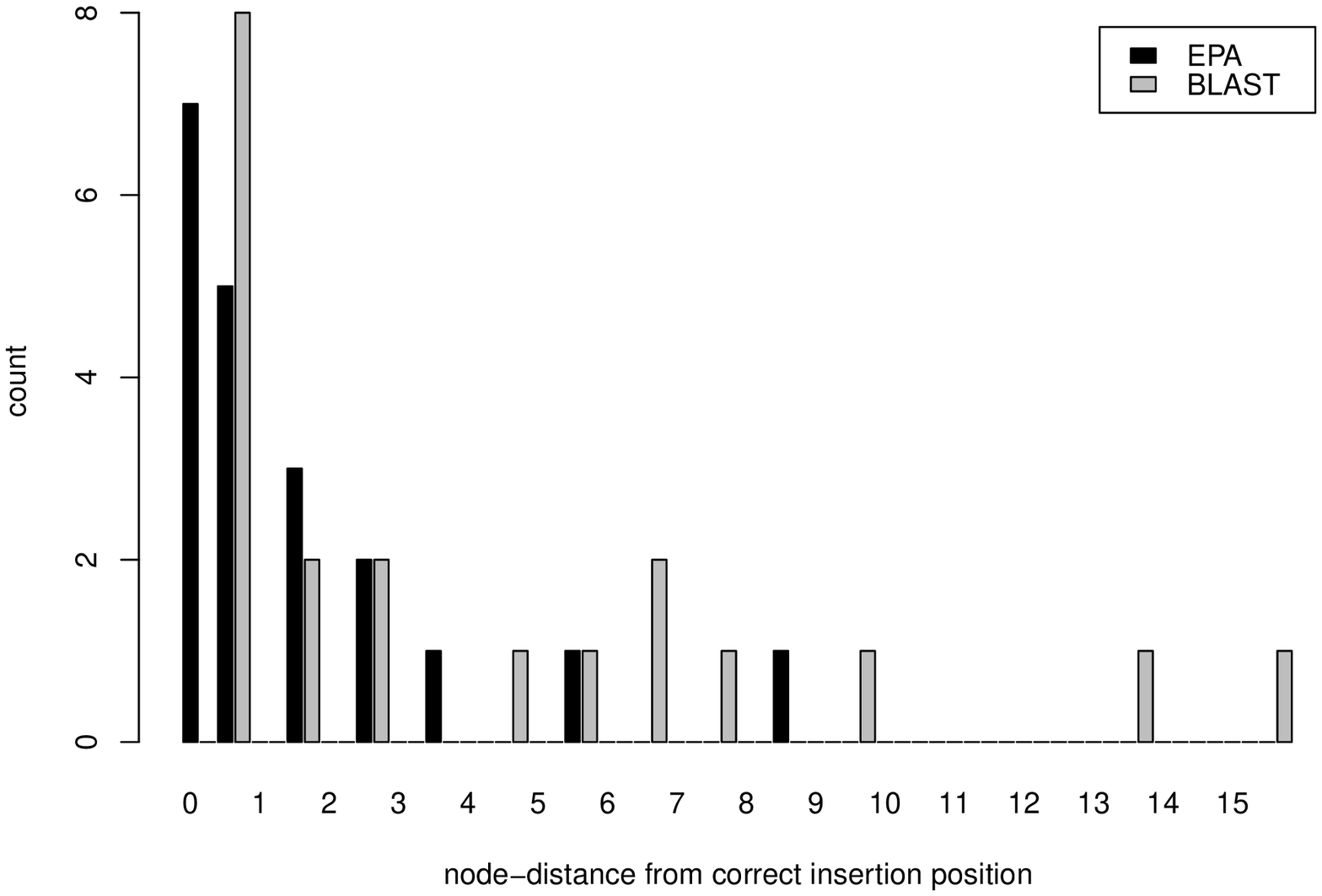}
\caption{Histogram plot of the prediction accuracies (Node Distance) for the placement of 2x50 BP paired-end reads on data set D628. The left plot comprises all QS, the right plot only inner QS.}
\end{figure}


\begin{figure}[ht]
\centering
\includegraphics[width=0.49\columnwidth]{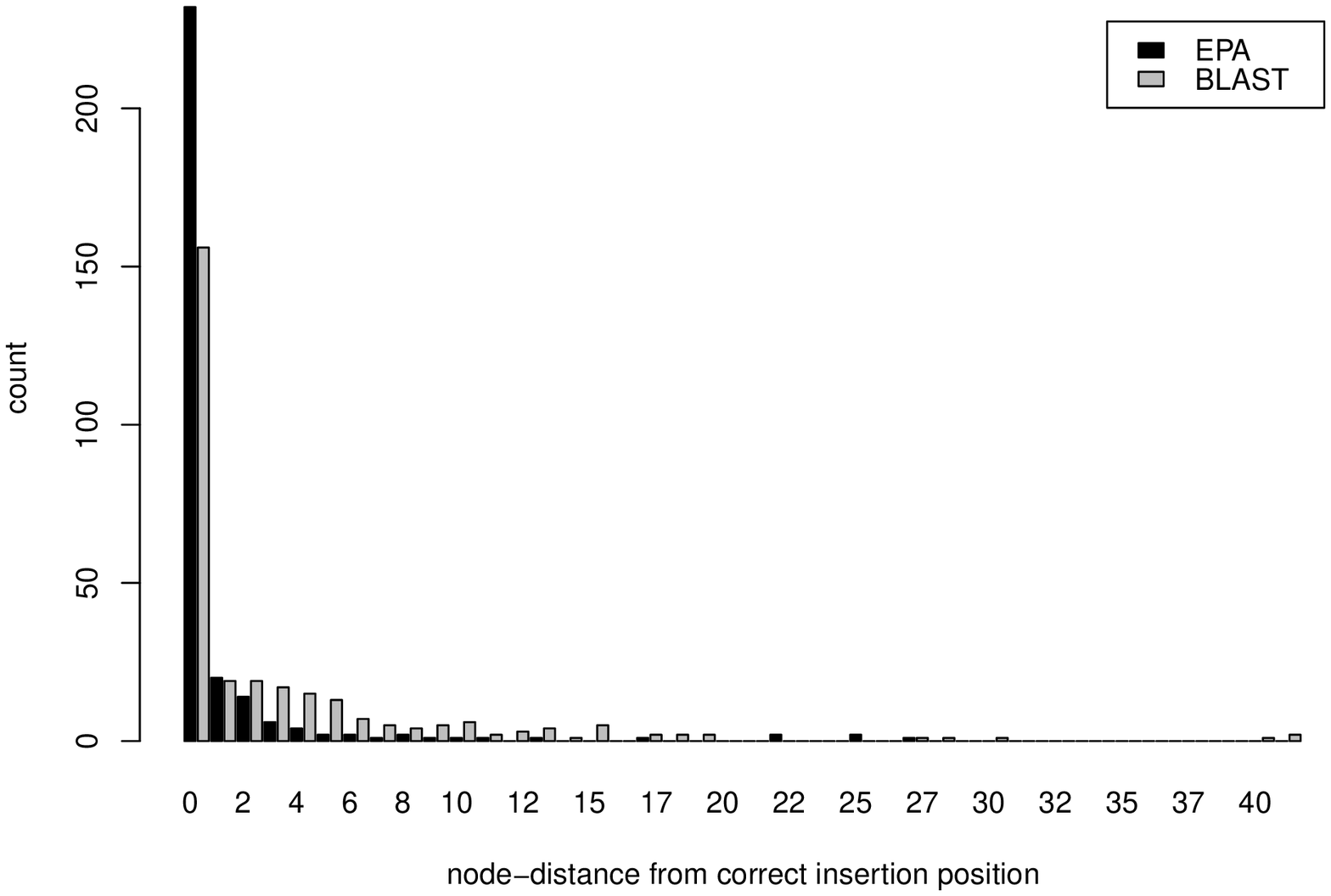}
\includegraphics[width=0.49\columnwidth]{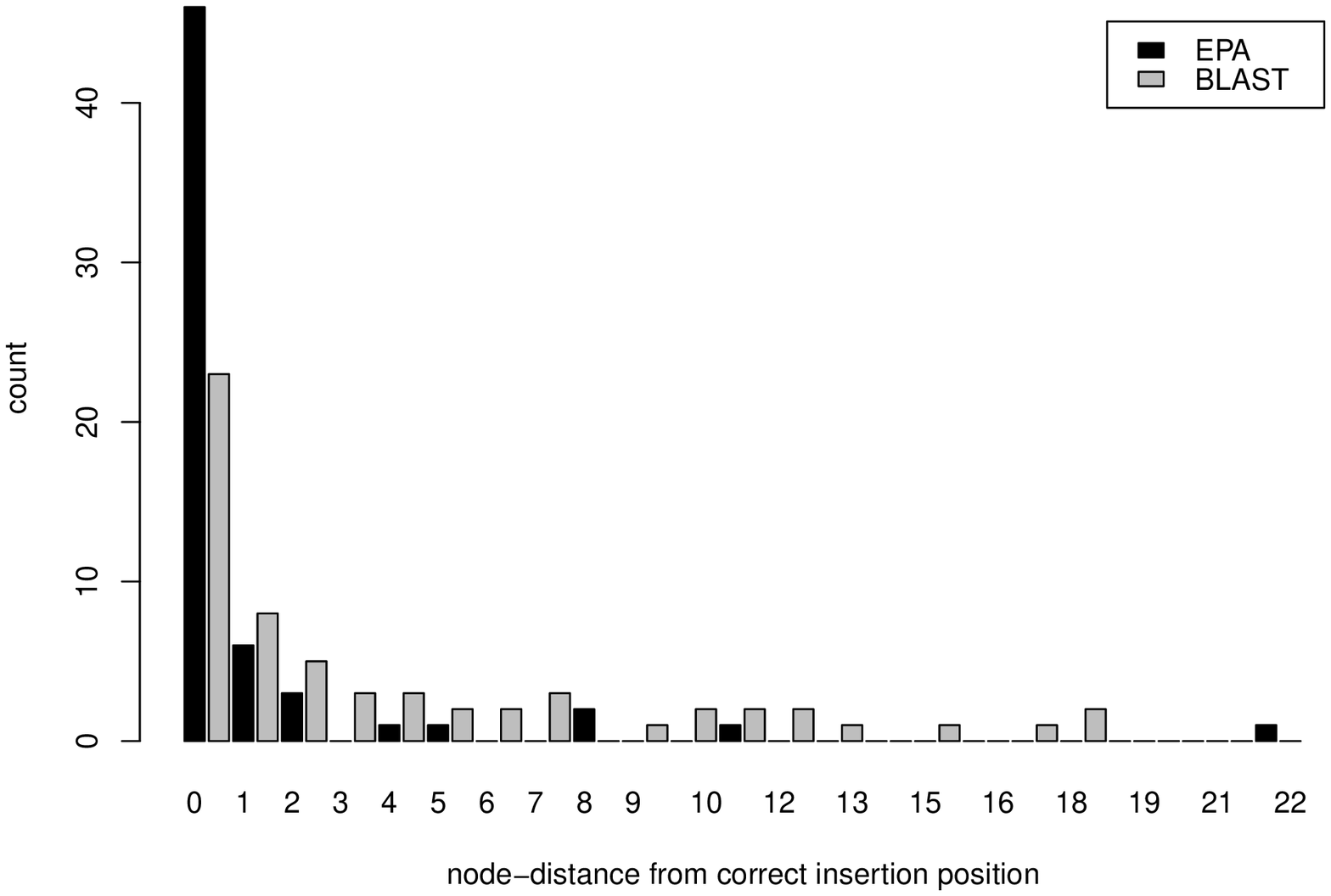}
\caption{Histogram plot of the prediction accuracies (Node Distance) for the placement of 2x100 BP paired-end reads on data set D714.The left plot comprises all QS, the right plot only inner QS.}
\end{figure}

\begin{figure}[ht]
\centering
\includegraphics[width=0.49\columnwidth]{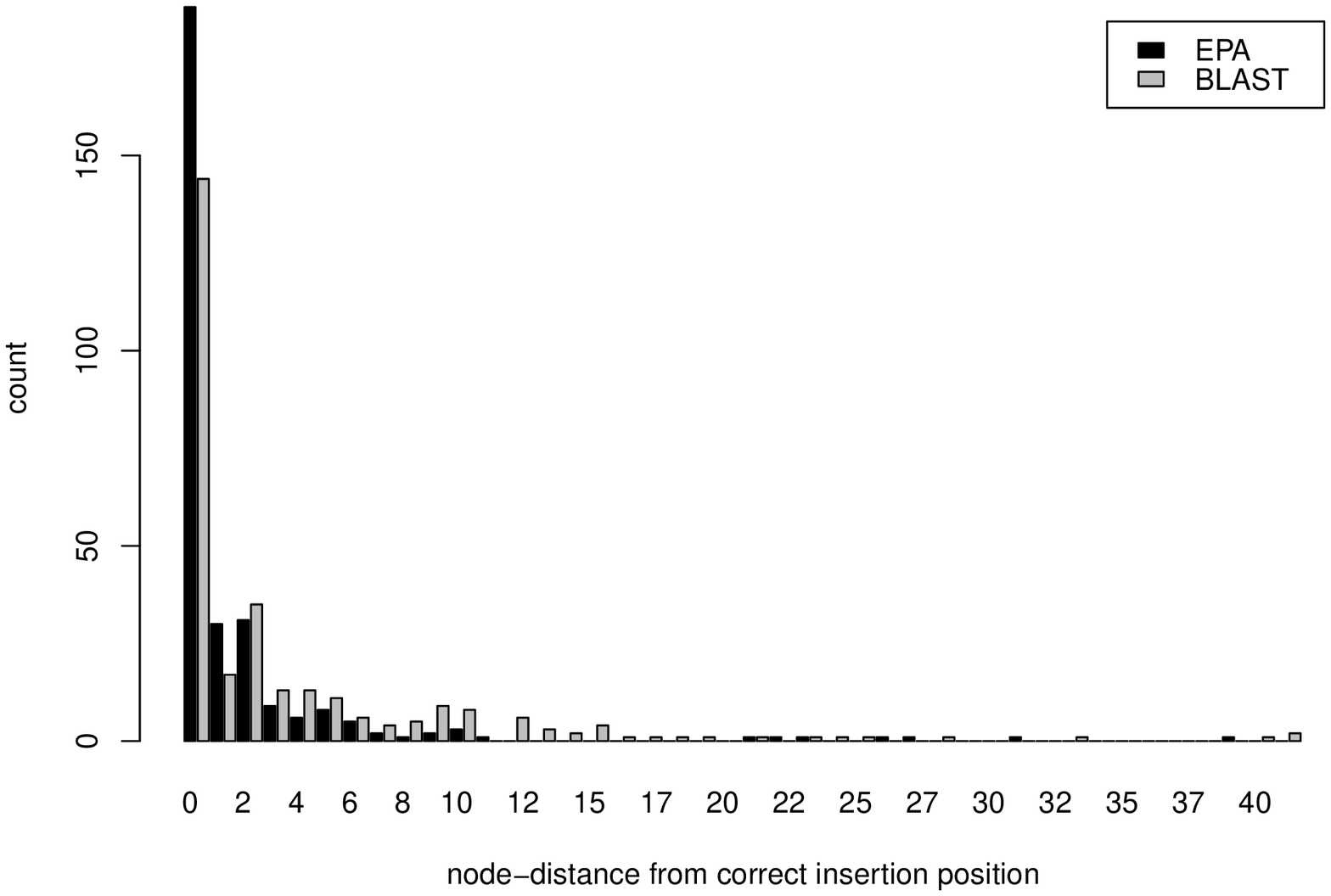}
\includegraphics[width=0.49\columnwidth]{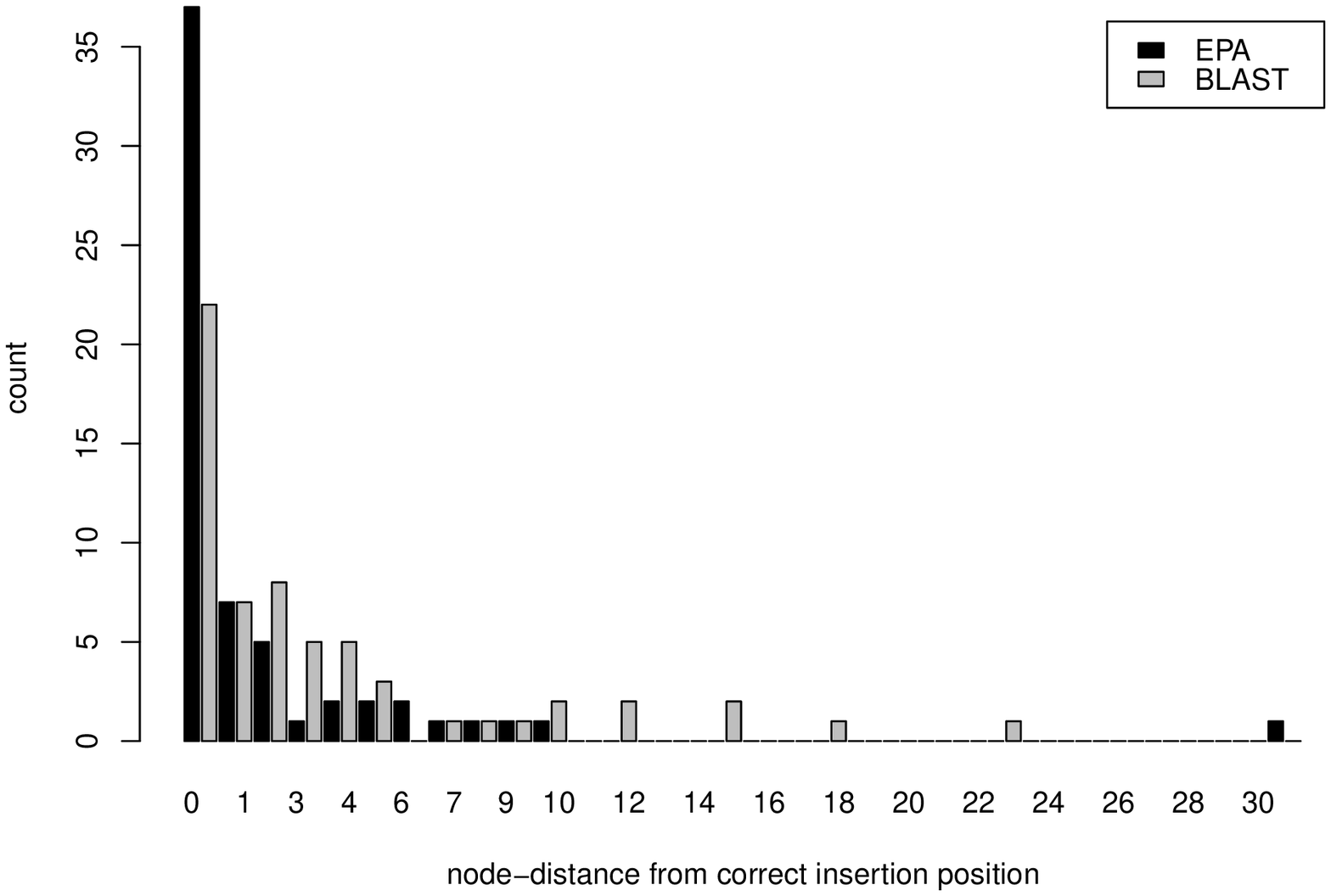}
\caption{Histogram plot of the prediction accuracies (Node Distance) for the placement of 2x50 BP paired-end reads on data set D714. The left plot comprises all QS, the right plot only inner QS.}
\end{figure}


\begin{figure}[ht]
\centering
\includegraphics[width=0.49\columnwidth]{plots_pairedend/raxml_vs_blast_pairedend_nd_855_100_all.eps}
\includegraphics[width=0.49\columnwidth]{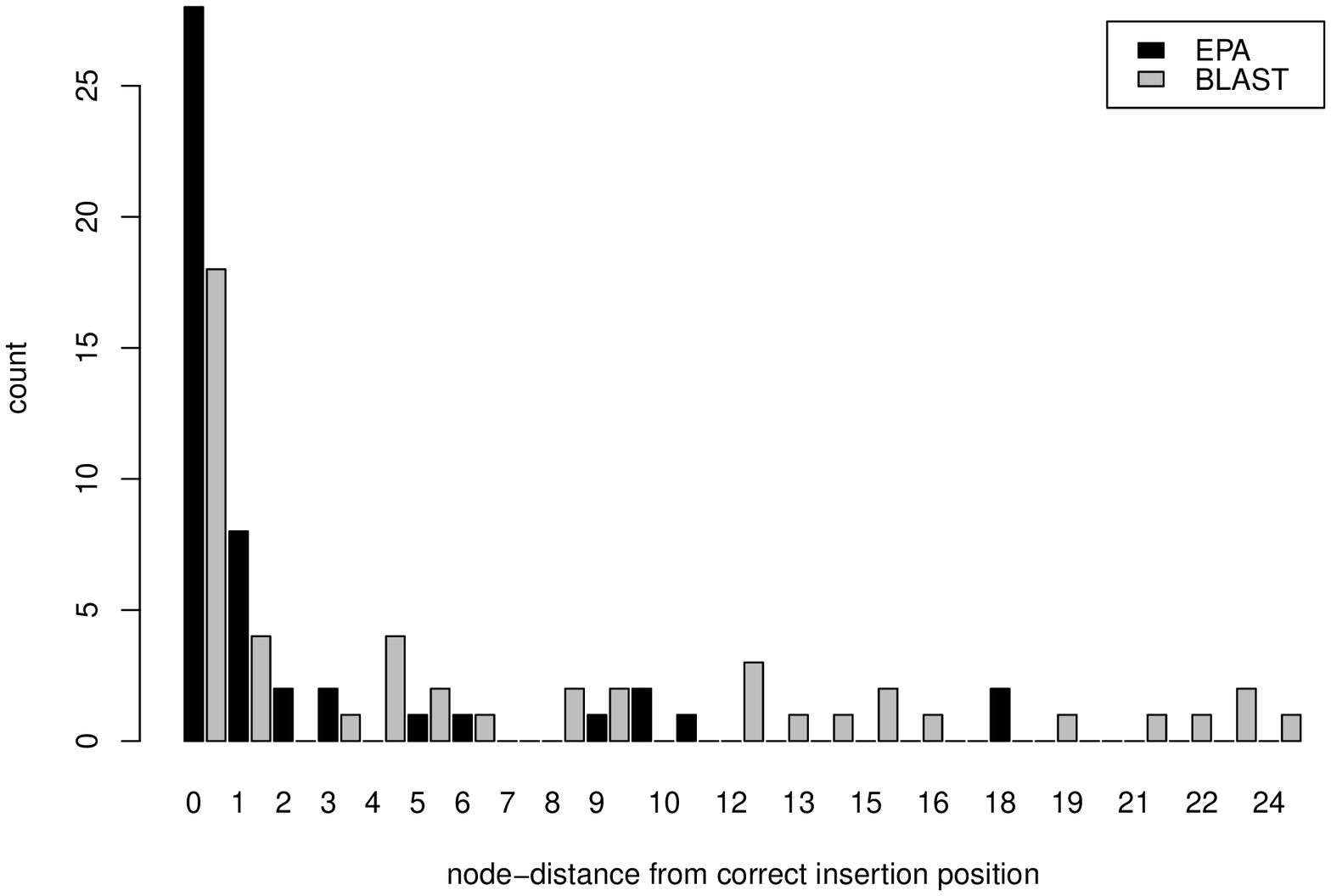}
\caption{Histogram plot of the prediction accuracies (Node Distance) for the placement of 2x100 BP paired-end reads on data set D855.The left plot comprises all QS, the right plot only inner QS.}
\end{figure}

\begin{figure}[ht]
\centering
\includegraphics[width=0.49\columnwidth]{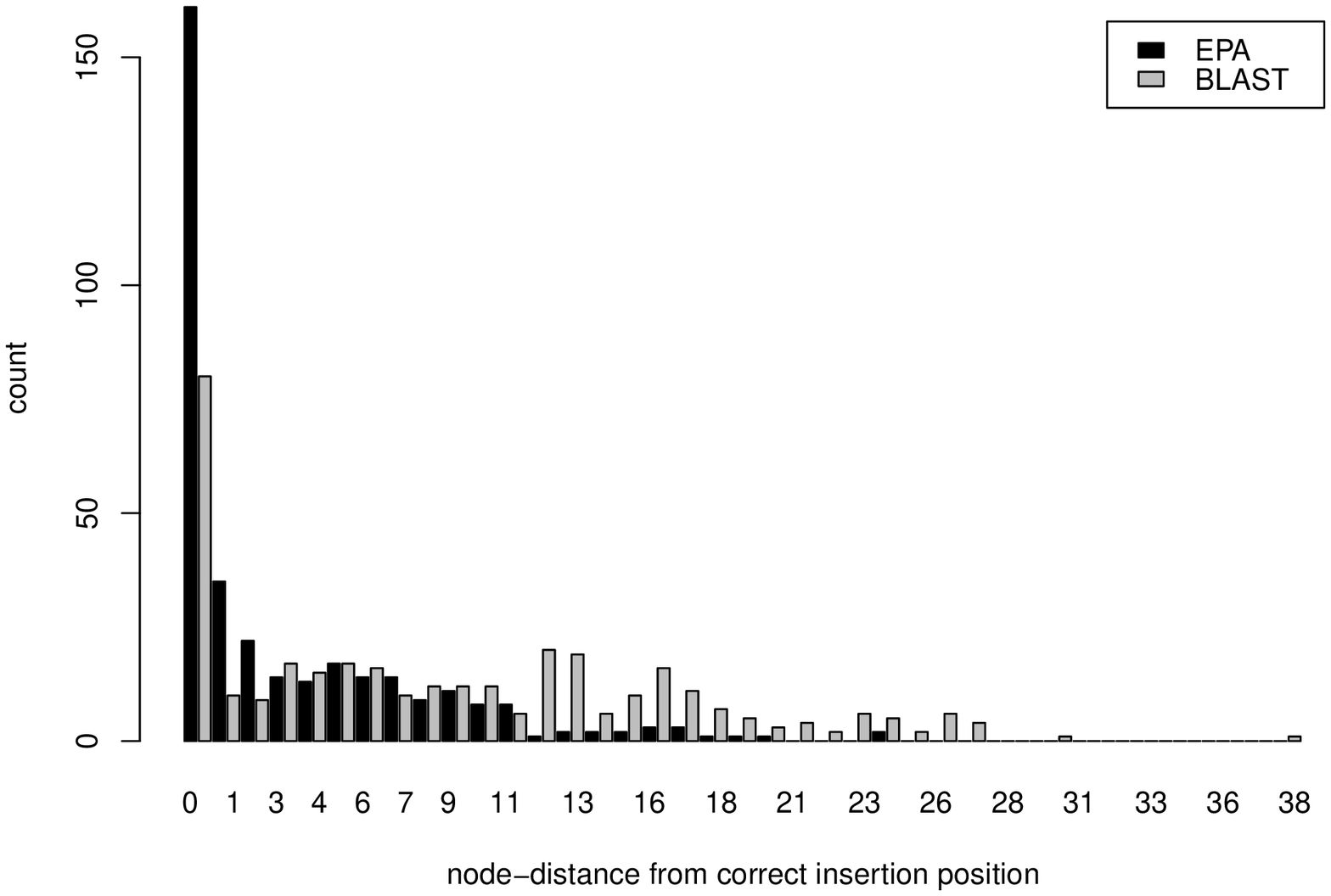}
\includegraphics[width=0.49\columnwidth]{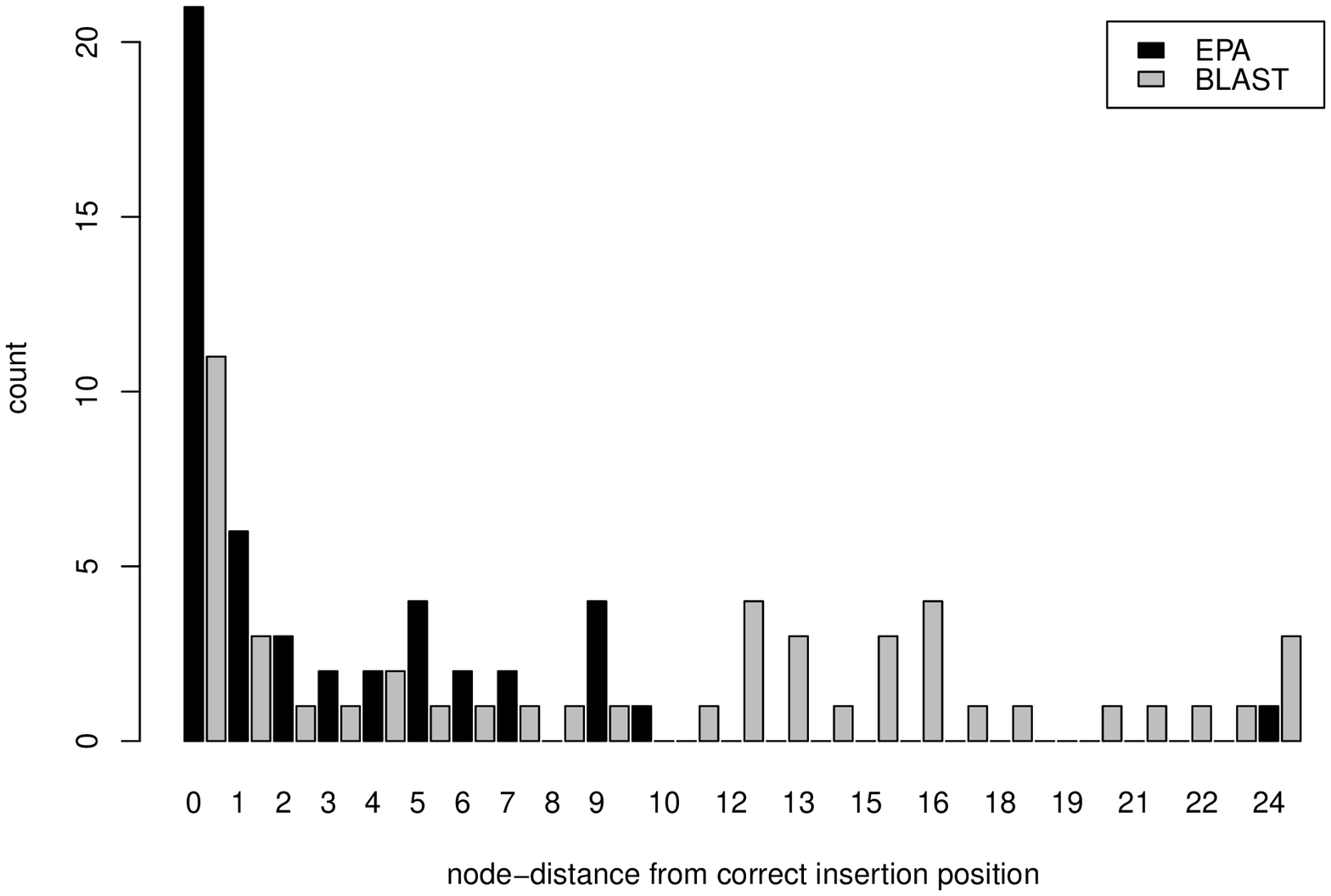}
\caption{Histogram plot of the prediction accuracies (Node Distance) for the placement of 2x50 BP paired-end reads on data set D855. The left plot comprises all QS, the right plot only inner QS.}
\end{figure}


\begin{figure}[ht]
\centering
\includegraphics[width=0.49\columnwidth]{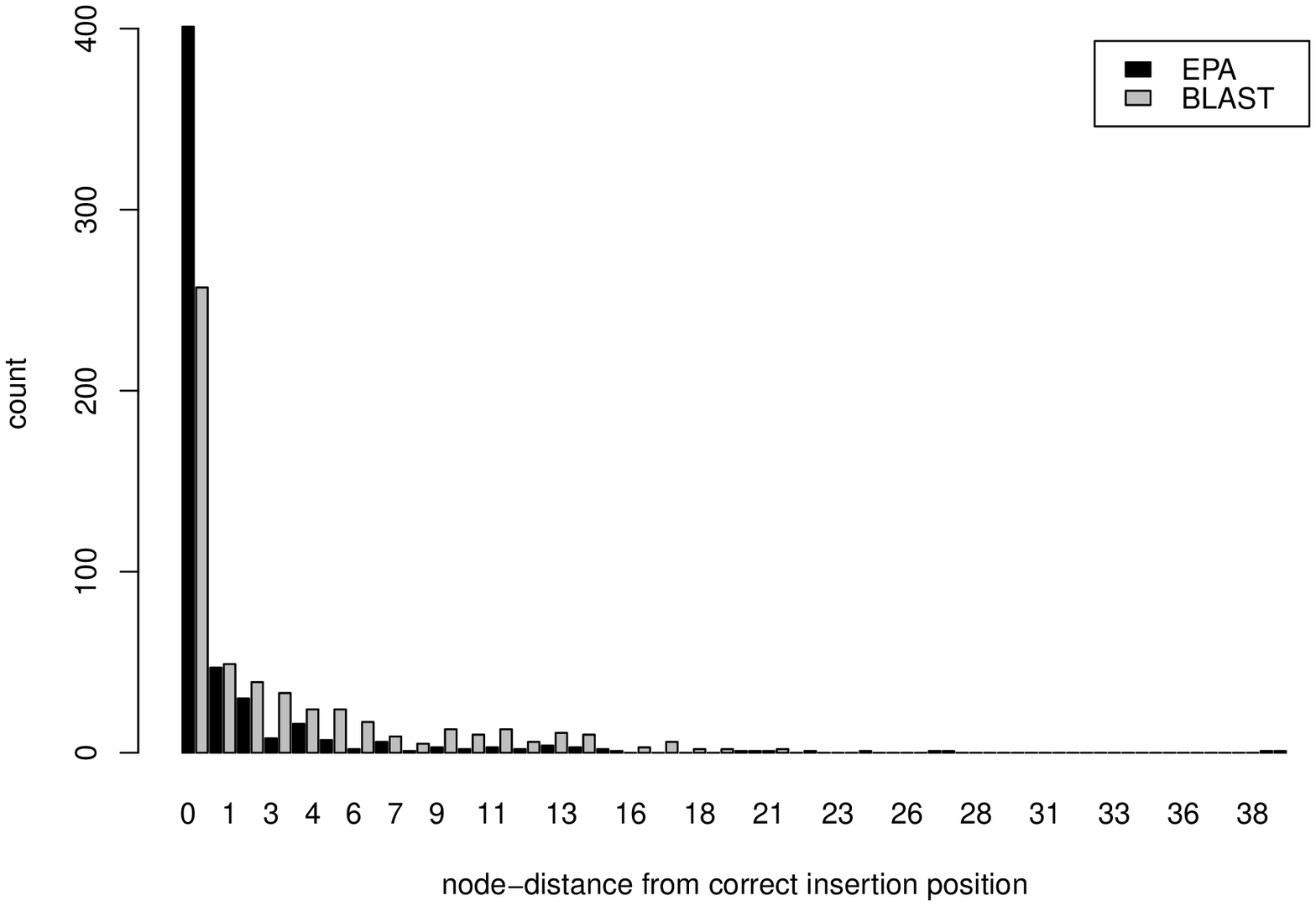}
\includegraphics[width=0.49\columnwidth]{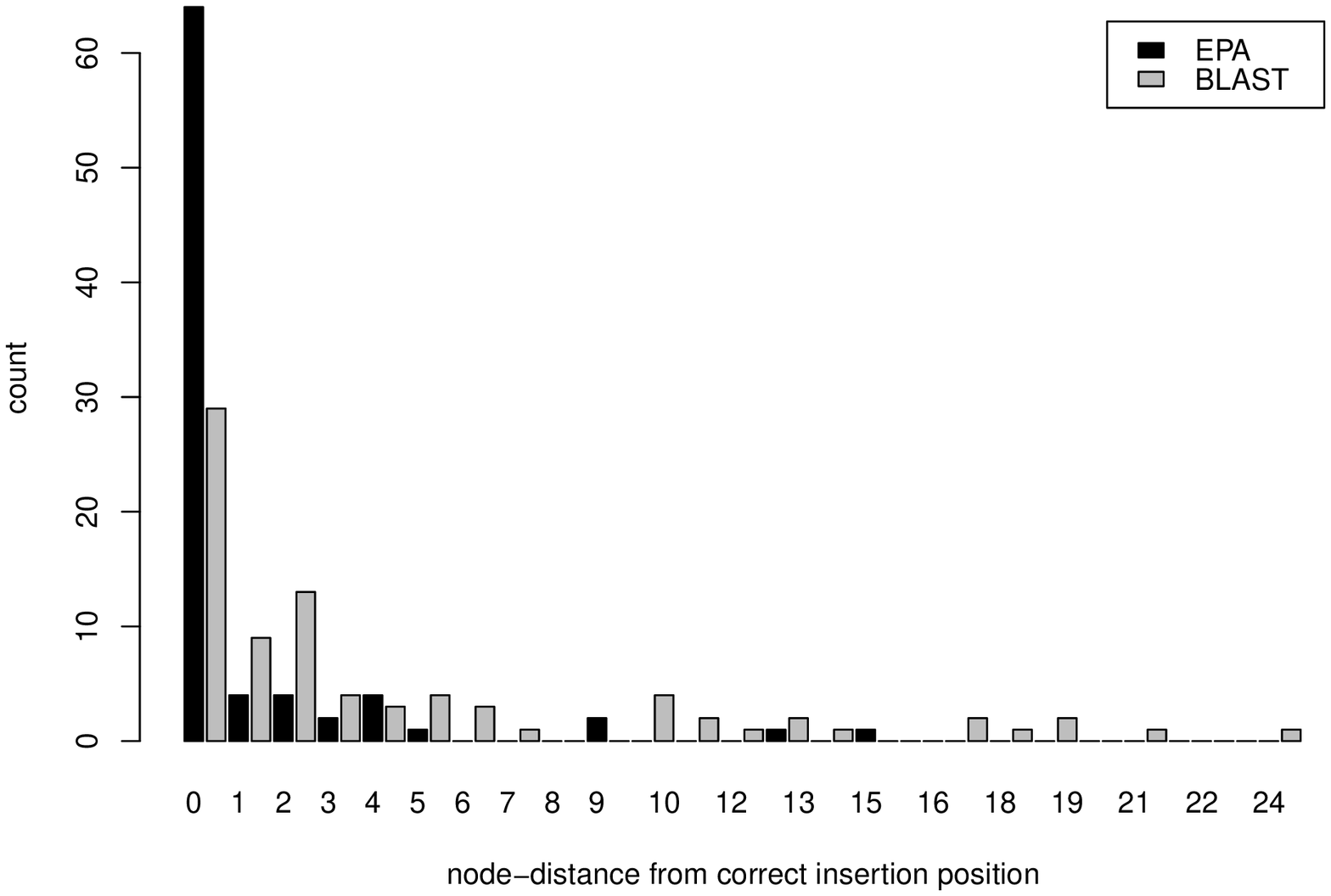}
\caption{Histogram plot of the prediction accuracies (Node Distance) for the placement of 2x100 BP paired-end reads on data set D1604.The left plot comprises all QS, the right plot only inner QS.}
\end{figure}

\begin{figure}[ht]
\centering
\includegraphics[width=0.49\columnwidth]{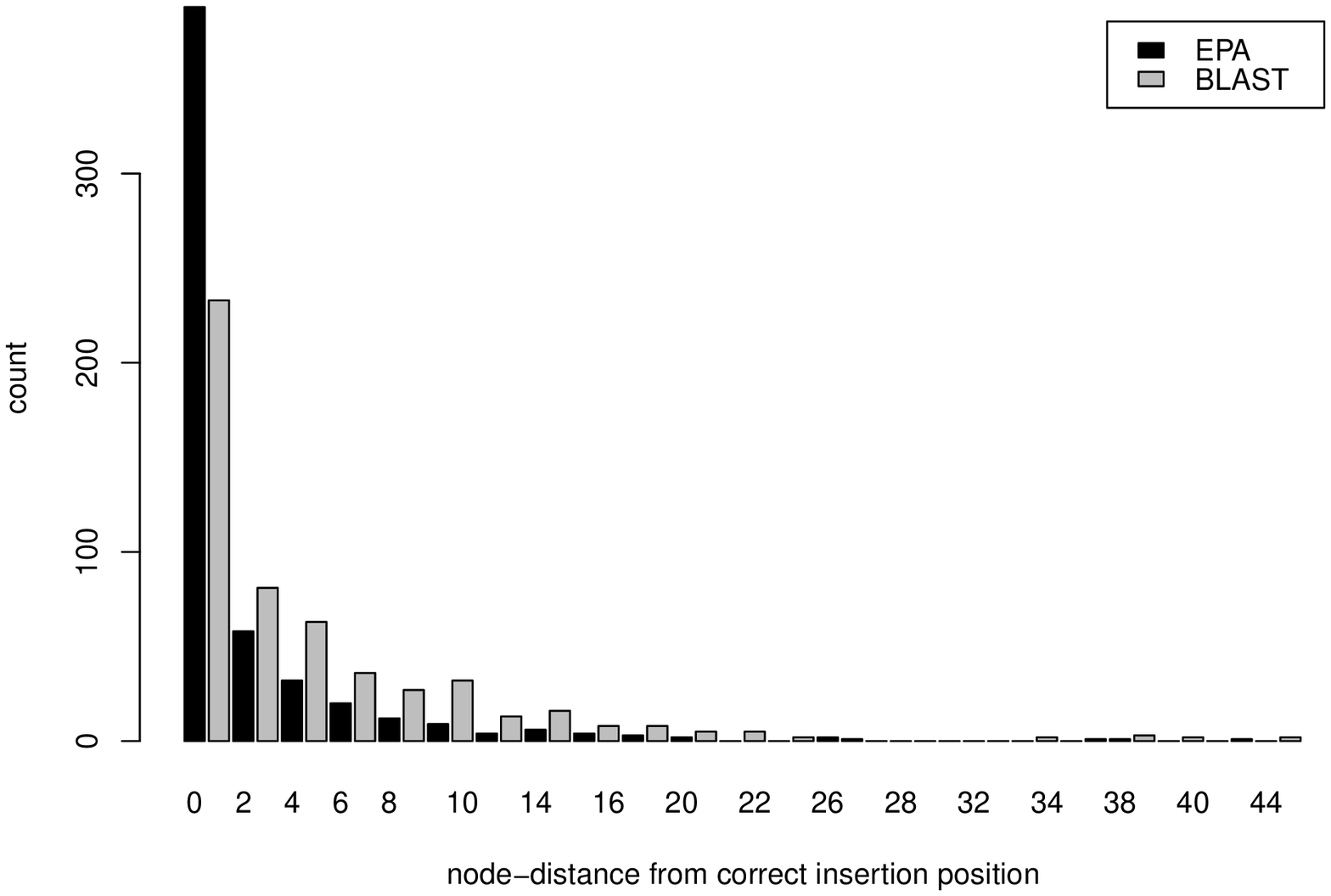}
\includegraphics[width=0.49\columnwidth]{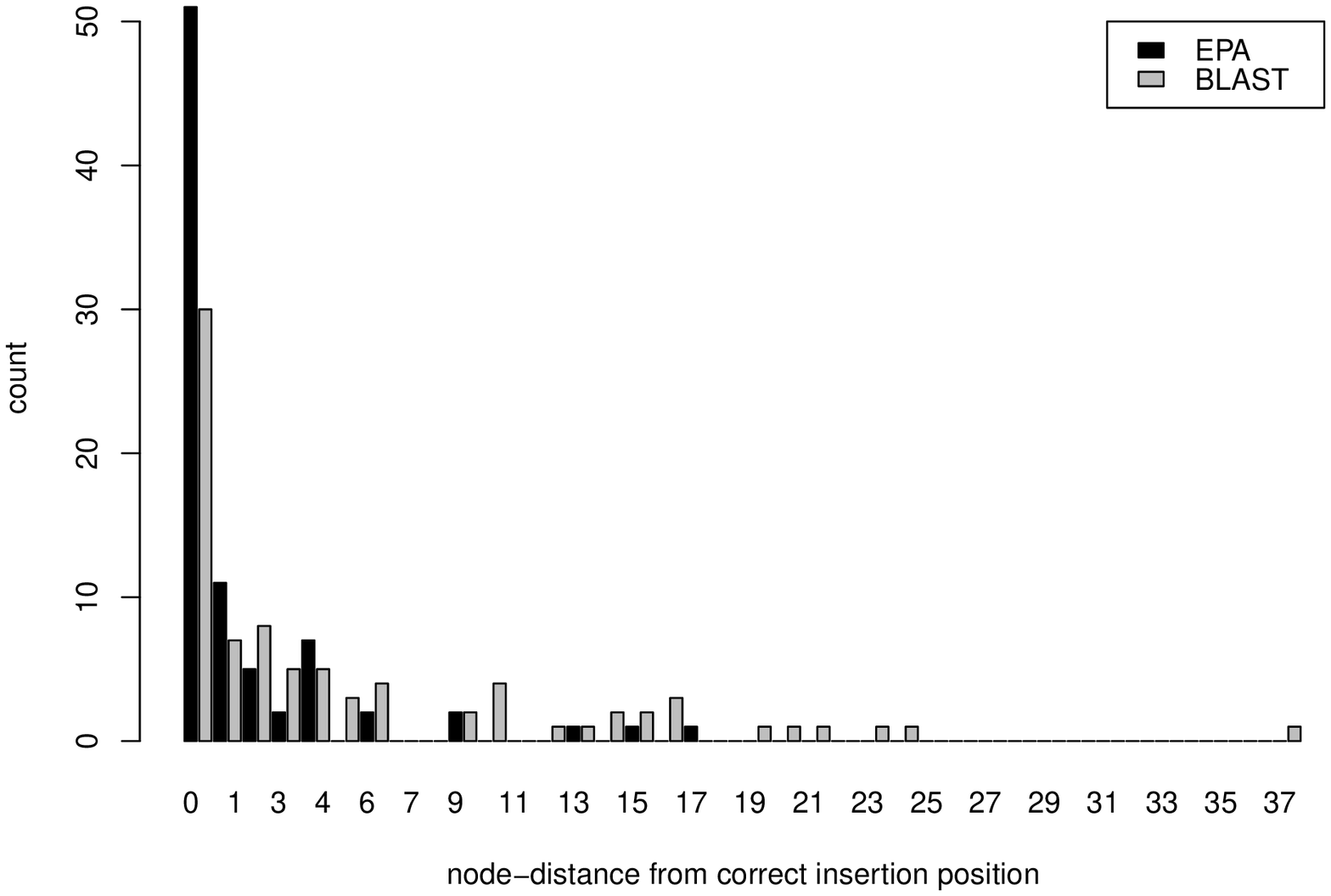}
\caption{Histogram plot of the prediction accuracies (Node Distance) for the placement of 2x50 BP paired-end reads on data set D1604. The left plot comprises all QS, the right plot only inner QS.}
\end{figure}
